\documentclass{aa}  
\usepackage{graphicx}
\usepackage{txfonts}
\usepackage{natbib}
\usepackage{float}
\usepackage{geometry}
\usepackage{afterpage}
\usepackage{pdflscape}
\usepackage{subfigure}
\pdfoutput=1
\begin{document}

   \title{Polarimetry of comets 67P/Churyumov–Gerasimenko, 74P/Smirnova–Chernykh, and 152P/Helin–Lawrence \thanks{Based on observations collected at the European Southern Observatory, Chile, under Programme IDs:089.C-0619 (PI=Tozzi), and 384.C-0115 (PI=Boehnhardt)}}


   \author{A. Stinson \inst{1}\inst{,2} \and 
          S. Bagnulo \inst{1} \and
          G.P. Tozzi \inst{3} \and
          H. Boehnhardt \inst{4} \and  
          S. Protopapa \inst{5}   \and
          L. Kolokolova \inst{5} \and  
          K. Muinonen \inst{6}\inst{,7} \and   
          G.H. Jones \inst{2}\inst{,8} 
          }

   \institute{Armagh Observatory, College Hill, Armagh BT61 9DG, Northern Ireland, UK. {\tt e-mail: ast@arm.ac.uk, sba@arm.ac.uk}
         \and
             Mullard Space Science Laboratory, University College London, Holmbury St. Mary, Dorking RH5 6NT, UK
         \and
             INAF - Osservatorio Astrofisico di Arcetri, Largo E. Fermi 5, I-50125 Firenze, Italy. {\tt e-mail: tozzi@arcetri.inaf.it}        	
         \and
             Max Planck Institute for Solar System Research, Justus-von-Liebig-Weg 3, 37077 G\"{o}ttingen, Germany
         \and
         	Department of Astronomy, University of Maryland, College Park, MD 20742, USA
         \and
         	Department of Physics, University of Helsinki, PO Box 64, Gustaf H\a"{a}llstr\"{o}min katu 2a, 00014, Helsinki, Finland
         \and
             Finnish Geospatial Research Institute FGI, Geodeetinrinne 2, FI-02430 Masala, Finland
         \and
         	The Centre for Planetary Sciences at UCL/Birkbeck, Gower Street, London WC1E 6BT, UK
             }

\date{Received: 2015-11-05 / Accepted: 2016-05-10}
 
  \abstract
   {}
{Polarimetric characteristics of comets at large heliocentric distances is a relatively unexplored area; we extend the idea by adding and analysing the data for three Jupiter family comets (JFCs).} 
{With the FORS2 instrument of the ESO VLT, we performed quasi-simultaneous photometry and polarimetry of three active JFCs 67P/Churyumov-Gerasimenko, 74P/Smirnova-Chernykh, and 152P/Helin-Lawrence.}
   {We obtained in total 23 polarimetric measurements at different epochs, covering a phase-angle range $\sim$1$^\circ$-16$^\circ$ and heliocentric distances from 3 to 4.6\,au. From our observations we obtained both colour and polarimetric maps to look for structures in the comae and tails of the comets.}
   {
   74P/Smirnova-Chernykh and 152P/Helin-Lawrence exhibit enough activity at large heliocentric distances to be detectable in polarimetric measurements. Polarimetric and colour maps indicate no evidence of dust particle evolution in the coma. From near-infrared spectra we find no evidence of water ice in the coma of comet 152P/Helin-Lawrence.
    }
	{}
	{}
	{}
	{}
	{}
   \keywords{Comets:individual: 67P/Churyumov-Gerasimenko - Comets:individual: 74P/Smirnova-Chernykh - Comets:individual: 152P/Helin-Lawrence - Polarization}
   \authorrunning{Stinson et al}
   \maketitle
%

\section{Introduction}

The light scattered by dust particles is linearly polarised by a small amount depending on the properties of the scattering media. By studying the polarised light reflected by cometary dust we can obtain information on the size, shape, and optical properties of particles in the cometary comae \citep[e.g.][]{Zubko2015, Markkanen2015} The method we use to study, from Earth, the scattered light from a solar system body is via photometric and linear polarimetric measurements obtained at different phase angles (the phase angle is angle between the Sun, the target, and the observer). The way the photometric and polarimetric measurements change as a function of phase angle can help us to determine the properties of the scattering medium, be it from a solid surface like an asteroid or the ejected dust from a comet \citep[e.g.][]{Muinonen2012, Muinonen2004, Jewitt2004}. Comets are of particular interest as they are believed to be some of the most primitive objects in the solar system and can give us information about the conditions in the early solar system when they were formed.\\ \indent
Since comets have eccentric orbits, the observable phase angle range is between 0 - 157$^\circ$. At phase angles $\la$ 20$^\circ$ the linear polarisation of cometary dust is usually negative. Negative polarisation means that the preferred direction of oscillation of the electric field vector is in the direction parallel to the scattering plane. This is in contrast to what is expected from the simple single Rayleigh scattering and Fresnel reflection model, which predicts polarisation perpendicular to the scattering plane. Additionally, at very small phase angles $\la$ 2$^\circ$ there may be an intrinsic increase in brightness above the linear brightening with phase angle. Both these phenomena are consequences of the micro-structure of the scattering media. \\ \indent
Polarimetry has been used to classify active comets into two main categories: high and low polarisation comets \citep{Levasseur-Regourd1996}. The distinction between these two categories occurs at phase angles beyond 40 degrees. At this point a fork in the polarimetric phase function occurs where dust-rich comets tend to show a much higher amount of linear polarisation and gas-rich comets tend to exhibit a low amount of linear polarisation. However it has been shown by \citet{Jockers2005} that the classification of comets by their polarisation needs to be considered very carefully as the size of the aperture used and the role of molecular emission in the observed wavelength can play a crucial role in the amount of polarisation measured.\\ \indent
At phase angle ranges $\la$ 20$^\circ$ the bulk of observational data consists of well-sampled data for comets 1P/Halley and C/1995 O1 Hale-Bopp.  From the database of comet aperture polarimetry \citep{Kiselev2006b} only comets 22P/Kopff \citep{Myers1985}, 47P/Ashbrook–Jackson (observations carried out by Jockers et al 1993, unpublished), 67P/Churyumov–Gerasimenko \citep{Myers1984} and more recently by \citet{Hadamcik2010}, 81P/Wild 2 \citep{Hadamcik2003}, and C/1990 K1 Levy \citep{{Renard1992}, {Rosenbush1994}} have been observed within this phase angle range. In addition, low phase-angle measurements of the nucleus of comet 2P/Encke have been presented by \citet{Boehnhardt2008}. Furthermore, numerous polarimetric maps have been obtained for comets e.g. 1P/Halley \citep{Eaton1988}, C/1995 O1 Hale-Bopp \citep{{Hadamcik1997}, {Furusho1999}}, 22P/Kopff, and 81P/Wild 2 \citep{Hadamcik2003}. \\ \indent
Almost all polarimetric observations of comets have been taken at heliocentric distances $<$ 2\,au when comets are more active, and hence brighter meaning they are much easier to observe. Beyond this heliocentric distance they become much harder to observe and it becomes difficult to investigate the properties of the dust. The exceptions to this are observations of comets C/1995 O1 Hale-Bopp at a heliocentric distance between 2.7 and 3.9\,au  \citep{Manset2000} and 47P/Ashbrook-Jackson at a distance of 2.3\,au \citep{Renard1996}.\\ \indent
In this paper we present photometric and linear polarimetric observations of three Jupiter family comets (JFCs): 67P/Churyumov–Gerasimenko (hereafter 67P), 74P/Smirnova–Chernykh (hereafter 74P), and 152P/Helin–Lawrence (hereafter 152P) at large heliocentric distances. Only 67P has been polarimetrically observed previously by \citet{Myers1984} and more recently by \citet{Hadamcik2010}. All three comets have been photometrically observed to varying extents. 67P has been observed and modelled by numerous authors in recent years as it is the target of the European Space Agency (ESA) Rosetta spacecraft. Photometry of 74P has been carried out by \citet{Lowry2001} and \citet{Lamy2011}. \citet{Lamy2011} observations were carried out with the Hubble Space Telescope as the comet was travelling outbound at  a heliocentric distance of 3.56\,au. From these measurements \citet{Lamy2011} were able to derive a nucleus radius of 2.25 $\pm$ 0.1 km, which exhibited an axis ratio $a/b = 1.14$ and a rotational period of 28 $\pm$ 6 hours. On the other hand, 152P has not been observed in as much detail; there is only one publication mentioning photometric observations \citep{Lowry1999}. From these observations \citet{Lowry1999} were able to find an upper limit on the size of the nucleus of 3.3 $\pm$ 0.9 km assuming a standard albedo of 0.04. The rotational period for 152P is unconstrained. 


\section{Observations}

\subsection{Optical photometry and broadband polarimetry}

\begin{table}
\caption{Observing log of SINFONI observations of comet 152P on the night 28 June 2012.}
\centering
\begin{tabular}{c c c c}

\hline\hline
UT start              &    Grating           &   Airmass start          &   Airmass end      \\
\hline
00:27:24 &	 J		& 1.129 	 & 1.117 \\
00:40:12 &	 J		& 1.102 	 & 1.092 \\
01:02:58 &	 J		& 1.063 	 & 1.056 \\
01:14:32 &	 J		& 1.048 	 & 1.042 \\
01:20:10 &	 J		& 1.042 	 & 1.036 \\
01:31:13 &	 J		& 1.031 	 & 1.027 \\
01:37:28 &	 J		& 1.026 	 & 1.022 \\
01:48:39 &	 J		& 1.019 	 & 1.016 \\
01:55:30 &	 J		& 1.015 	 & 1.013 \\
02:06:29 &	 J		& 1.012 	 & 1.011 \\
02:12:06 &	 J		& 1.011 	 & 1.01  \\
02:23:09 &	 J		& 1.01  	 & 1.01  \\
02:37:21 &	 H+K		& 1.012 	 & 1.014 \\
02:48:34 &	 H+K		& 1.017 	 & 1.019 \\
02:54:12 &	 H+K		& 1.019 	 & 1.023 \\
03:05:19 &	 H+K		& 1.027 	 & 1.032 \\
03:10:57 &	 H+K		& 1.032 	 & 1.037 \\
03:22:00 &	 H+K		& 1.043 	 & 1.049 \\
03:27:36 &	 H+K		& 1.049 	 & 1.056 \\
03:38:43 &	 H+K		& 1.064 	 & 1.072 \\
03:44:20 &	 H+K		& 1.073 	 & 1.081 \\
03:55:25 &	 H+K		& 1.091 	 & 1.101 \\
04:01:02 &	 H+K		& 1.102 	 & 1.112 \\
04:12:04 &	 H+K		& 1.125 	 & 1.137 \\

\hline
  \label{tab:sinfoni}%

\end{tabular}
\tablefoot{Integration time 300 seconds, heliocentric distance 3.12\,au, and geocentric distance is 2.23\,au.}
\end{table}

The comets were observed in service mode in February - March 2010 for comet 67P and April-September 2012 for comets 74P and 152P using the FORS2\footnote{http://www.eso.org/sci/facilities/paranal/instruments/fors/} instrument installed on Unit Telescope 4 (UT4) of the ESO Very Large Telescope (VLT) \citep{Appenzeller1998}. The visual UV FOcal Reducer and low dispersion Spectrograph (FORS) is a multi-purpose instrument capable of imaging and spectroscopy, equipped with polarimetric optics that follow the design of \citet{Appenzeller1967}. \\ \indent
The observations for comets 67P, 74P, and 152P consisted of both quasi-simultaneous photometric and linear polarimetric measurements. The photometric observations for 67P consisted of two 60\,s exposures in the \textit{R}-Special filter ($\lambda _0$ = 655\,nm, FWHM = 165\,nm), whereas for 74P and 152P it consisted of four 60\,s exposures, using both the \textit{R}-Special and \textit{v}-high filters ($\lambda _0$ = 557\,nm, FWHM = 123.5\,nm). Owing to the exposure time, differential autoguiding of the telescope at the apparent velocity of the objects was applied to the observations. Between each exposure a different offset was applied to the telescope to ensure that the image of the comet did not fall on the same pixels during each exposure.\\ \indent
The photometric observations were immediately followed by the linear polarisation measurements using the \textit{R}-Special filter only. In our initial set of polarimetric observations we obtained a series of frames with the half-wave plate set at eight different position angles 0 - 157.5$^\circ$ in steps of 22.5$^\circ$ each with an exposure time of  380\,s for 67P, 270\,s for 74P, and 300\,s for 152P. All of the comets' photocentres were found to be brighter than expected, and for 152P we eventually reduced the exposure time and increased the number of exposures to avoid saturation.\\ \indent
67P was observed at seven different epochs giving us access to a phase angle range of 2-15$^\circ$. 74P was photometrically observed at six different epochs and polarimetrically observed at eight different epochs giving us access to a phase angle range of 2-11$^\circ$. 152P was photometrically observed at six different epochs and polarimetrically observed at seven different epochs giving us access to a phase angle range of 3-15$^\circ$. 

\subsection{Infrared integral field spectroscopy}
In addition to the photometric and polarimetric observations of comets 74P and 152P we also obtained infrared spectroscopic observations using the integral field spectrometer, SINFONI, installed at UT4 of the VLT \citep{Eisenhauer2003,Bonnet2004}. The spectra were obtained over the whole night of 28 June 2012 in visitor mode (see Table \ref{tab:sinfoni} for the observing log). For these observations we used the \textit{J} and \textit{H+K} grating corresponding to the wavelength range 1.1 - 2.45 $\mu$m. We also chose the largest field of view of 8$\times$8 arcseconds, which gave us a spatial resolution of 125$\times$250mas per pixel. Adaptive optics were not used for these observations as it is not an option for a moving target. The science observations were carried out using a fixed offset template, that shifted between the sky and the comet in an ABBA pattern where A is the sky and B is the target position of the telescope. The exposure time used for both comets and the on-sky observations was 300\,s. A total of 30 exposures were obtained for comet 152P, 16 in the \textit{J} grating and 14 in the \textit{H+K} grating. For comet 74P a total of 19 exposures were obtained, 12 with the \textit{J} grating and 7 with the \textit{H+K} grating. A number of telluric standard stars prescribed by the SINFONI calibration plan and our chosen solar analogue, Land (SA) 110-361, were observed using the same instrumental settings and at an airmass as close as possible to the comet's airmass to provide the necessary calibration.

\section{Data analysis}
\subsection{FORS data pre-processing}

The polarimetric and photometric images were reduced in a similar way. All images were bias-subtracted by removing a master bias image obtained from a series of five frames taken the morning after the observations. The science images were then flat-field corrected by dividing the images by a flat-field obtained from a series of five sky flat-fields taken during twilight. This process outlines the reduction steps that were applied to all the science images; we detail the different techniques used to analyse the data using this as the starting point.

\subsection{FORS photometry}
\label{sec:phot_meth}
The comets displayed coma and dust tails in the images indicating recent or current activity; this activity made it impossible to distinguish between the signals from the coma and nuclei of the comets. We used an aperture of between 7 - 13 pixels which covers a diameter of 10,000\,km around the coma. This aperture included flux from most of the coma and some of the tail of the comet. The background sky was estimated using a detached annulus at a location close to the comet but free from contamination from the coma, tail, and background stars. \\ \indent
In general, it is not possible to give accurate night-by-night values for the photometric zero point or extinction coefficients because photometric standard stars are not taken by default under the FORS calibration plan for polarimetric observations. Therefore, we used the nightly zero point and extinction coefficient available on the ESO quality control and data processing web page\footnote{http://www.eso.org/observing/dfo/quality/index$\_$fors2.html}. These were calculated using all the photometric standard stars observed over a period of about 28 nights centred at each night under consideration. We assigned an error of 0.05 mag to the magnitude measurements, which is consistent with the uncertainties of the zero points. The errors due to photon noise and background subtractions are negligible in comparison with those of the zero points.\\ \indent
Finally the apparent magnitudes for different epochs were magnitude corrected for Sun and Earth distances of the comet using

\begin{eqnarray}
H=m-5\log _{10}(r \times \Delta)\; ,
\label{eq:abso}
\end{eqnarray}
\noindent
where $m$ is the apparent magnitude, $r$ is the heliocentric distance, and $\Delta$ is the geocentric distance. This corrected magnitude $H$ allows us to investigate how the brightness of the comet changes independently of its distance from the observer while it is still dependent on the phase angle.
We can also use the apparent magnitudes in the \textit{V} and \textit{R} filters to calculate a \textit{V-R} colour magnitude for the comets. This allows us to see whether there are any fluctuations in the colour that would suggest a change in the dust particles emitted. 

In addition to colour fluctuations, we can use the flux from the comets to measure the relative dust production rate. This is done using the quantity $Af\rho$ first defined by \citet{Ahearn1984}; $Af\rho$ is a slightly aperture-dependent quantity roughly proportional to the dust production rate of a comet. It is typically measured in cm and is determined from the observations using

\begin{eqnarray}
Af\rho = \frac{\left(2\Delta r\right)^{2}}{\rho}\, \frac{F_c}{F_\odot}\; ,
\label{eq:afrho}
\end{eqnarray}
\noindent
where $A$ is the geometric albedo of the cometary dust grains, $f$ is the filling factor (i.e. the percentage area covered by the dust), $\rho$ the radius of the field of view used to measure the flux from the comet measured in cm, $r$ is the heliocentric distance measured in au, $\Delta$ is the geocentric distance measured in cm, $F_C$ is the flux measured from the comet, and $F_\odot$ is the solar flux taken using the same filter as the flux measured for the comet. Since we are using a fixed aperture our value for $\rho$ is $5 \times 10^8$\,cm. The photometric results for comets 152P, 74P, and 67P are presented in Table \ref{tab:phot} and the results will be discussed in Sect. \ref{sec:152P_phot}, \ref{sec:74P_phot}, and \ref{sec:67P_phot}.

\subsection{FORS intensity maps}

FORS intensity maps are constructed by stacking all the photometric images taken using the same filter together. We can also use the polarimetric images to create intensity maps; however, owing to the limited field of view of a FORS polarimetric strip, it is usually better to use photometric images if available. Analysis of the coma and tail to search for structure was performed using a combination of numerical techniques and visual inspection. The first of these techniques was the Laplace filtering, which highlights regions of intensity change that can be used to search for localised  activity such as jets \citep{Boehnhardt1994}. The second technique is radial renormalisation  which highlights deviations in the mean coma brightness \citep{Ahearn1986}. The analysis of the structure of each comet is discussed in Sects. \ref{sec:152P_struct}, \ref{sec:74P_struct}, and \ref{sec:67P_struct}.\\

\subsection{Colour maps}
\label{sec:cmap_method}

Aperture photometry gives us information over a large portion of the active region of the comet. To inspect the colour of the coma and search for small-scale features such as jets in the coma we created \textit{V-R} colour maps using the photometric frames analysed in Sects. \ref{fig:cmap_152P} and \ref{fig:cmap_74P}. Here we give details on how these colour maps were obtained.\\ \indent
To reach the highest S/N, for each filter we combined all four photometric maps. We note that each image was obtained with a slightly different telescope offset. Owing to a residual second-order flat-field effect present in all FORS images (a common feature in all focal reducer instruments; \citep{Moehler2010}),  the background spatial behaviour is more regular on each individual image than in the combined image. Therefore, we preferred performing background subtraction on each individual frame prior to image stacking rather than removing the background from the combined image. \\ \indent
To remove the background sky from the photometric images we created a `full resolution' background map using the Source Extractor\footnote{http://www.astromatic.net/software/sextractor}  (SE) software. Not only does this background map estimate the background sky, but it also fits the second-order flat-field effect. We prefer this method of background removal over a simple offset annulus estimation as the annulus only gives an estimation on a very localised spot on the CCD, and the colour map will extend for a much larger distance. The background maps were carefully checked to ensure they contained no contribution from the coma and tail of the comets and then they were simply subtracted from the photometric images. The comet in each photometric image was then shifted to the same position using the photometric centre of the comet. This was done with both the \textit{v}-high and \textit{R}-Special which gave us four \textit{R}-Special and four \textit{v}-high images all centred on the same location. At this point we carefully checked that both the \textit{V}-high and \textit{R}-Special images had similar seeing conditions so as not create artificial structures in the coma. Once this check was made we simply added each set of images together and converted each pixel into an apparent magnitude. Then the \textit{V} and \textit{R} magnitude images were subtracted from each other to create the final \textit{V-R} colour maps which are discussed in Sects. \ref{sec:152P_cmap} and \ref{sec:74P_cmap}. 

\subsection{Polarimetry}

Linear polarimetric images were typically taken after the photometric images. These polarimetric images are the same as the photometric ones except that a retarder wave plate and Wollaston prism are introduced into the optical path, and the light beam is split into two components. Beam overlapping is avoided by introducing a Wollaston mask consisting of nine 22\arcsec\ strips. The resulting image consists of nine pairs of 22\arcsec\ strips. \\ \indent
To calculate Stokes \textit{Q} and \textit{U} from these images we employed the beam swapping technique which significantly reduces the impact of instrumental polarisation \citep{Bagnulo2009}. Reduced Stokes parameters $P_X$ (where \textit{X=Q} or \textit{X=U}) are obtained using

\begin{center}
\begin{eqnarray}
P_X =
\frac{1}{2N}\displaystyle\sum\limits_{j=1}^N\left[\left(\frac{f^\parallel
- f^\perp}{f^\parallel + f^\perp} \right)_{\alpha_j} -
\left(\frac{f^\parallel - f^\perp}{f^\parallel + f^\perp}
\right)_{\alpha_{j+45}}\right]\; ,
\label{eq:pol}
\end{eqnarray}
\end{center}
where $f_\parallel$ is the flux in the parallel beam, $f_\perp$ the flux in the perpendicular beam,  \textit{N}  the number of pairs of exposures, and $\alpha_j$ denotes the position angle of the retarder waveplate. For $\alpha$ = 0$^\circ$, 90$^\circ$, 180$^\circ$, and 270$^\circ$ Eq.(\ref{eq:pol}) gives $P_Q$; for $\alpha$ = 22.5$^\circ$, 112.5$^\circ$, 202.5$^\circ$, 295.5$^\circ$ Eq.(\ref{eq:pol}) gives $P_U$.\\ \indent
Normally during the observations the vertical direction of the instrument field of view is aligned with the north celestial meridian. We thus transform the Stokes parameters into a reference system whose reference direction is perpendicular to the scattering plane, using

\begin{center}
\begin{eqnarray}
P'_Q = \rm{cos}\left(2\left(\varphi+\frac{\pi}{2}\right)\right)\mathit{P_Q} +
\rm{sin}\left(2\left(\varphi+\frac{\pi}{2}\right)\right)\mathit{P_U} 
\label{eq:pol_rot}
\end{eqnarray}
\end{center}
\begin{center}
\begin{eqnarray}
P'_U = -\rm{sin}\left(2\left(\varphi+\frac{\pi}{2}\right)\right)\mathit{P_Q} +
\rm{cos}\left(2\left(\varphi+\frac{\pi}{2}\right)\right)\mathit{P_U}\; ,
\end{eqnarray}
\end{center}
where $\varphi$ is the angle between the direction object-north pole and the object Sun direction \citep{Bagnulo2006}.\\ \indent
In a system composed of randomly oriented particles $P_U$ should be zero for symmetry reasons. However, if the particles are aligned along a certain direction, $P_U$ could deviate from zero by a small amount. Otherwise, the consistency of $P_U$ with zero should be used as a good quality check.

\subsubsection{Aperture polarimetry}
Aperture polarimetry is carried out in a similar way to that described in Sect \ref{sec:phot_meth}. The $f_\parallel$ and $f_\perp$ from Eq.(\ref{eq:pol}) are measured from the polarimetric FORS images by using aperture photometry. As mentioned in Sect. \ref{sec:phot_meth}, the background sky level was estimated using an offset annulus that was close to the comet but far enough away so that influence from the coma, tail, and background stars was at a minimum. The size of the aperture chosen to measure the flux on a given night was based on the error on the measured $P_Q$ and $P_U$, and if the values of $P_Q$ and $P_U$ were not varying with size of aperture used \citep{Bagnulo2011}. The polarimetric results for comets 67P, 74P, and 152P are discussed in  Sect. \ref{sec:pol_phase}.

\subsubsection{Polarimetric map}

\afterpage{
\begin{landscape}
\begin{center}
\begin{table}
\caption{Photometric and polarimetric results obtained for comet 152P, 74P, and 67P. Each magnitude has a standard error of 0.05 magnitude.}
\begin{tabular}{c c c c c c c c c c c c c}

\hline\hline
Target & Date  & Time  & Phase angle & $r$     & $\Delta$  & $R$ & $V$ & $H_{R}$ & $H_{V}$& $V-R$ & $P'_Q$   & $P'_U$\\

    &(dd/mm/yyyy) & hh:mm & (degrees) & au    & au     &  & &  &  && (\%)  & (\%) \\ \hline
    152P & 05/04/2012 & 08:30 & 15.37 & 3.158 & 2.496 & 17.97  & 18.74  & 13.49 & 14.24  & 0.74 & - 0.922 $\pm$ 0.066 & 0.235 $\pm$ 0.065 \\ 
     & 30/04/2012 & 06:15 & 9.80 & 3.139 & 2.239 & 17.51& 18.02 & 13.29 & 13.79 & 0.57 & -1.483 $\pm$ 0.039 & 0.006 $\pm$ 0.039 \\ 
     & 21/05/2012 & 05:45 & 3.38 & 3.127 & 2.126 &  16.95 & 17.49 & 12.84 & 13.37  & 0.54& -1.121 $\pm$ 0.031 & 0.029 $\pm$ 0.031\\ 
     & 23/05/2012 & 03:15 & 2.86 & 3.126 & 2.122 & 16.87& 17.47 &12.87 & 13.36 & 0.60& -1.209 $\pm$ 0.031 & 0.035 $\pm$ 0.031\\ 
     & 15/07/2012 & 03:30 & 14.54 & 3.116 & 2.368 & 17.49& 18.15 & 13.15 &13.81 & 0.66& -1.096 $\pm$ 0.039 & 0.054 $\pm$ 0.040\\
     & 16/07/2012 & 00:57 & 14.91 & 3.116 & 2.368 & -& - & - &- & -& -1.089 $\pm$ 0.038 & 0.117 $\pm$ 0.039 \\ 
     & 24/07/2012 & 02:00 & 16.31 & 3.117 & 2.466 & 17.79& 18.12& 13.16 & 13.69  & 0.54& -0.926 $\pm$ 0.038 &0.040 $\pm$ 0.038\\[2mm] 
    74P & 21/06/2012 & 07:05 & 7.74 & 4.557 & 3.705 & 19.34 & 20.12& 13.20 &13.99 & 0.79&-&-\\ 
    & 26/06/2012 & 09:04 & 6.70  & 4.561 & 3.666 &19.37 & -& 13.26 & - & -&-1.384 $\pm$ 0.087 & 0.182 $\pm$ 0.094\\ 
     & 17/07/2012 & 06:24 & 2.35  & 4.579 & 3.576 &19.00 & 19.68& 12.93& 13.61 & 0.68& -1.231 $\pm$ 0.088 &      0.002 $\pm$ 0.088 \\ 
     & 24/07/2012 & 06:11 & 1.36  & 4.585 & 3.574& 18.89 & 19.58& 12.82 & 13.51 & 0.70& -0.800 $\pm$ 0.078  &  0.031 $\pm$ 0.101\\ 
     & 18/08/2012 & 06:25 & 5.75 & 4.606 & 3.680& 19.41 & 20.11& 13.27 & 13.97 & 0.70&-&-\\ 
     & 19/08/2012 & 02:17 & 5.92 & 4.602 & 3.688 & 19.09& 19.77& 12.94 & 13.62  & 0.68& -1.222 $\pm$ 0.037 & 0.150 $\pm$ 0.037\\
    & 10/09/2012 & 03:55 & 9.72 & 4.602 & 3.920  & -& -& - & -  & -&-1.160 $\pm$ 0.114 &  0.210 $\pm$ 0.115\\
    & 16/09/2012 & 03:56 & 10.47 & 4.627 & 3.999 & -& -& -& -& -&-1.400 $\pm$ 0.132  & 0.001 $\pm$ 0.131 \\[2mm]
    67P & 09/02/2010 & 06:54 & 15.37 & 3.447 & 2.951 & 21.08 & -& 16.05 &- & -&   -0.793 $\pm$ 0.53    &   -0.621 $\pm$ 0.53\\ 
     & 22/02/2010 & 05:41 & 13.14  & 3.523 & 2.850 &21.12 & -& 16.11 & - & -& -1.067 $\pm$ 0.49 & 0.178 $\pm$ 0.49\\ 
     & 06/03/2010 & 08:42 & 10.33  & 3.593 & 2.780& 20.39& -& 15.39 & - & -& -1.116 $\pm$ 0.57 &      -0.227 $\pm $ 0.59\\ 
     & 07/03/2010 & 04:30 & 10.11 & 3.598 & 2.776& 20.95 & -& 15.95 & - & -& -1.959 $\pm$ 0.50  &  -0.621 $\pm$ 0.49 \\ 
     & 09/03/2010 & 06:08 & 9.56 & 3.610 & 2.768 & 20.84 & -& 15.84 & -  & -& -2.031 $\pm$ 0.41      & -0.178 $\pm$ 0.42\\
     & 16/03/2010 & 04:45 & 7.61 & 3.649 & 2.747 & 20.76 & -& 15.76 & -  & -& -1.428 $\pm$ 0.37 & 0.258 $\pm$ 0.38 \\ 
     & 29/03/2010 & 03:50 & 2.72 & 3.721 & 2.744 & 19.85 & -& 14.81 & -  & -& -0.568 $\pm$ 1.17 &  -1.683 $\pm$ 1.18\\
 \hline
  \label{tab:phot}%

\end{tabular}
\end{table}
\end{center}
\end{landscape}
}

Any change in polarimetric characteristics would be due to changes in characteristics of the scattering medium, in our case, in size or composition of the dust particles in the coma. These polarimetric changes can then be compared to the \textit{V-R} colour maps where variations are also related to variations in size or composition of the dust particles.\\ \indent
Creating the polarimetric maps is a much more laborious task than creating the colour maps of Sect. \ref{sec:cmap_method}; we cannot easily remove the background sky from an entire polarimetric FORS image because the background sky is polarised causing a discontinuity of the flux counts in the parallel and perpendicular beam. The SE interpolation algorithm cannot create an accurate estimation of the background sky owing to this striped nature. Depending on the settings used, the interpolation will either contain an averaged out flux count from both strips or will create a transition area where the background flux gradually changes from one extreme to another. Even if we separate the strips into parallel and perpendicular strips the SE interpolation algorithm has trouble creating a background map without containing a contribution from the coma. Hence we have used an offset annulus to calculate the background sky. This annulus should ideally be placed at a pixel location that is free of coma contribution. However FORS suffers from instrumental errors the further you travel from the centre to the edge of the CCD due to a stressed element in the optical train. For this reason the annulus was placed in a location that minimised the instrumental effects and the contribution of the coma. The location of the annulus varied from epoch to epoch due to changes in the extent of the coma. Ideally, we only need the strips in which the target appears; however, we use all the strips to see if the coma extends beyond the strip in which the comet primarily resides. Once the background sky has been removed from these strips, we use Eq.(\ref{eq:pol}) and (\ref{eq:pol_rot}) to create $P_Q$ and $P_U$ maps for each comet. The disadvantage of this method is that in order to utilise all frames obtained at each retarder wave plate position we have to remove the background sky 16 times. If we assume that we incur a small error each time we create a background estimate, many of these small errors could become significant to the accuracy of our results. A better method is to combine the images using Eq.(\ref{eq:pol}) and remove the background sky at the end. This is outlined in the equation

\begin{eqnarray}
P_X = \frac{\left(P_{X}^{\text{tot}} \times I_{X}^\text{{tot}}\right) - \left(P_{X}^\text{{sky}} \times I_{X}^\text{{sky}}\right)}{\left(I_{X}^\text{{tot}} - I_{X}^{\text{sky}}\right)}\; ,
\end{eqnarray}
\noindent
where $P_{X}^{\text{tot}}$ is the total Stokes parameter without any background sky subtraction, $I_{X}^{\text{tot}}$ the total flux counts of the images used to calculate the Stokes parameter, $P_{X}^{\text{sky}}$ the background estimate of $P_{X}^{\text{tot}}$, and $I_{X}^{\text{sky}}$ the background estimate of  $I_{X}^{\text{tot}}$. This method is numerically the same as Eq.(\ref{eq:pol}), but here we only have to create 2 background maps instead of 16 as was done previously, hence the error is reduced.\\ \indent
In the case of comet 74P  we applied a $3 \times 3$ boxcar to our original flux images to increase the S/N of the polarimetric maps. The spatial resolution for 152P is 1668\,km per arcsecond (417\,km per pixel) and 2640\,km per arcsecond (660\,km per pixel) for 74P. We present the polarisation maps of comets 152P and 74P in Sects. \ref{sec:152P_polmap} and \ref{sec:74P_polmap}. We note that we examined the other strips above and below the target and found no evidence of the coma or tail so we only show the strips containing the photometric centre of the comets.

\subsection{IR spectrophotometry}

The infrared spectra obtained by the SINFONI instrument were reduced using the ESO SINFONI reduction pipeline (version 2.3.2), with all the relevant calibration files provided by ESO. The pipeline was also used to extract all the spectra from the data cubes using a six-pixel aperture centred around the approximate photometric centre of the comet. 

The individual spectra extracted in the \textit{J}- and \textit{H+K}-bands were corrected for the exposure time and combined performing a resistant mean  with a threshold of 2.5\,$\sigma$. The same data reduction steps were applied to the solar analogue spectra. The target spectrum was divided by that of the solar analogue  Land (SA) 110-361 observed at similar airmass to correct for telluric lines and remove the Sun's contribution, obtaining in this way the comet relative reflectance spectrum. The latter was normalised to unity at 1.2 micron.  Unfortunately, we do not have near-infrared photometric data to verify the adjustment of the separate spectra taken in the J and H+K wavelength bands. We therefore need to rely on our data processing. In Sect. \ref{sec:SI} we present the relative reflectance spectrum of 152P only, since the signal-to-noise ratio of the comet 74P spectrum was too low to yield any useful information.

\section{Results}
The results of the photometry and polarimetry for the comets are listed in Table \ref{tab:phot}. Section \ref{sec:152P} is dedicated to the results of comet 152P, Sect. \ref{sec:74P} to comet 74P, and Sect \ref{sec:67P} to comet 67P. In Sects. \ref{sec:152P}, \ref{sec:74P}, and \ref{sec:67P} we present the results of aperture photometry, $Af\rho$, colour maps  and polarimetric maps. In Sect. \ref{sec:pol_phase} we present the aperture polarimetry for 152P, 74P, and 67P.

\subsection{Comet 152P/Helin-Lawrence}
\label{sec:152P}
\subsubsection{Aperture photometry}
\label{sec:152P_phot}
\begin{figure}[h]
\resizebox{\hsize}{!}{\includegraphics[scale=1.0]{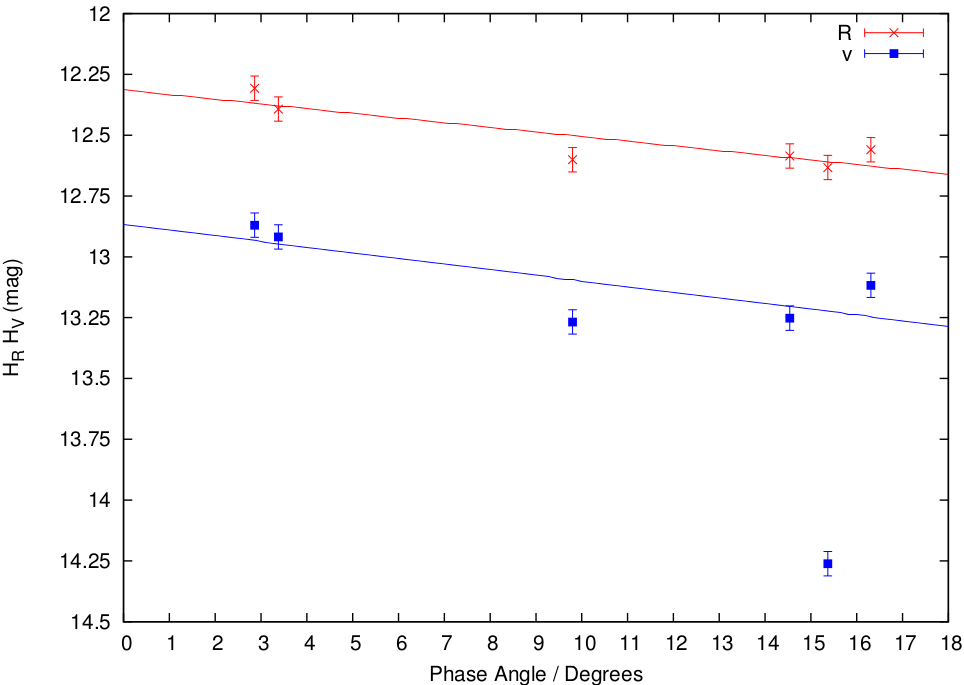}}
\caption{Magnitude corrected for the Sun and Earth distances of comet 152P as a function of phase angle.}
\label{fig:abso_152P}
\end{figure}

In Fig. \ref{fig:abso_152P} we plot the magnitude corrected for the Sun and Earth distances of comet 152P as a function of phase angle. If we ignore the \textit{v}-high result on the night of 5 April 2012 and use a straight line fit, the extrapolated average brightness at zero phase angle are 12.79 $\pm$ 0.13 and 13.32 $\pm$ 0.12 in the \textit{R} and \textit{V} filters assuming no opposition surge. This results in an average \textit{V-R} colour of 0.53$\pm$ 0.18, which is equivalent to a spectral gradient of 18.03\,\%/100\,nm. \\ \indent
\citet{Lowry1999} photometrically observed 152P in the \textit{V} and \textit{R} filters and found that 152P had a very red colour of 0.77 $\pm$ 0.12 (47\,\%/100\,nm) when the comet was at 4.6\,au from the Sun. Within the error our results are consistent with both measurements.\\ \indent
Using Eq. (\ref{eq:afrho}), the flux extrapolated back to zero phase in the \textit{R}-Special filter and the average $r$ and $\Delta$ distances to the comet yields an $Af\rho$ value of 194.9 $\pm$ 22.0\,cm. This compares to the value of 56.4 $\pm$ 3.6 cm measured by \citet{Lowry1999}, which suggests the comet exhibits less activity beyond 4\,au.



\subsubsection{Intensity maps}
\begin{table}
\begin{center}
\caption{\label{tab:152P_FIN} Comparison between measured position angle and Finson-Probstein synchrone (sync) analysis at 30, 60, 120, 240, and 360 days for comet 152P.}
\begin{tiny}
\begin{tabular}{ccccccc}

\hline \hline

Date &	PA 	& PA  &	PA  &	PA  &	PA  &	PA  \\ 
	& Tail & Sync & Sync & Sync & Sync & Sync \\
	& measured	& 30 d 	& 60 d 	& 120 d &	240 d &	360 d \\ 
(UT)&	(deg) &	(deg)&	(deg)&	(deg)&	(deg)&	(deg)\\ \hline

2012-Apr-05	& 279-283	& 276	&276	&280	&282	&283 \\
2012-Apr-30	& 278-283	& 273	&275	&280	&282	&283 \\
2012-May-21	& 278-283	& 267	&274	&280	&283	&285 \\
2012-May-23	& 278-282	& 266	&274	&280	&283	&285 \\
2012-Jul-15	& 290-95	    & 102	&99	&338	&295	&293 \\
2012-Jul-24	& 290-100	& 101	&96	&24	&297	&294 \\ \hline

\end{tabular}
\end{tiny}
\end{center}
\end{table}

\label{sec:152P_struct}
\begin{figure}[htp]
\centering
\subfigure{\includegraphics[scale=0.415]{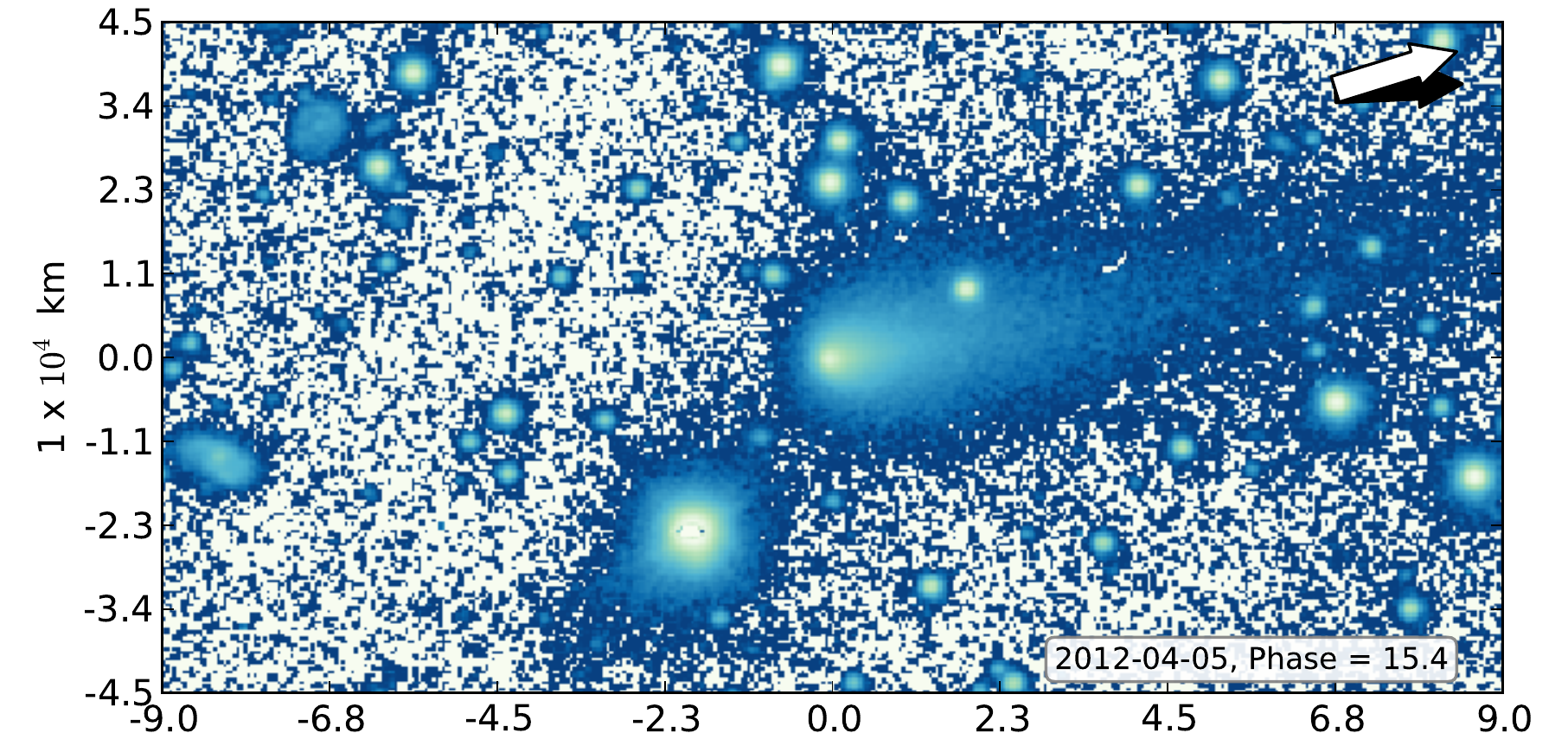}}
\subfigure{\includegraphics[scale=0.415]{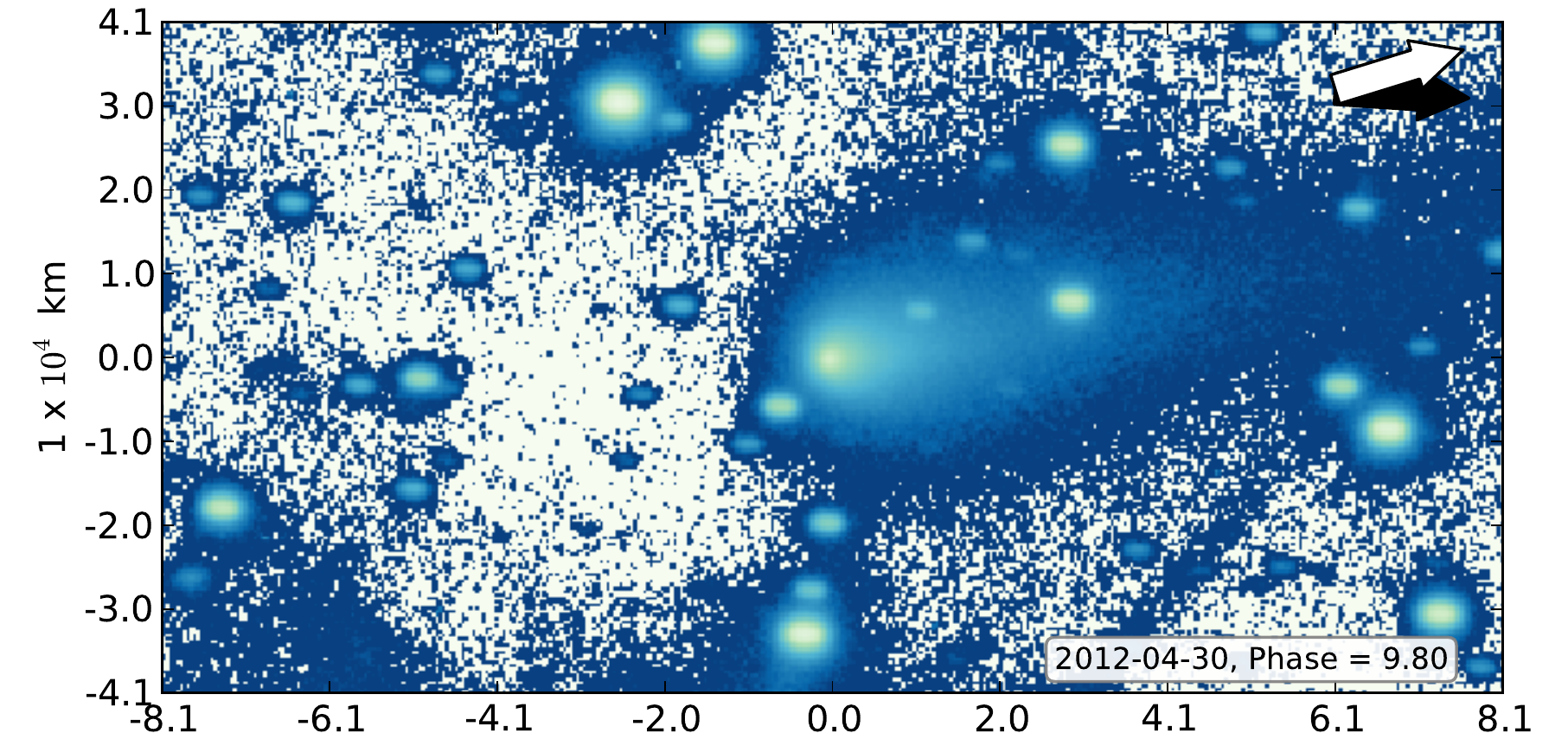}}
\subfigure{\includegraphics[scale=0.415]{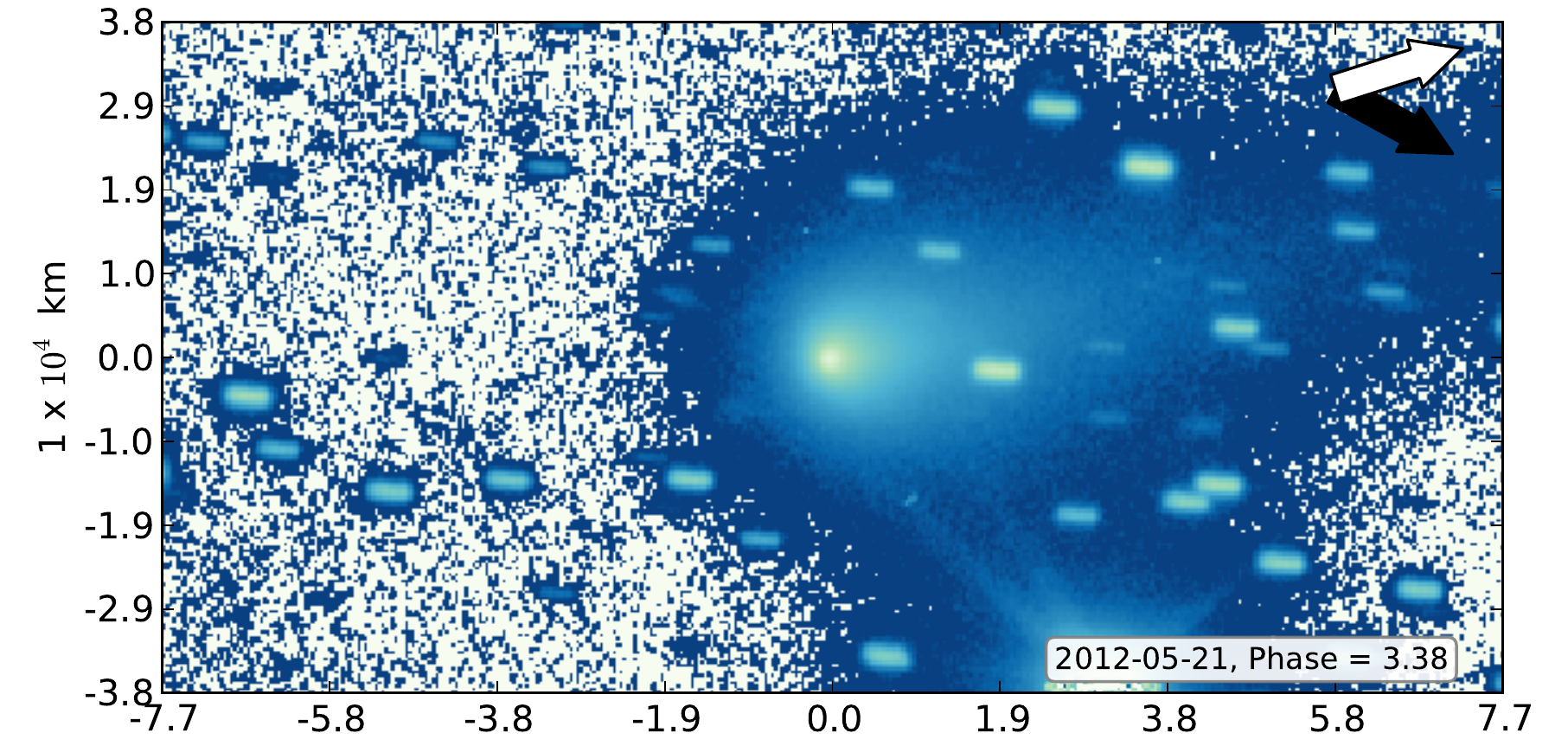}}
\subfigure{\includegraphics[scale=0.415]{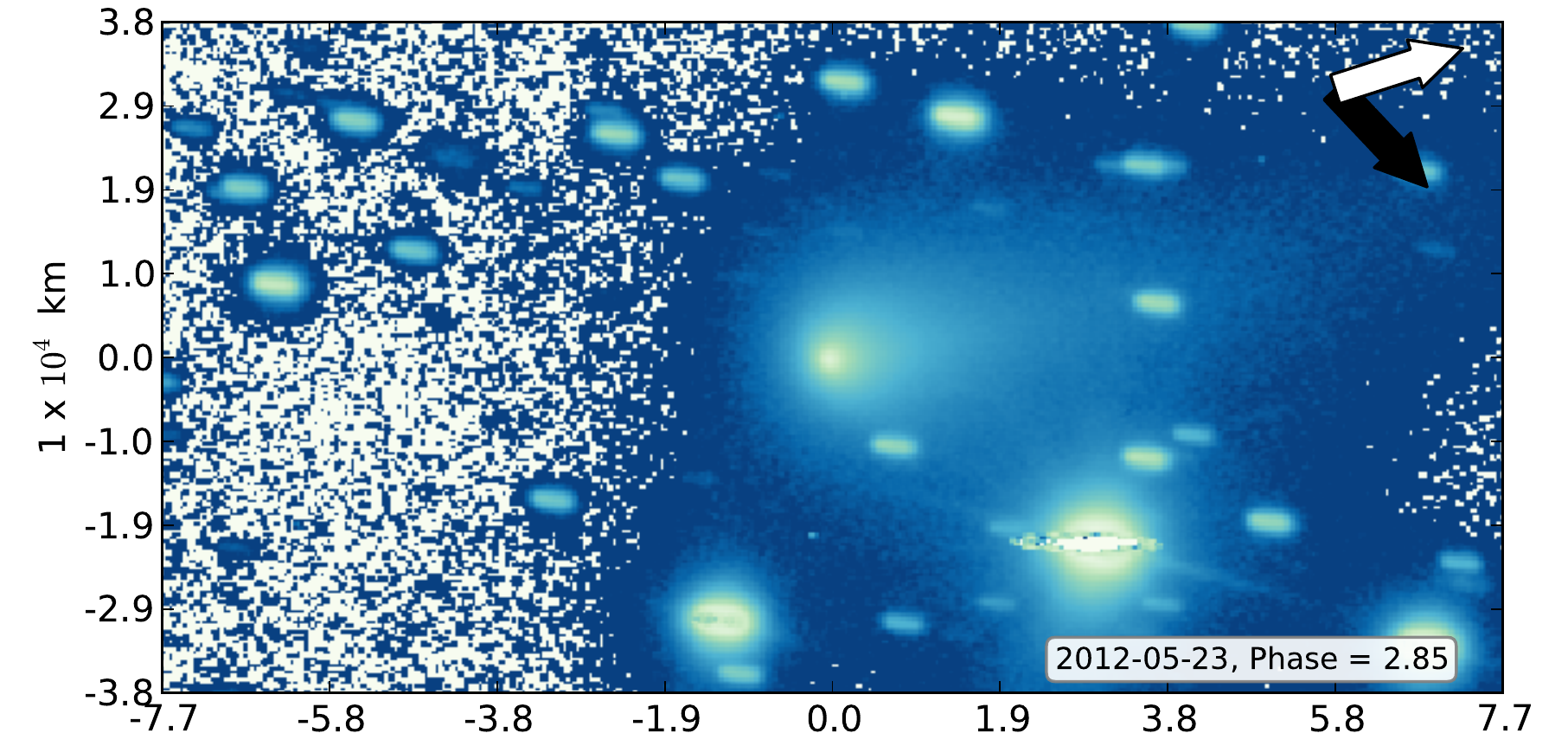}}
\subfigure{\includegraphics[scale=0.415]{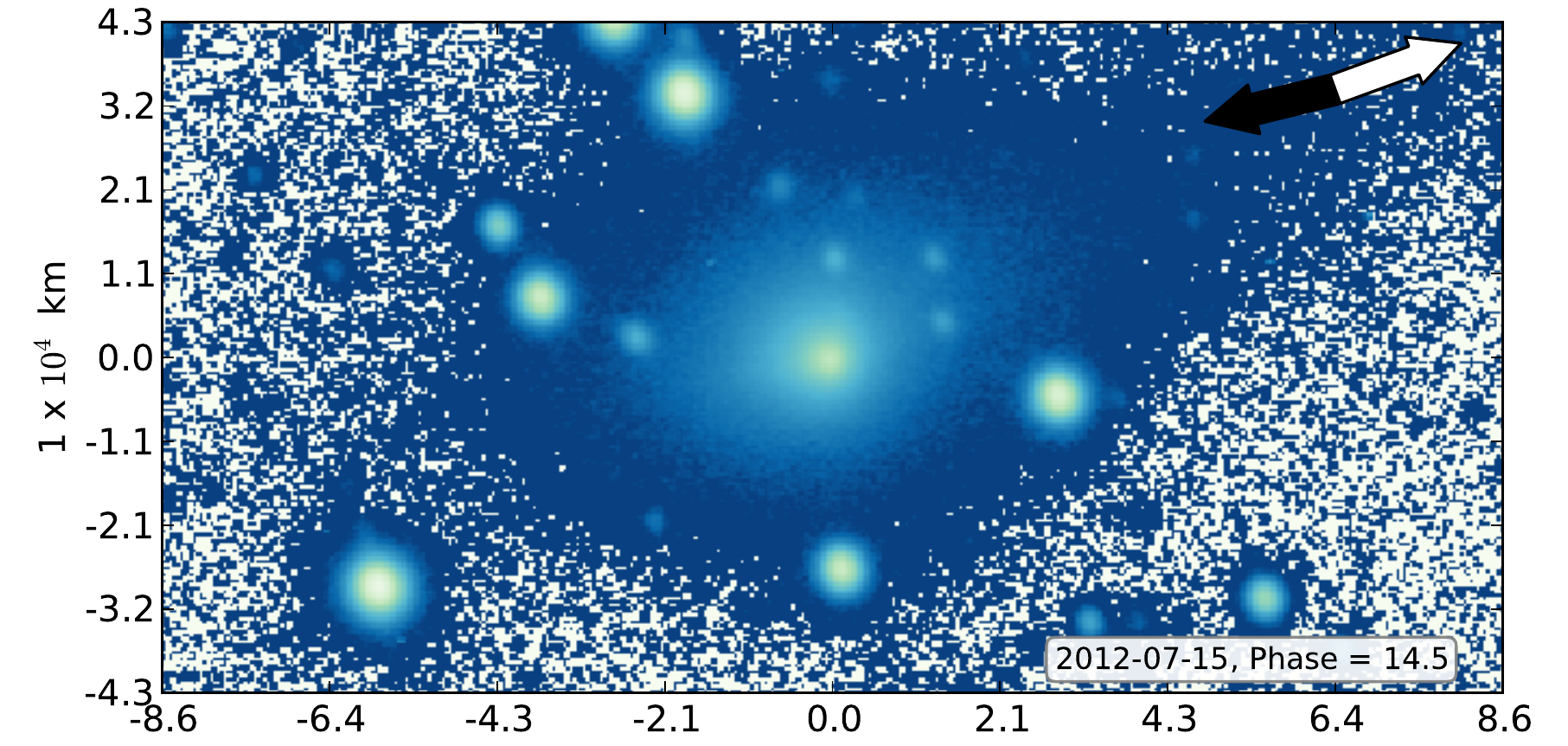}}
\subfigure{\includegraphics[scale=0.415]{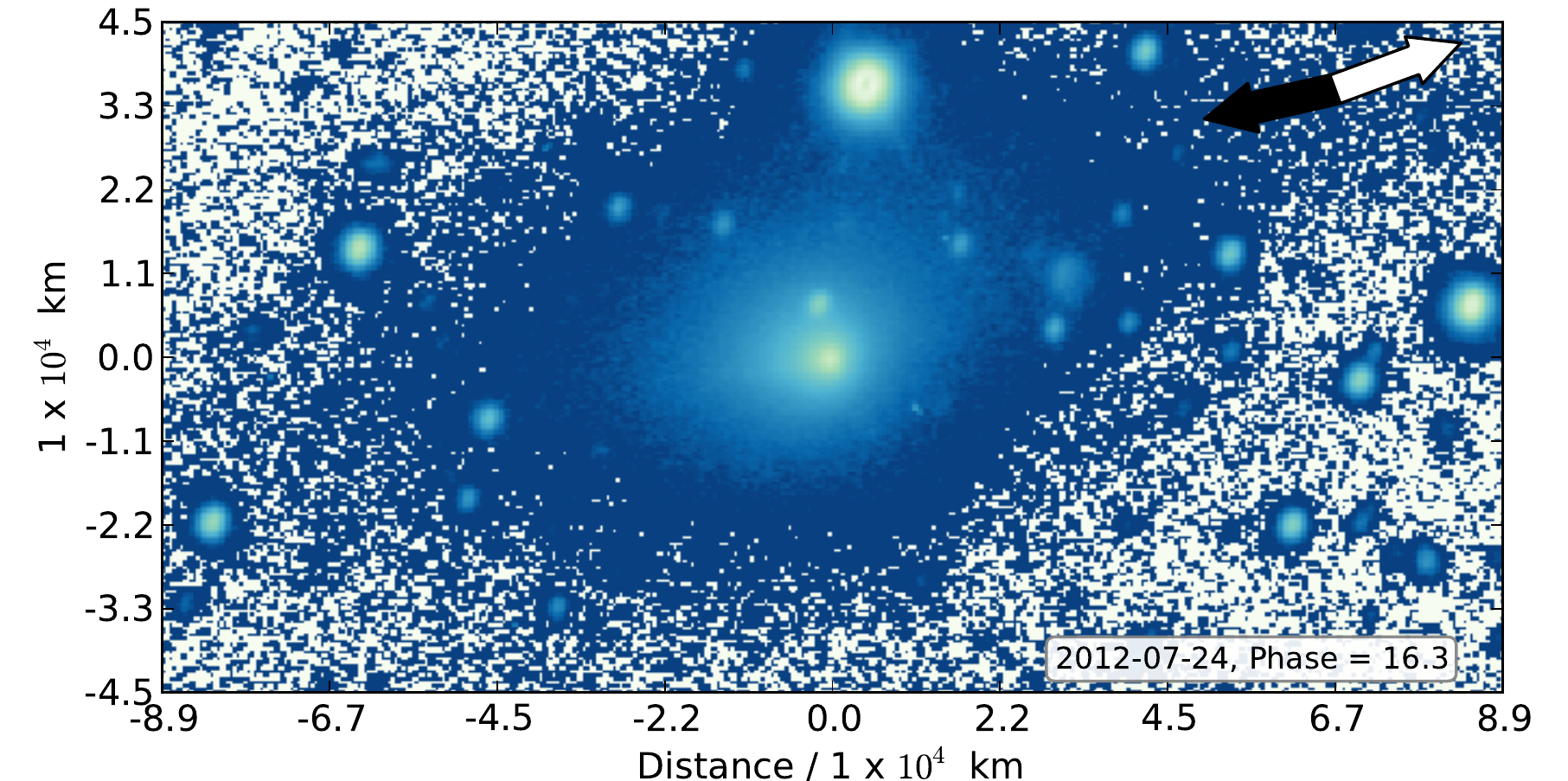}}
\caption{Intensity maps of comet 152P. the white arrow points in the direction of the negative target velocity as seen in the observer's plane of sky. The black arrow is the direction of the anti-solar direction. North is up and east is to the left.}
\label{fig:Imap_152P}
\end{figure}
The intensity images for comet 152P are presented in Fig. \ref{fig:Imap_152P}. Apart from slightly asymmetric coma extensions in the northern and north-western directions and from the main tail axis no distinct coma structure (jet, fan, or shell) was found in the processed images. Given the wide wavelength range of the \textit{R}-Special and \textit{v}-high filters used, the coma and tail are mainly represent the dust distribution around the nucleus. Features that could be attributed to gas and plasma, for instance tail rays, are not seen in the images. Tail-like extensions of the coma pointed westward in April and May 2012 and appeared in two parts during the second half of July, one towards west-north-west and one towards the east.\\ \indent
Finson-Probstein calculations \citep{Finson1968,Beisser1992} (Table \ref{tab:152P_FIN}) for the dust distribution show that the dust tail is orientated westward during April and May 2012. The appearance of two dust tail features in the second half of July 2012 indicates that young dust grains, i.e. typically released less than 2 months before the observing epoch, project into the eastern sector between angles of 90 and 100$^\circ$, while much older dust, typically released more than 8 months before the observing date, is found in the west-north-western coma region. Dust produced by the nucleus between 2 and 8 months projects into the northern coma hemisphere (as seen from Earth) and creates the asymmetric appearance in the coma.

\begin{figure}[htp]
\resizebox{\hsize}{!}{\includegraphics[scale=0.5]{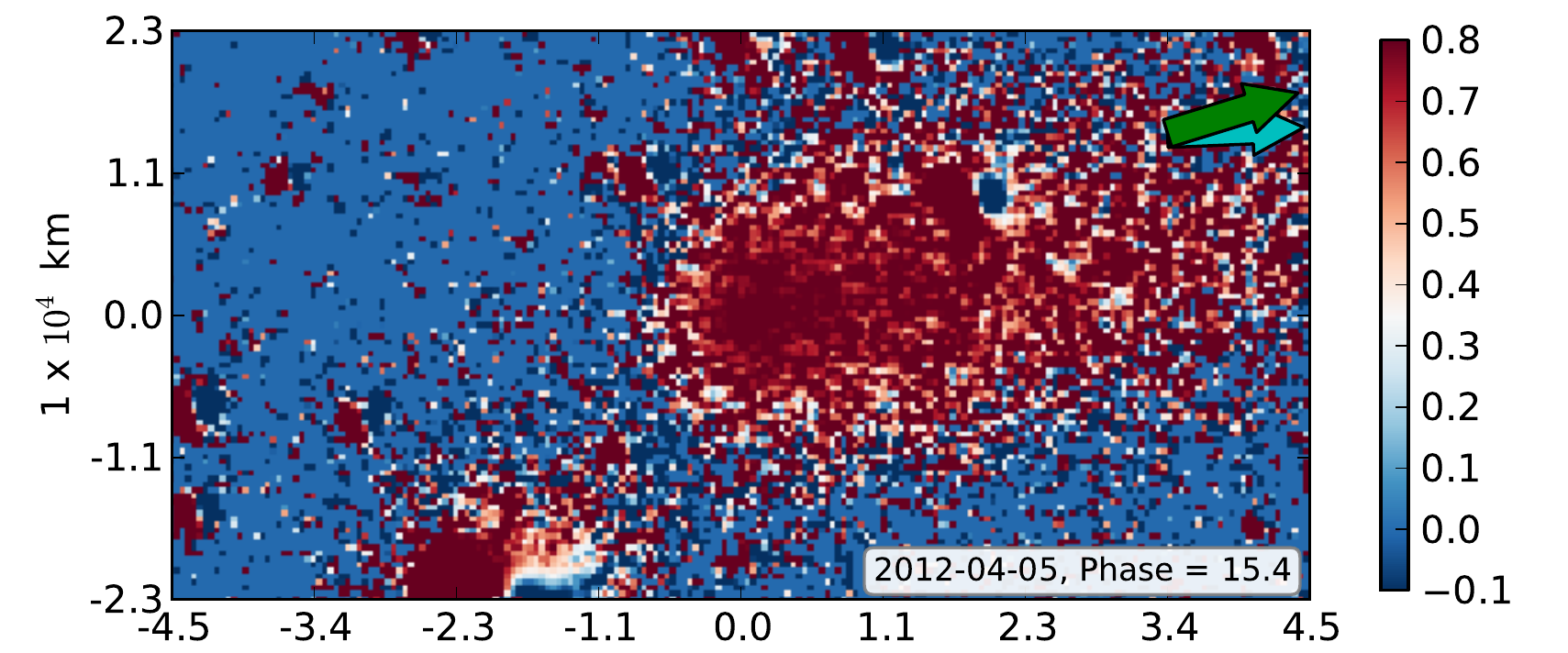}}
\resizebox{\hsize}{!}{\includegraphics[scale=0.5]{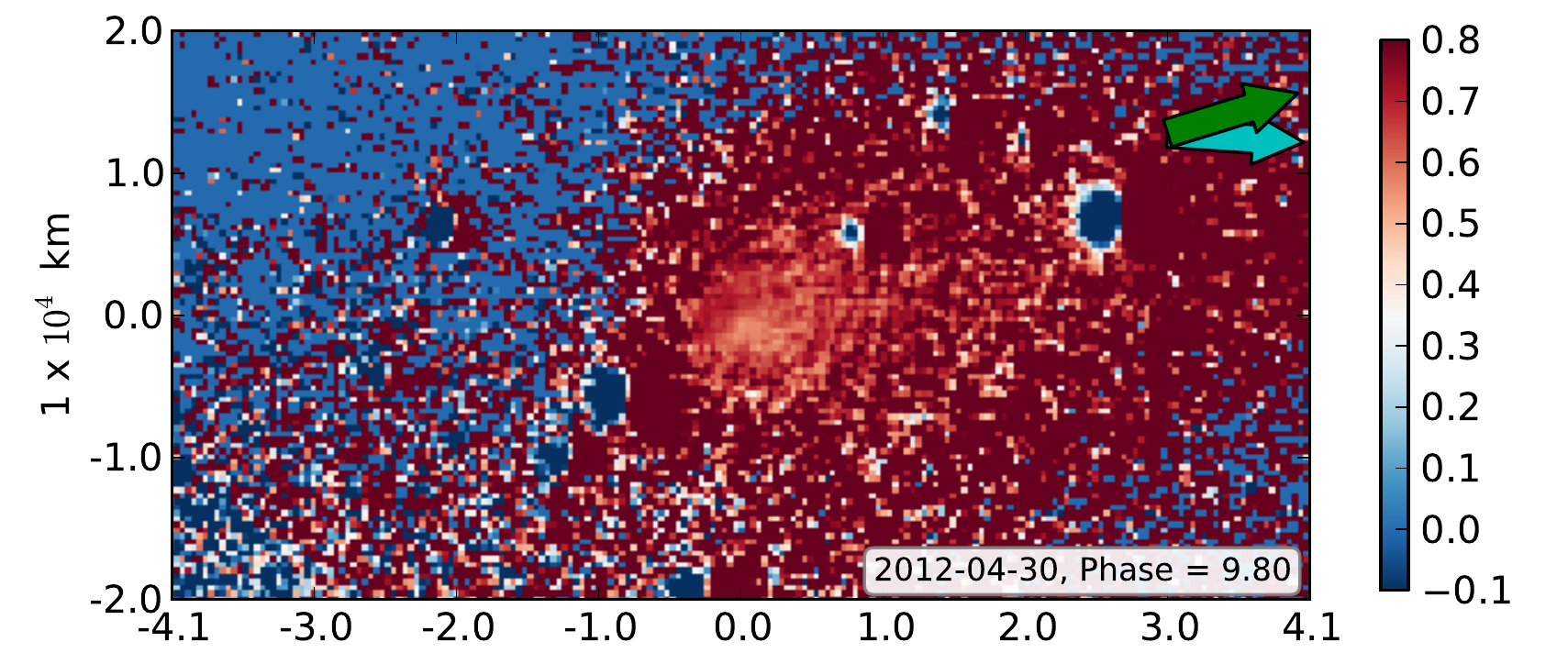}}
\resizebox{\hsize}{!}{\includegraphics[scale=0.5]{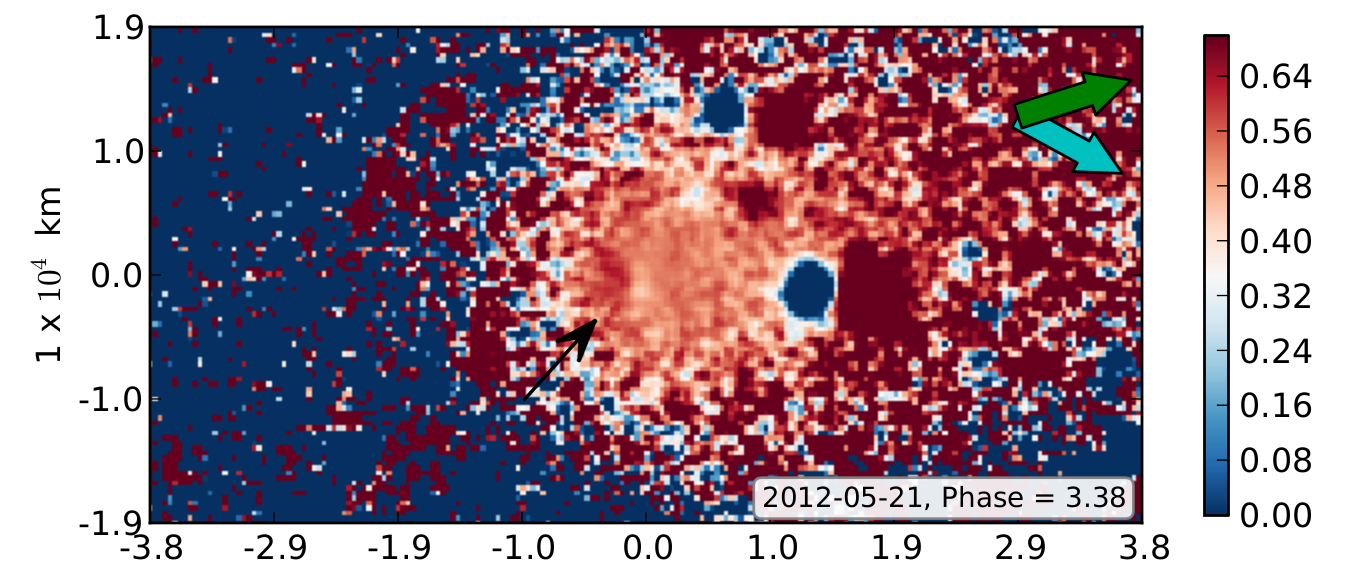}}
\resizebox{\hsize}{!}{\includegraphics[scale=0.5]{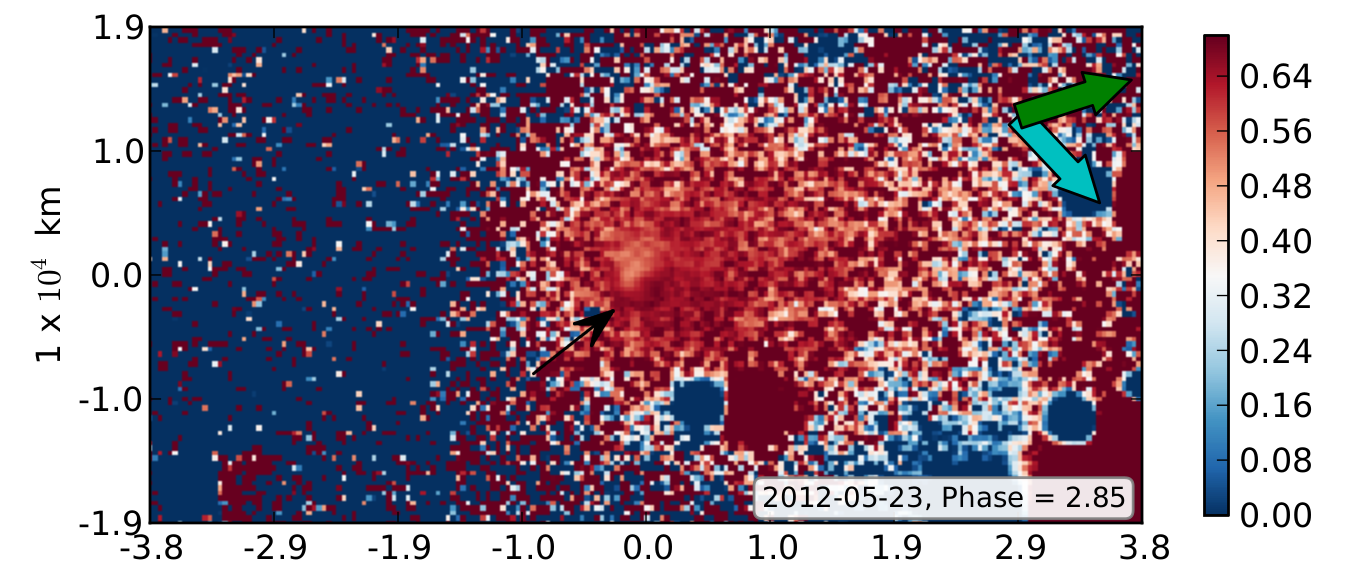}}
\resizebox{\hsize}{!}{\includegraphics[scale=0.5]{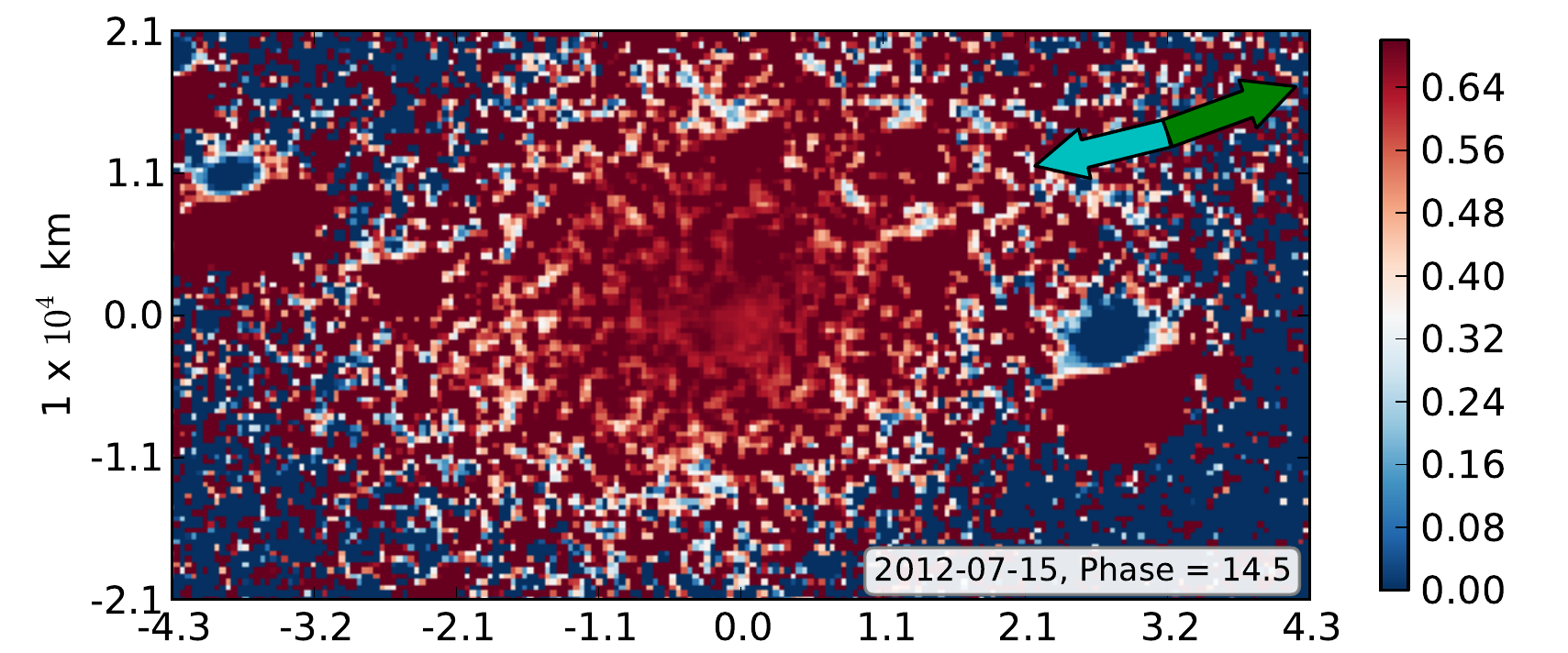}}
\resizebox{\hsize}{!}{\includegraphics[scale=0.5]{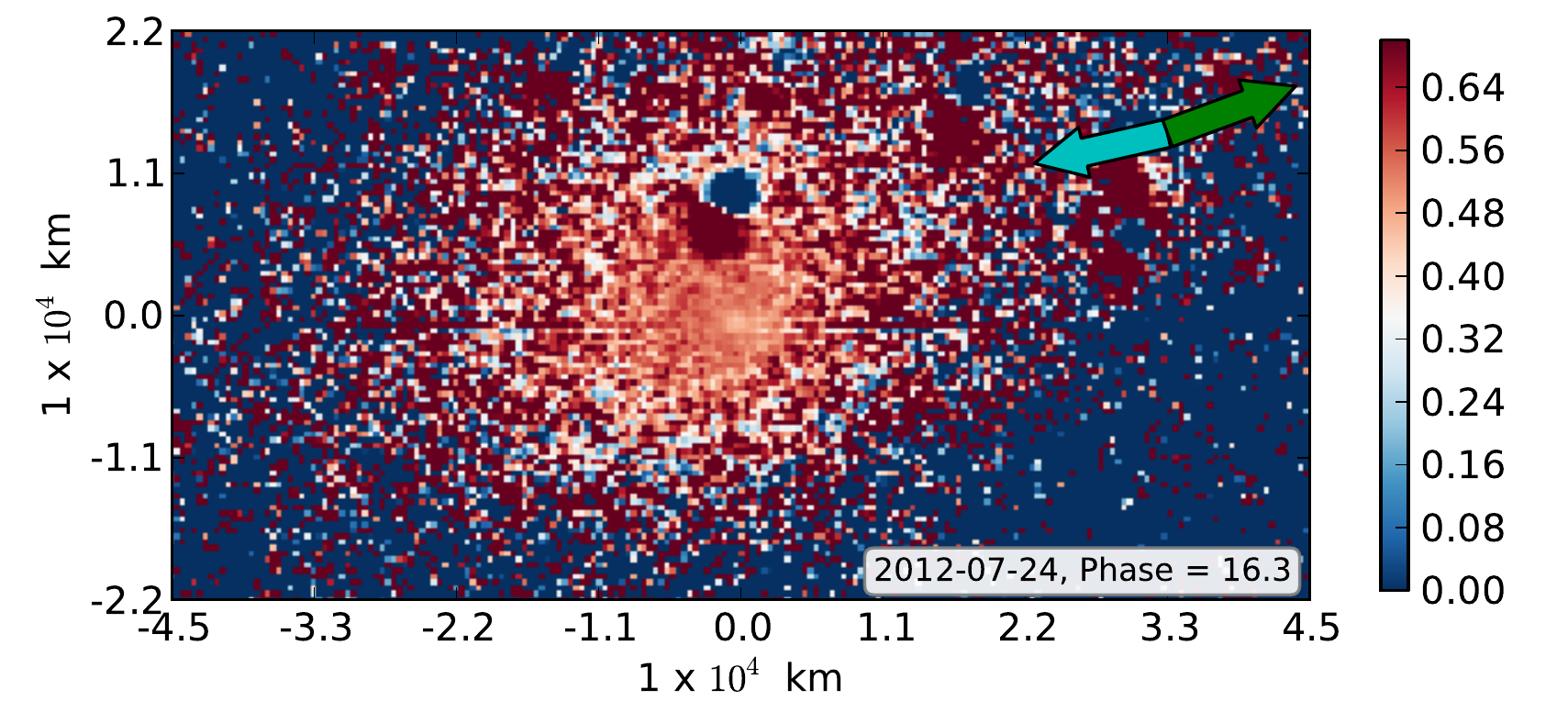}}

\caption{\textit{V-R} colour map of comet 152P. The green arrow points in the direction of the negative target velocity as seen in the observer's plane of sky. The cyan arrow is the direction of the anti-solar direction. North is up and east is to the left.}
\label{fig:cmap_152P}
\end{figure}
\subsubsection{Colour maps}
\label{sec:152P_cmap}

Figure \ref{fig:cmap_152P} shows the colour maps of comet 152P. The colour scale is centred about the solar reflectance colour of 0.35 for \textit{V-R}. Any feature redder or bluer than this value appears in the figure as red or blue, respectively. We note the occurrence of some red-blue ``dipole'' features, especially in the background. These are artefacts due to the presence of background stars that, owing to the differential tracking of the telescope, appear offset in the \textit{R} and \textit{V} images.
The exceptions are the top two maps in Fig. \ref{fig:cmap_152P} where the colour scale has been extended to accommodate their slightly redder colour compared to the other data; however, the colour scale is still centred about 0.35. \\ \indent
The colour maps created show little to no structure apart from the nights of 30 April, 21 and 23 May 2012. On 30 April 2012 the photometric centre of the comet is bluer than that 5 April 2012. This suggests either the comet was producing a redder material or larger particles on 5 April 2012 than on 30 April 2012 or was producing bluer material or particles of a smaller size on the 30 April 2012. On 21 May 2012 the average coma colour is $\sim$ 0.54; however, there is a small spot at the photometric centre with a colour of $\sim$0.43, and north-west of this there is a much redder feature with a colour of $\sim$0.63 (marked by a black arrow). This could be an indication of activity, jet or outburst, in the inner coma. Two days later on 23 May 2012 the average colour of the coma increased to $\sim$ 0.60, perhaps indicating the results of the activity noticed on 21 May 2012. On 23 May 2012 there is a feature north-west of the photometric centre of the coma as was the case on 21 May; however, the colour of this feature has changed to a value of $\sim$0.43 (marked with a black arrow), which could be due to particle movement and nucleus rotation. These features are not a simple case of image misalignment as this would result in the dipole feature seen in the background of all these colour maps which are due to stars moving in the background  with respect to the comet. In addition, the images used to calculate these colour maps were carefully selected to ensure that they had very similar seeing conditions. After 23 May 2012 the coma continues to increase in colour to an average colour of $\sim$ 0.66, which is fairly uniform across the coma. Unfortunately the  \textit{V} and \textit{R} images are taken under different seeing conditions making it impossible to see if there was any activity to cause the colour to drop to $\sim$ 0.54 in the coma.
\subsubsection{Polarimetric maps}
\label{sec:152P_polmap}
\begin{figure*}[htp]
\resizebox{\hsize}{!}{{\includegraphics[scale=0.5]{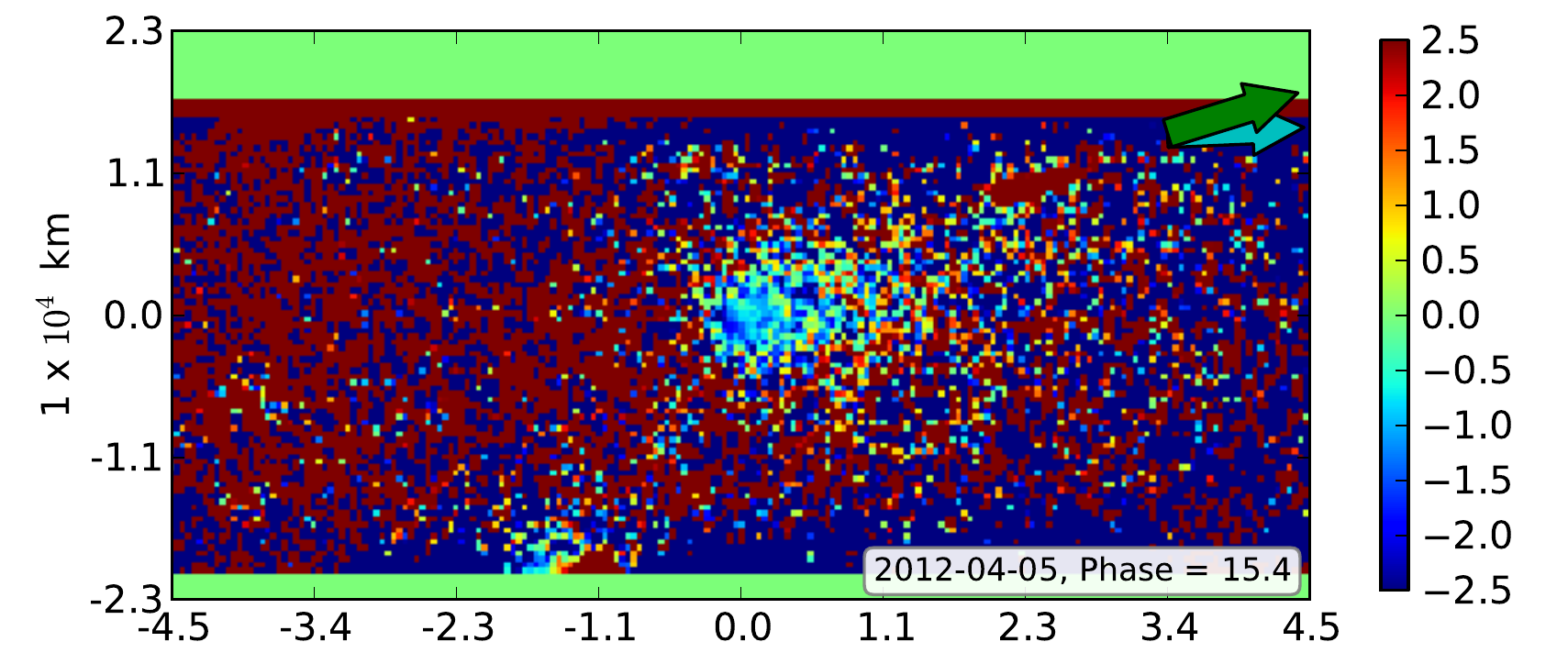}}{\includegraphics[scale=0.5]{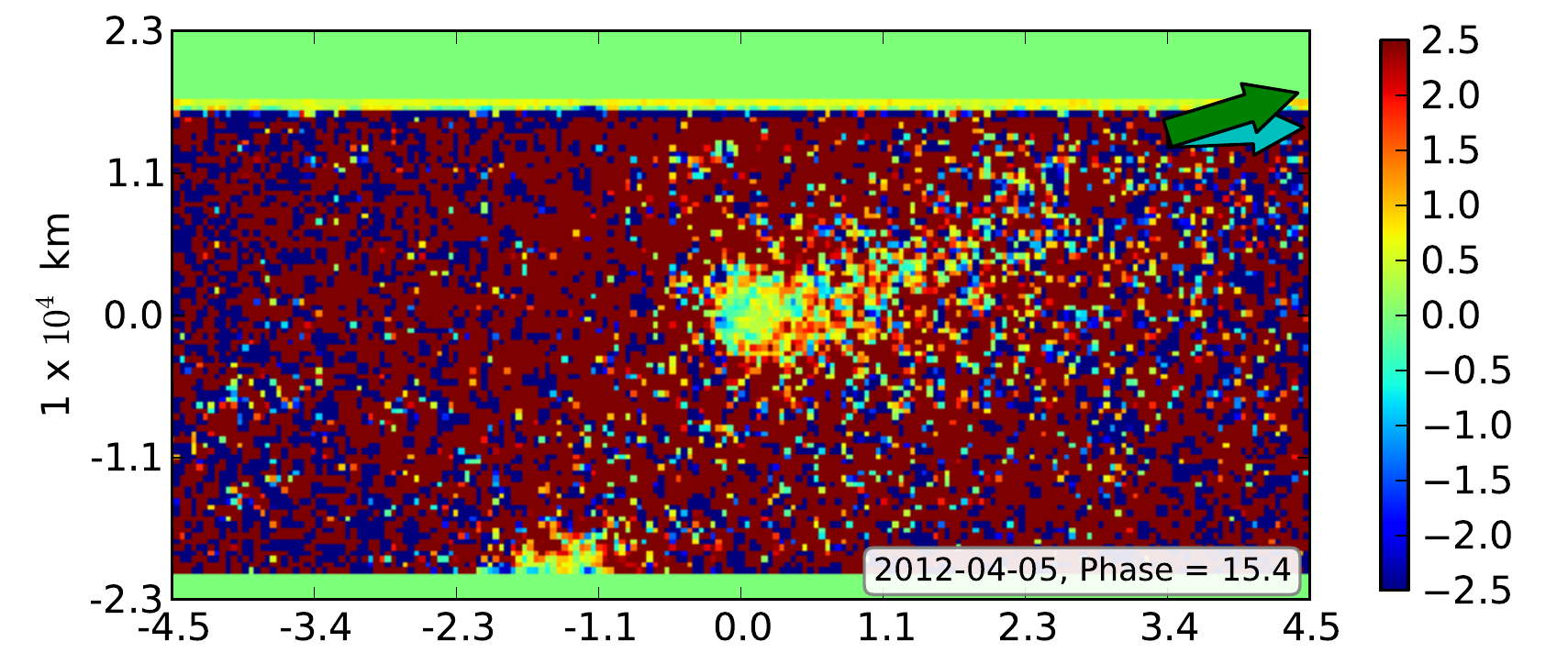}}}
\resizebox{\hsize}{!}{{{\includegraphics[scale=0.5]{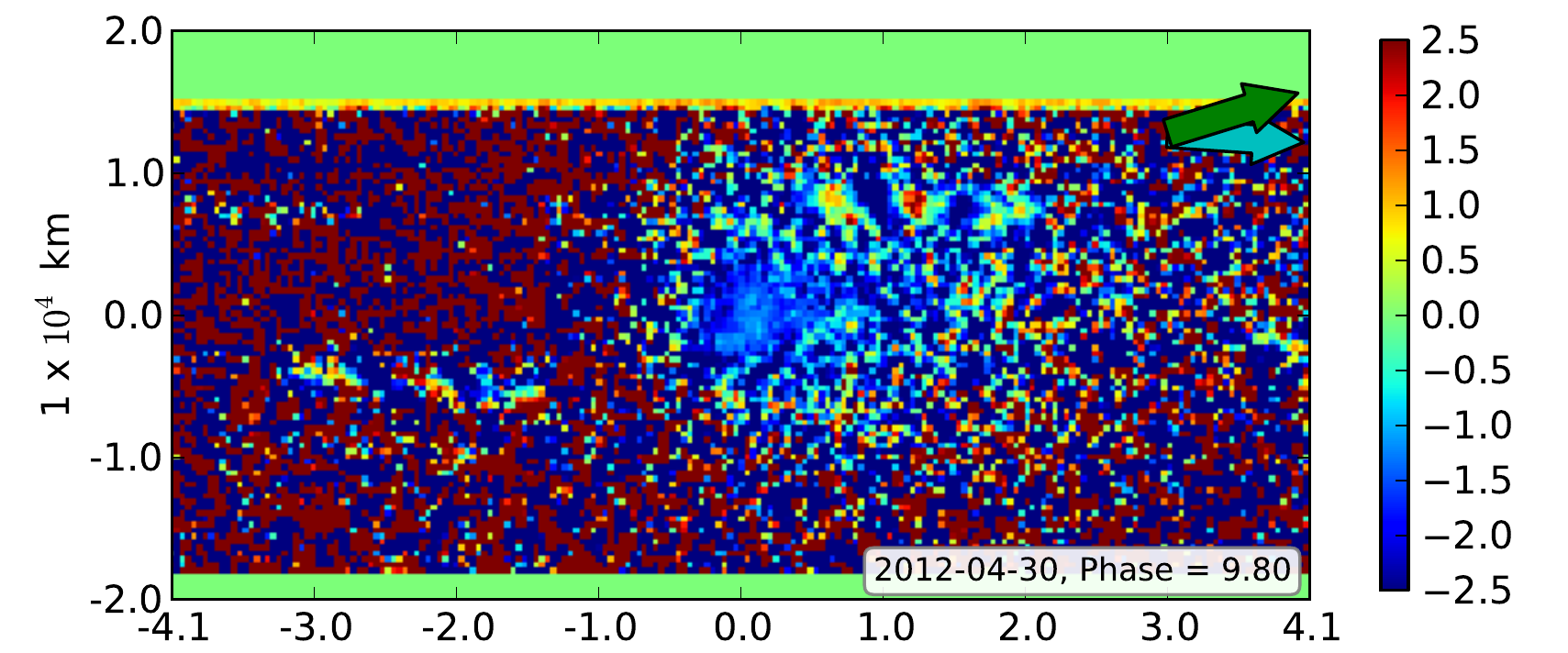}}}{{\includegraphics[scale=0.5]{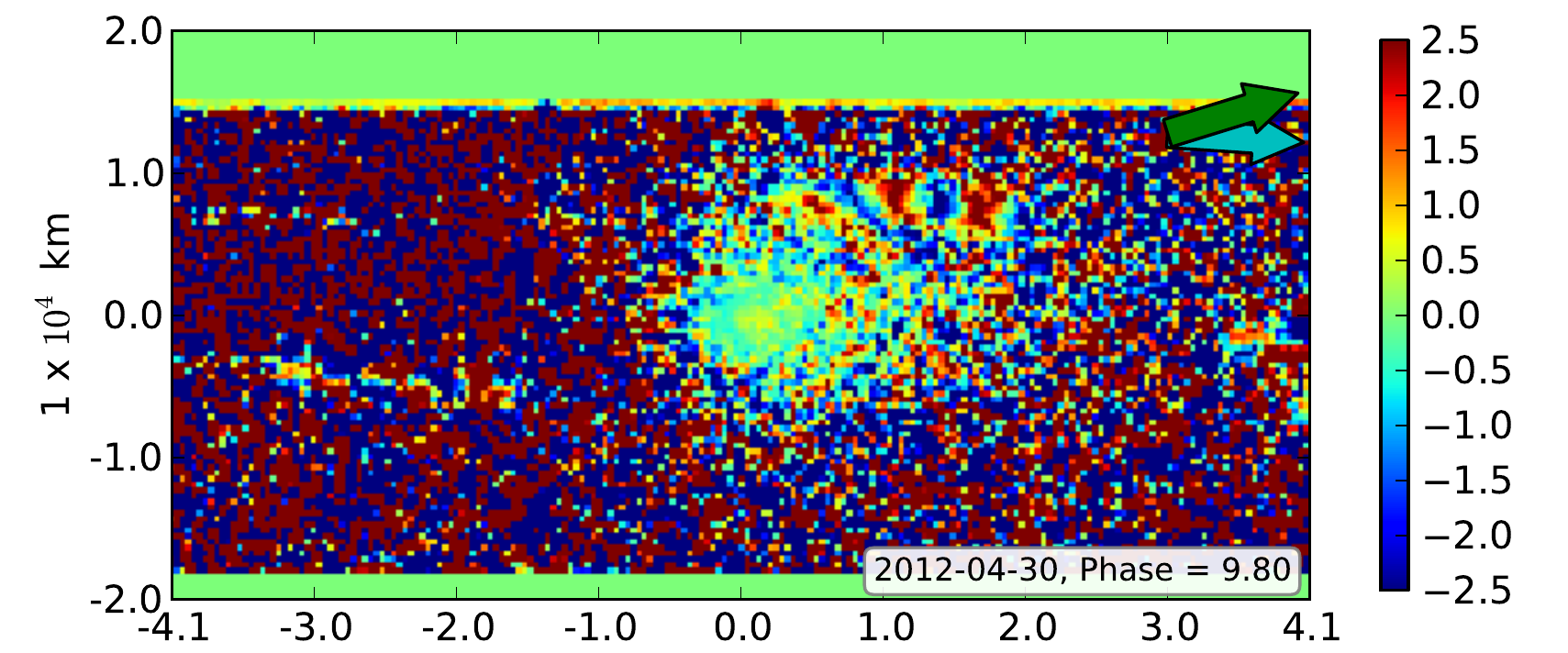}}}}
\resizebox{\hsize}{!}{{\includegraphics[scale=0.5]{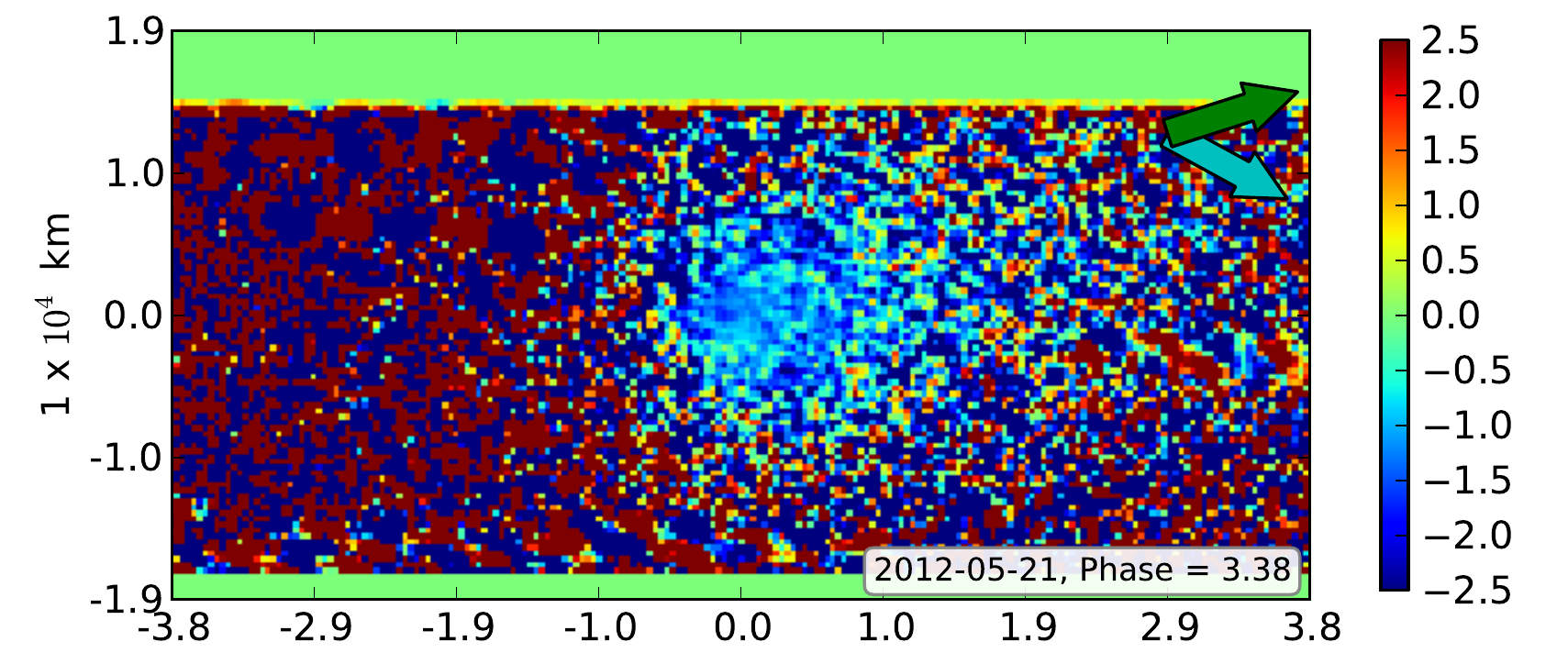}}{\includegraphics[scale=0.5]{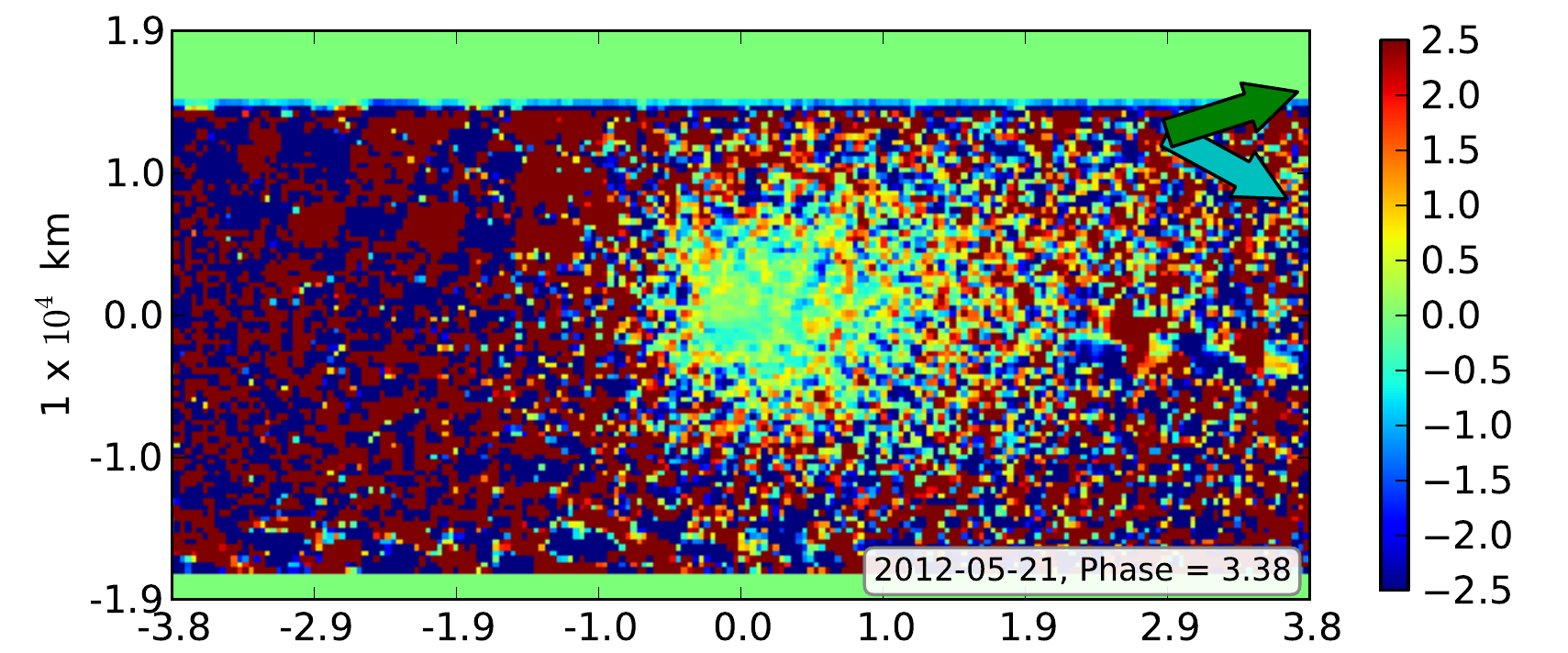}}}
\resizebox{\hsize}{!}{{\includegraphics[scale=0.5]{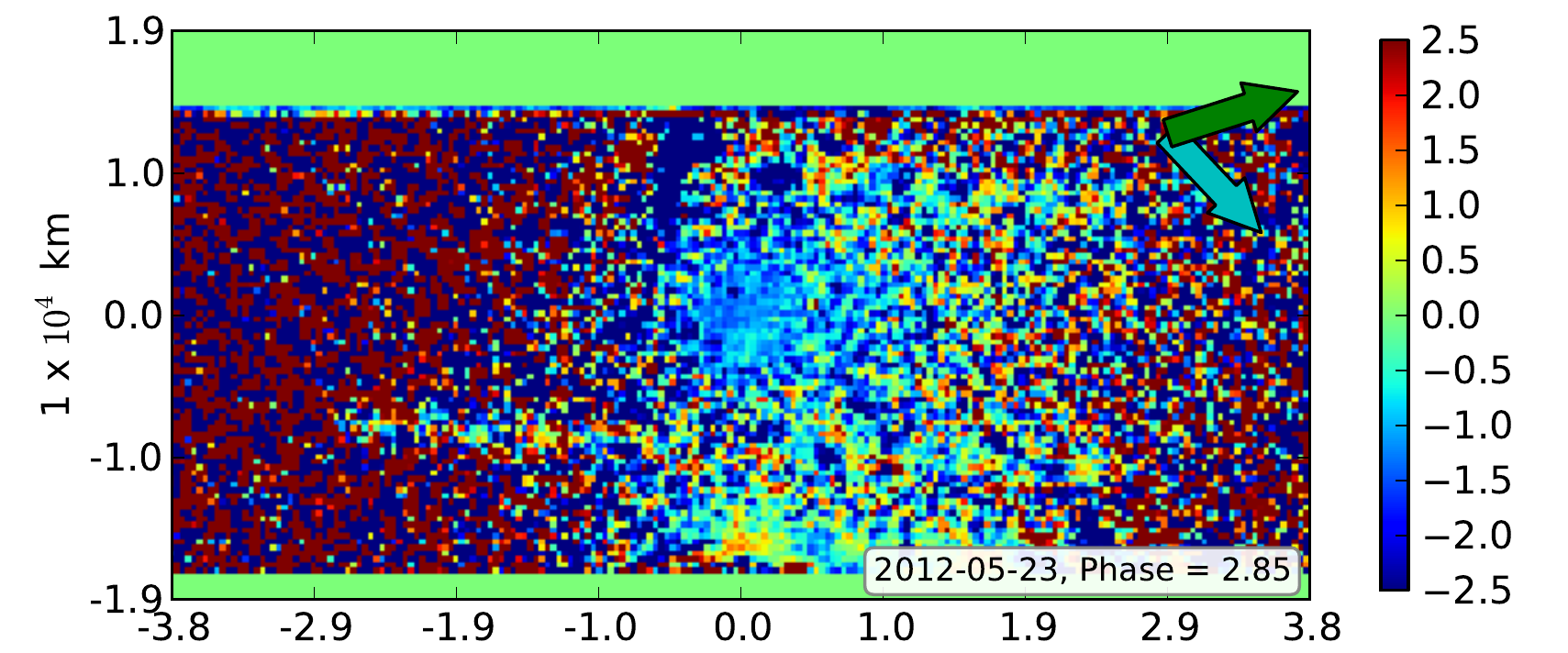}}{\includegraphics[scale=0.5]{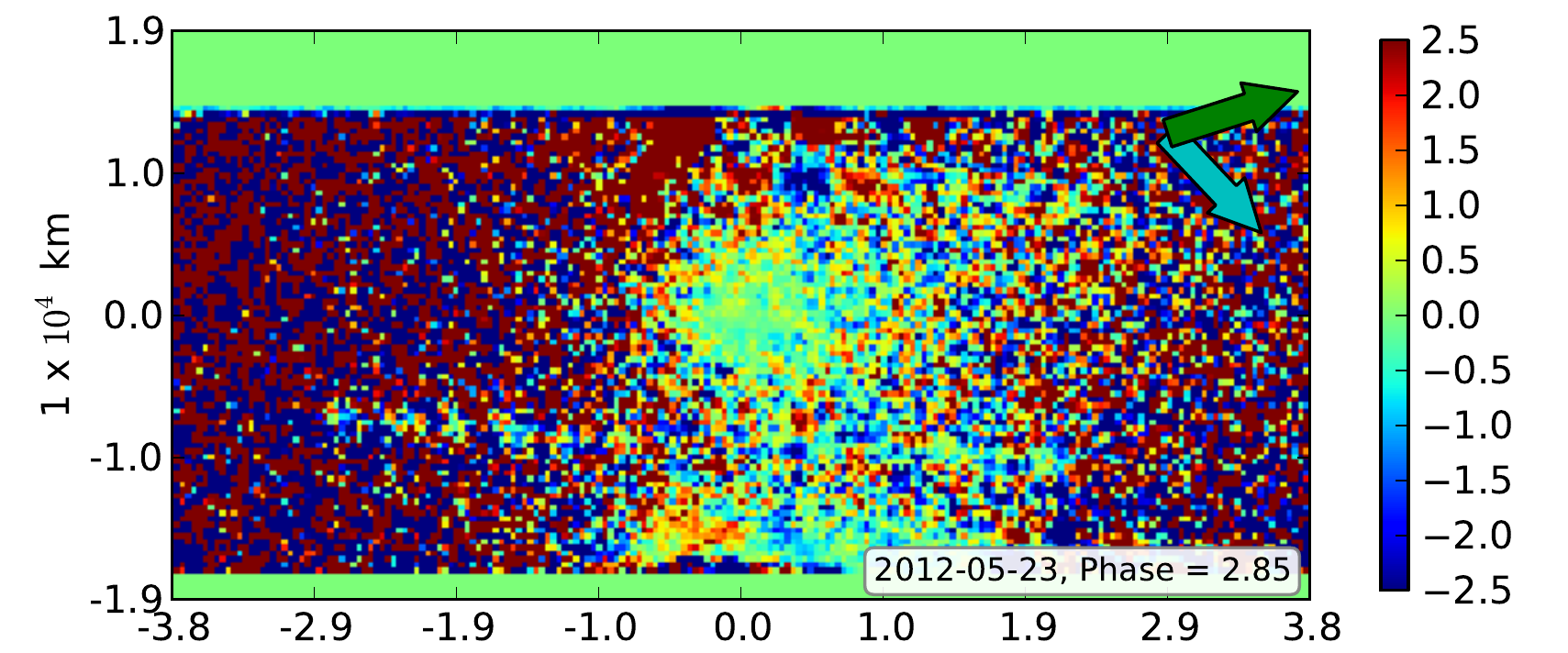}}}
\resizebox{\hsize}{!}{{\includegraphics[scale=0.5]{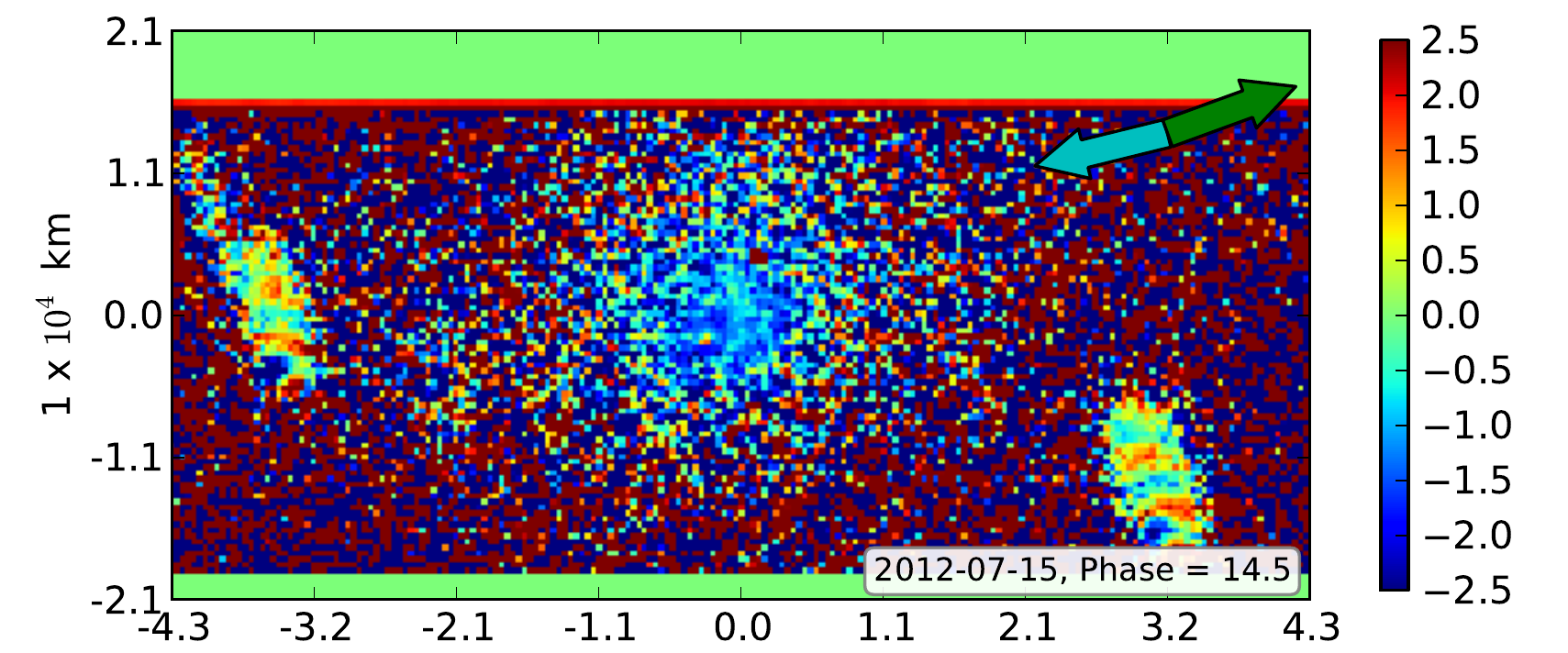}}{\includegraphics[scale=0.5]{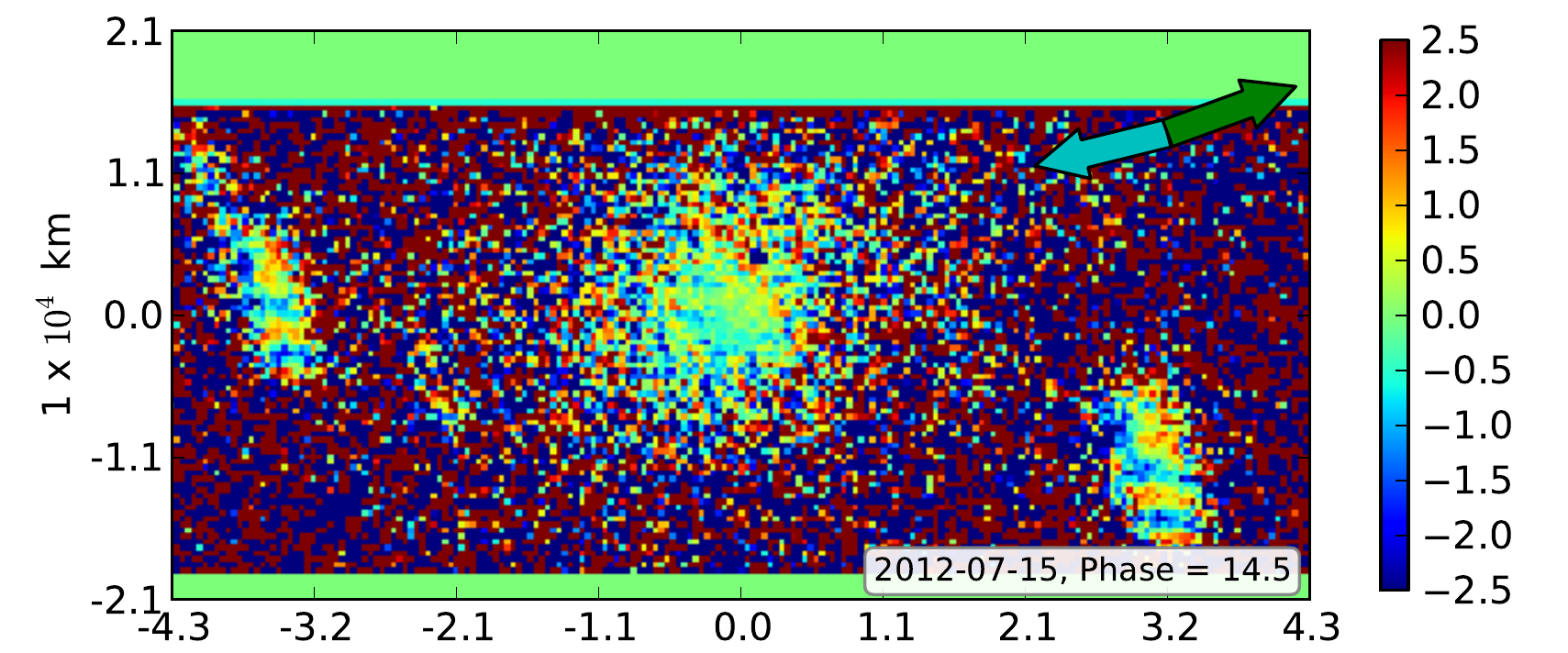}}}
\resizebox{\hsize}{!}{{\includegraphics[scale=0.5]{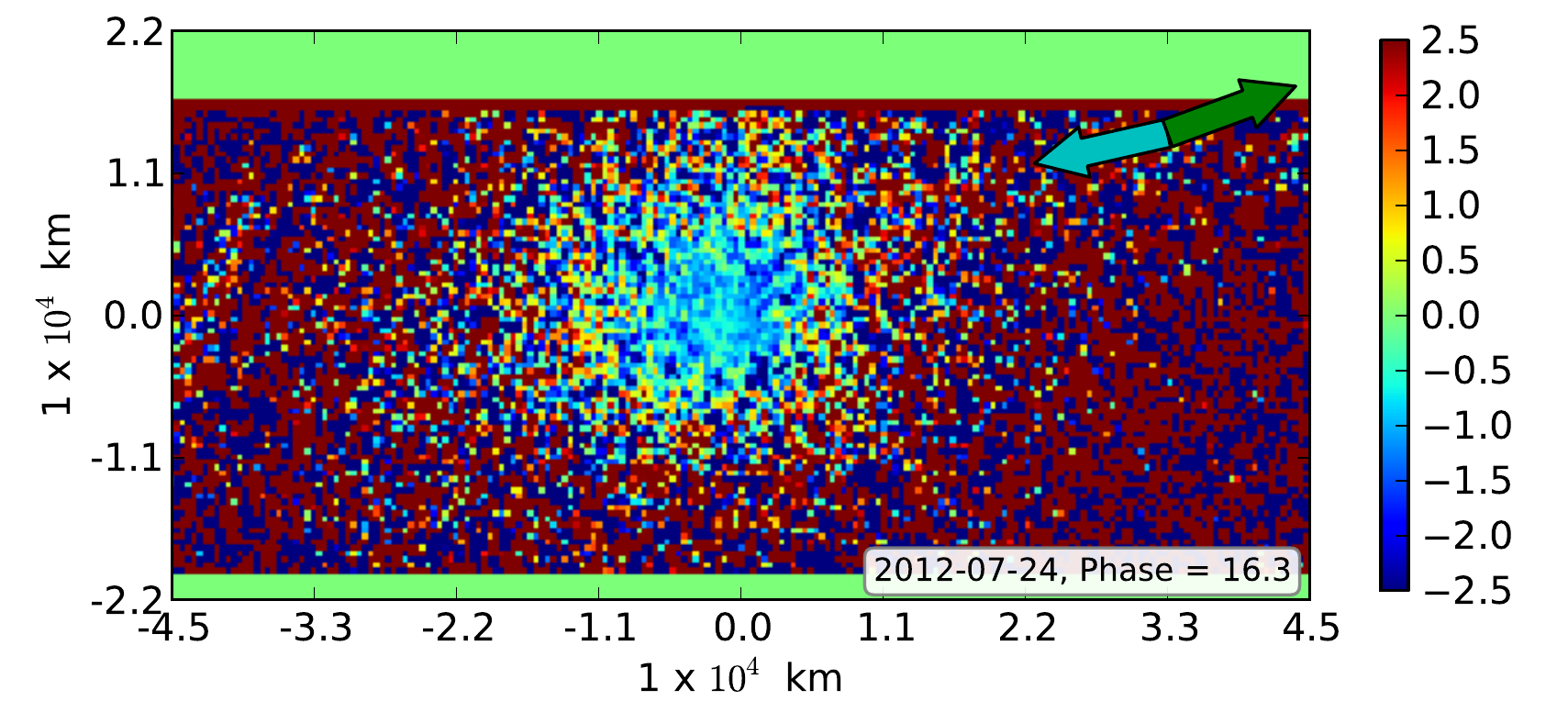}}{\includegraphics[scale=0.5]{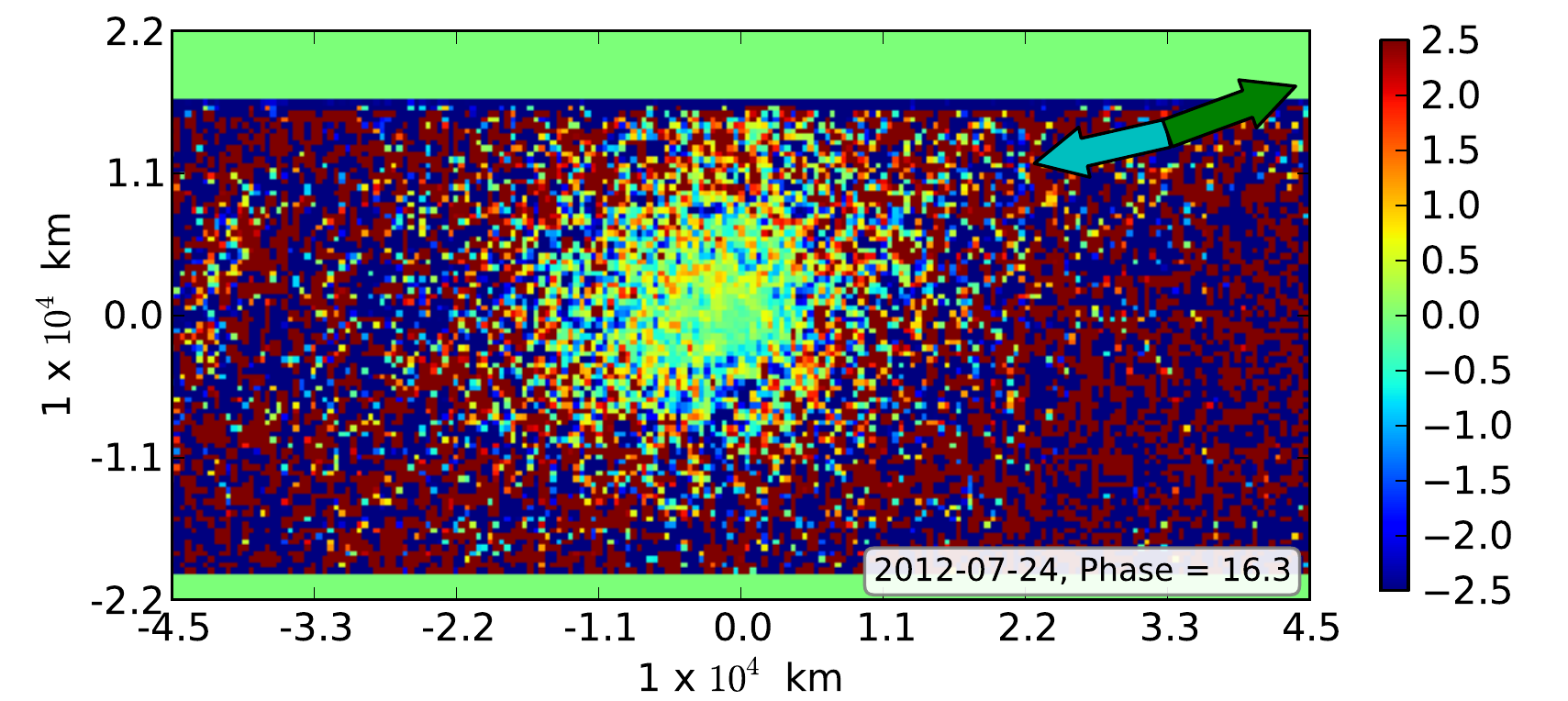}}}
\caption{$P'_Q$ (left) and $P'_U$ (right) polarimetric maps for comet 152P. Both $P'_Q$  and $P'_U$ are measured in per cent. The green arrow points in the direction of the negative target velocity as seen in the observer's plane of sky. The cyan arrow is the direction of the anti-solar direction. North is up and east is to the left.}
\label{fig:pol_map_152P}
\end{figure*}

The polarimetric maps for comet 152P are presented in Fig. \ref{fig:pol_map_152P}; on the left side are the $P'_Q$ maps and on the right side are the $P'_U$ maps. On the night of April 2012 the background sky was highly polarised, which made getting an accurate background sky estimate difficult, as can be seen in the background features in the polarimetric maps. In spite of this difficulty, there are no unusual polarimetric values that coincide with the red colour seen in the colour maps in Fig. \ref{fig:cmap_152P}.\\ \indent
The polarimetric maps created for the nights of 21 and 23 May 2012 can be compared to the colour maps created in Fig. \ref{fig:cmap_152P} to determine whether the same features are present. Figure \ref{fig:pol_map_152P} shows that on the night of 21 May 2012 the amount of polarisation becomes more negative $\sim$ $-3\%$ in the direction of the outburst seen in the colour map in Fig. \ref{fig:cmap_152P}. In the $P'_U$ map for the same night we see an increase in $P_U$ in the location of the outburst region of $\sim$ 0.5$\%$. Again, this suggests that the outburst is composed of a different material or has a different morphology to the surrounding coma.  \\ \indent
On 23 May 2012 there is a slight hint of structure in the coma in the $P'_Q$ map for this night. In the north-western direction from the photometric centre there is a slightly more negative polarisation, $\sim$ $-1.4\%$, compared to $\sim$ $0.7\%$ in the south-eastern direction. This polarisation difference occurs in the same location as the colour feature seen in the colour maps of Fig. \ref{fig:cmap_152P}.\\ \indent
In the last two maps in July as seen in Fig. \ref{fig:pol_map_152P} both maps in $P'_Q$ show very few features of note. In the $P'_U$ maps on the same nights there are small fluctuations in polarisation of around $\pm$ 0.1$\%$, consistent with noise.

\begin{figure}[tbp!]
\resizebox{\hsize}{!}{\includegraphics[scale=1.0]{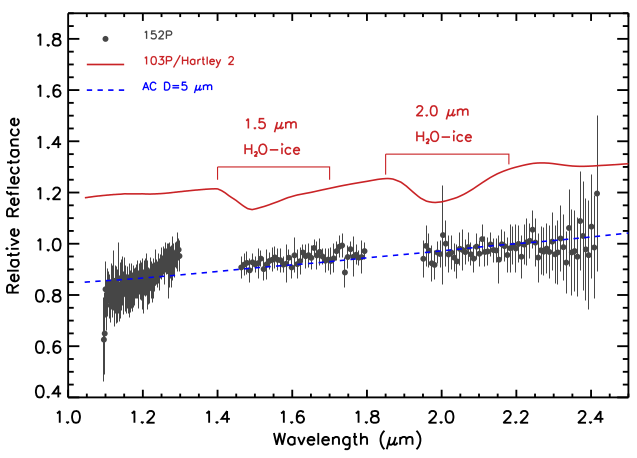}}
\caption{Relative reflectance spectrum of comet 152P (grey dots). For comparison, a synthetic spectrum of amorphous carbon (AC) grains (particle diameter of 5 $\mu$m, blue line) and the synthetic spectrum used to represent the Hartley 2 coma (solid red line) composed of 1 $\mu$m water-ice grains and dust not in thermal equilibrium \citep{Protopapa2014}.}
\label{fig:sinfoni}
\end{figure}

\subsubsection{SINFONI infrared spectra}
\label{sec:SI}

Figure \ref{fig:sinfoni} shows the relative reflectance spectrum of comet 152P. The target spectrum presents a positive spectral slope of 13.51 $\pm$ 0.7\,$\%$/100\,nm similar to the spectral slope from the \textit{V-R} measurements. We looked for the presence of water ice absorption features at 1.5 and 2.0 $\mu$m, displayed by other JFC spectra such as that of Hartley 2 \citep{Ahearn2011} and the outbursting comet P/2010 H2 (Vales) \citep{Yang2010}. However, the 152P spectrum does not display clear water-ice absorptions and resembles, at first order, the spectrum of a dark and featureless refractory component (e.g. amorphous carbon, dashed blue line). The comparison of our spectral slope to other active comets is difficult as they are influenced by the presence of water ice. Water ice is blue in the near infrared spectrum which tends to make the spectral slope shallower.

\subsection{Comet 74P/Smirnova–Chernykh}
\label{sec:74P}
\subsubsection{Aperture photometry}
\label{sec:74P_phot}

\begin{figure}[tbp!]
\resizebox{\hsize}{!}{\includegraphics[scale=1.0]{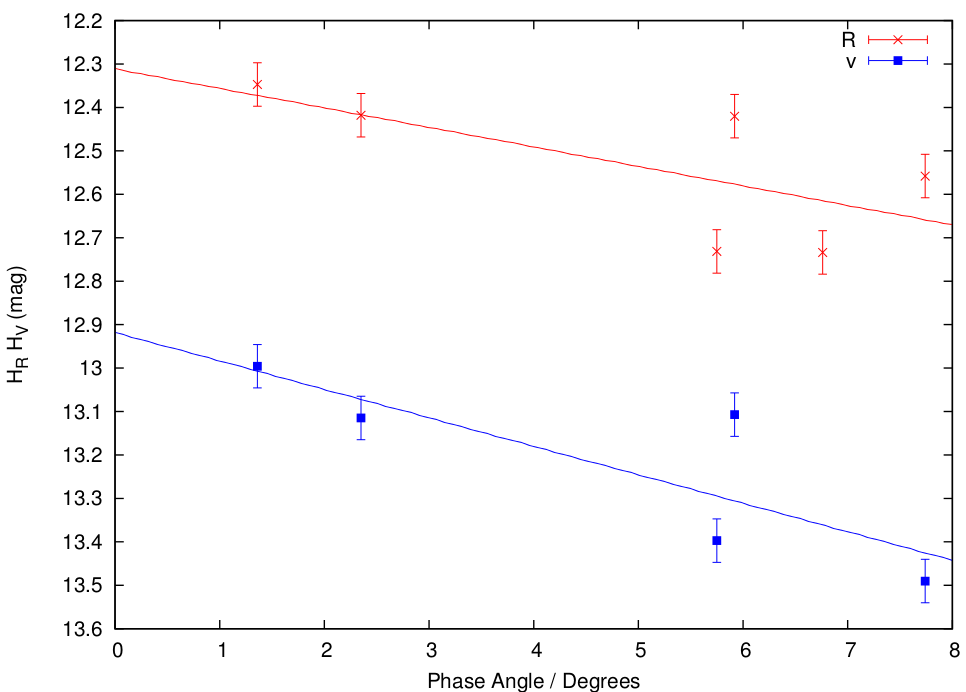}}
\caption{Magnitude corrected for the Sun and Earth distances of comet 74P as a function of phase angle.}
\label{fig:abso_74P}
\end{figure}

74P shows a constant solar \textit{V-R} colour throughout all our observations. The only exception is the night of 21 June 2012, which was contaminated by two nearby saturated stars particularly when the \textit{R} filter observations were carried out.\\ \indent
In Fig. \ref{fig:abso_74P} we plot the magnitude corrected for the Sun and Earth distances of comet 74P as a function of phase angle. If we ignore the result on the night of 21 June 2012 and use a straight line fit, the extrapolated average brightnesses at zero phase angle are 12.76 $\pm$ 0.13 and 13.43 $\pm$ 0.15 in the \textit{R} and \textit{V} filters assuming no opposition surge. This results in an average \textit{V-R} colour of 0.67 $\pm$ 0.20, which is equivalent to a spectral gradient of 34.28\,$\%$/100\,nm. \citet{Lowry2001} observed 74P and found that it had a \textit{V-R} colour of 0.44 $\pm$ 0.10, which is not quite as red as suggested by our findings.

Using Eq. (\ref{eq:afrho}) and the flux extrapolated back to zero phase in the \textit{R}-Special filter and the average $r$ and $\Delta$ distances to the comet yields an $Af\rho$ value of 200.8 $\pm$ 22.7\,cm. This compares to the values of 228.8 $\pm$ 11.4 cm and 298.9 $\pm$ 11.3 cm measured by \citet{Lowry1999} and \citet{Lowry2001} at a heliocentric distance of 4.2 and 4.6\,au. The large amount of activity shown by 74P at heliocentric distances beyond 4\,au could suggest that this comet constantly shows signs of activity throughout its orbit.

\subsubsection{Intensity maps}
\label{sec:74P_struct}

\begin{figure}[htpb!]
\centering
\subfigure{\includegraphics[scale=0.45]{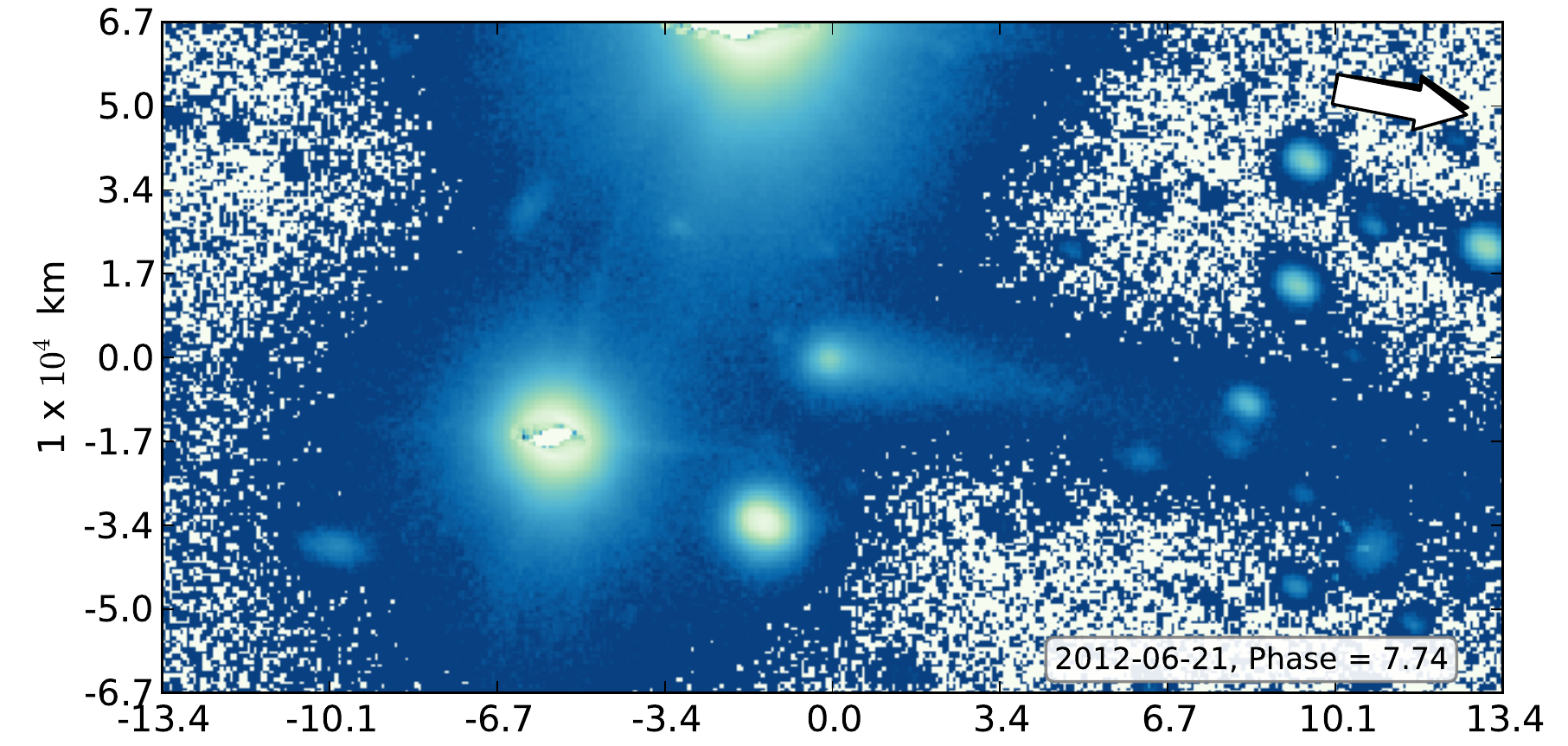}}
\subfigure{\includegraphics[scale=0.45]{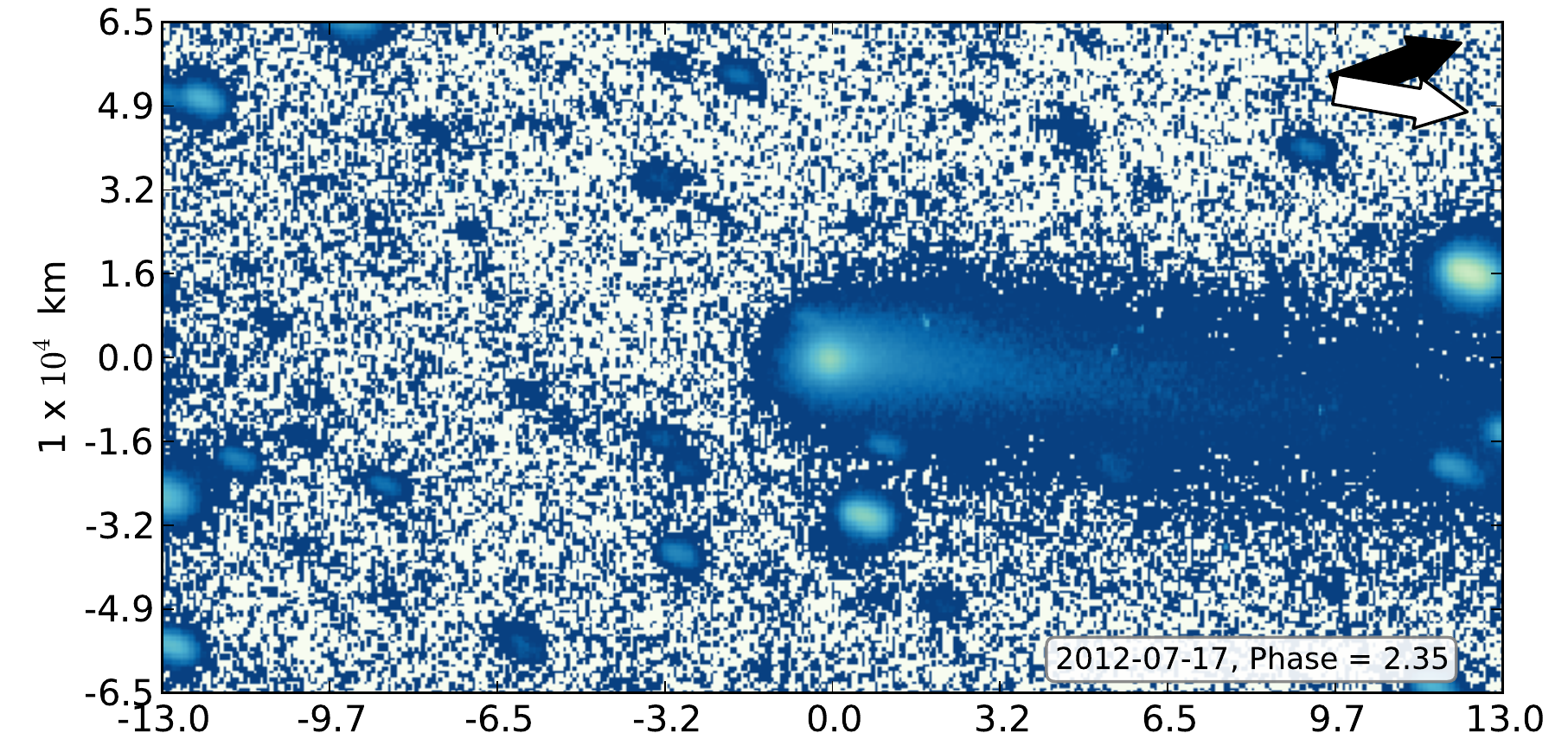}}
\subfigure{\includegraphics[scale=0.45]{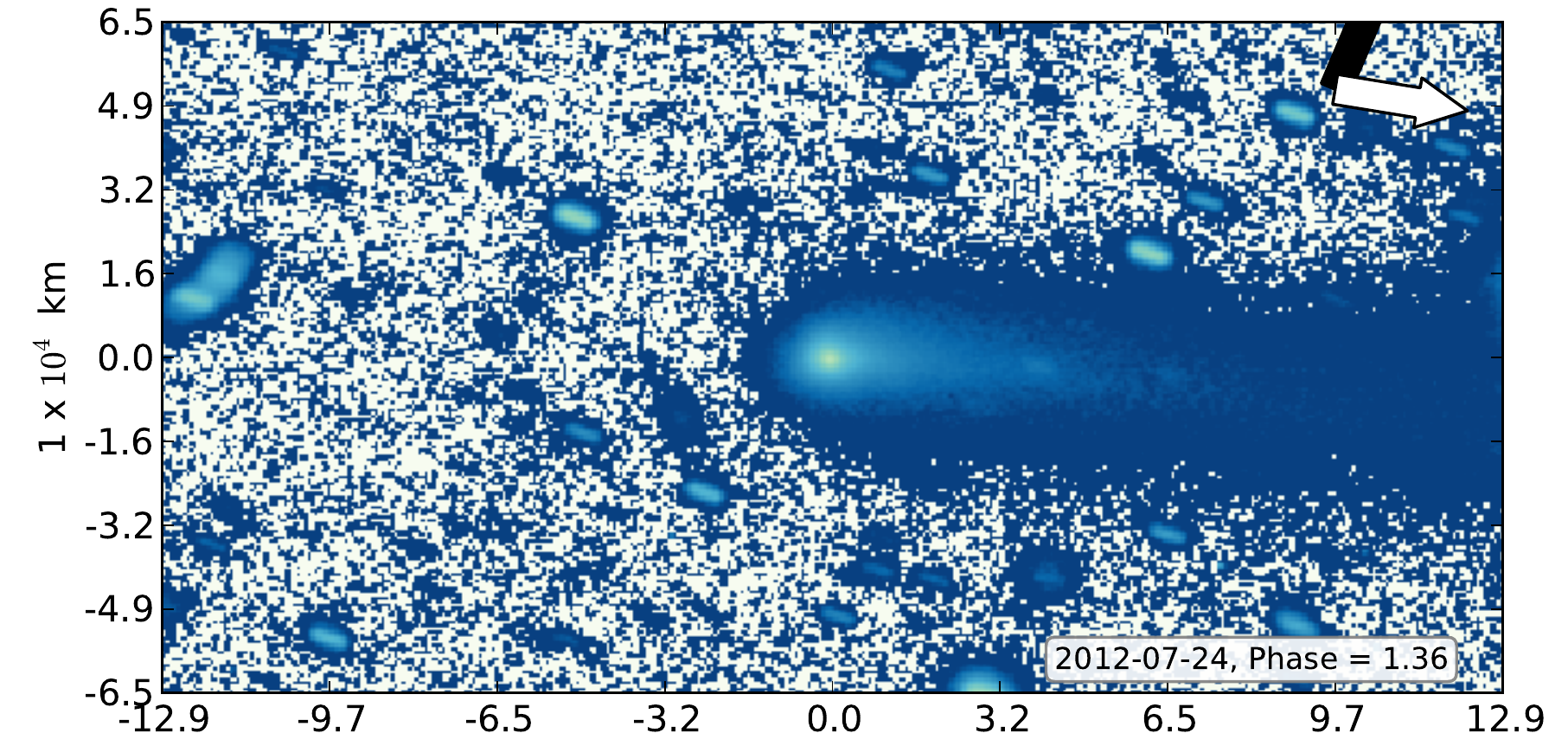}}
\subfigure{\includegraphics[scale=0.45]{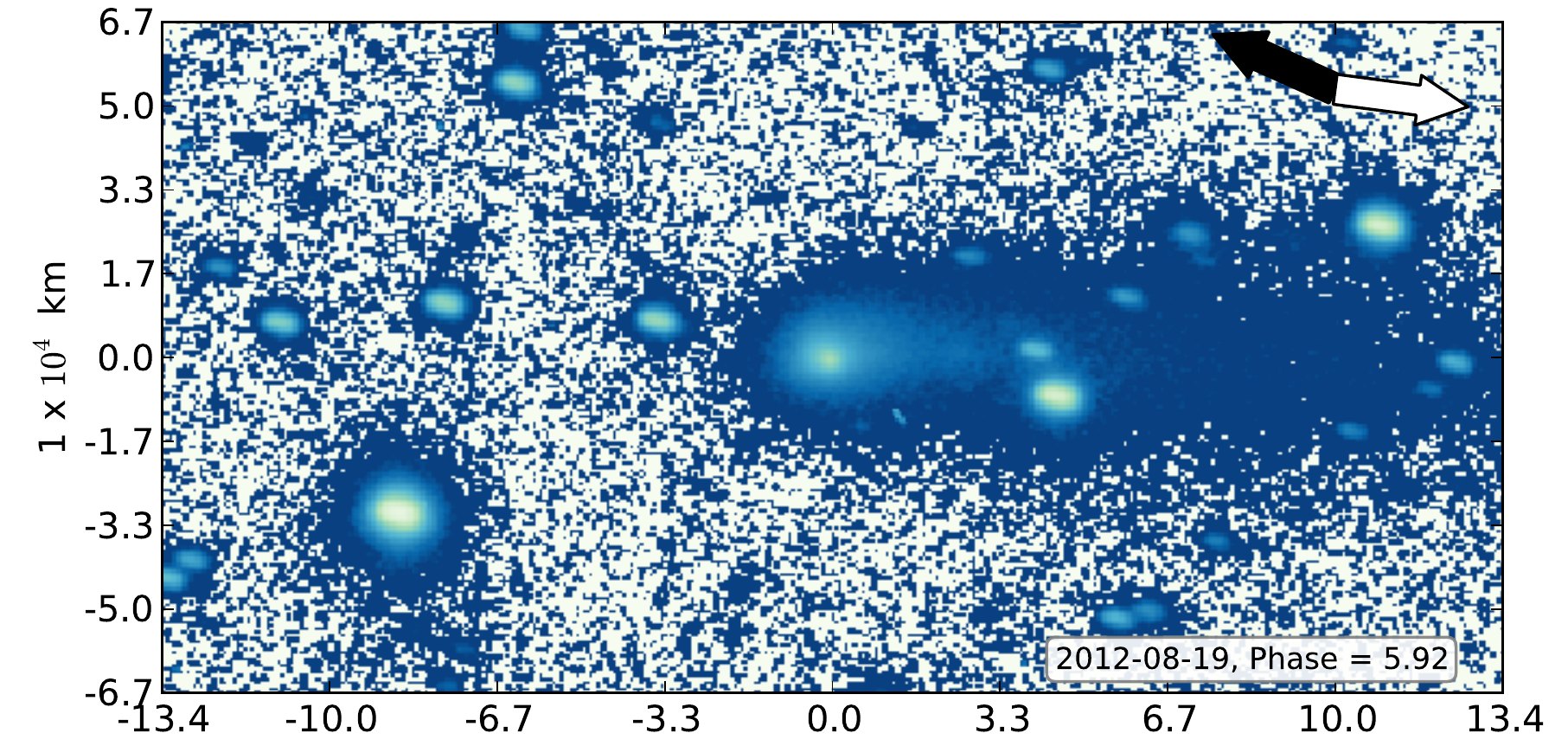}}
\subfigure{\includegraphics[scale=0.45]{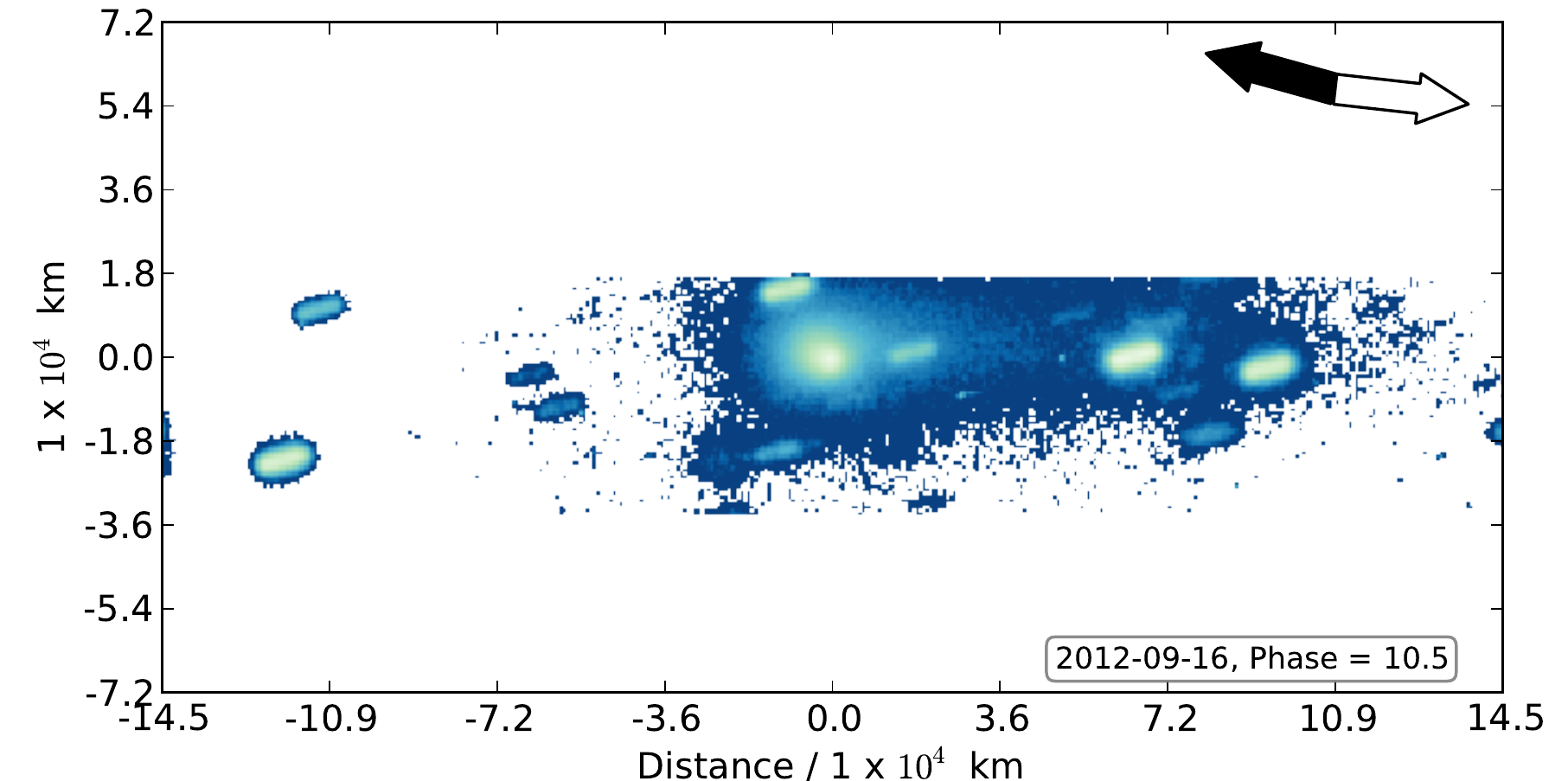}}
\caption{Intensity maps of comet 74P. The white arrow points in the direction of the negative target velocity as seen in the observer's plane of sky. The black arrow is the direction of the anti-solar direction. North is up and east is to the left.}
\label{fig:Imap_74P}
\end{figure}

\begin{table}
\begin{center}
\caption{\label{tab:74P_FIN} Comparison between measured position angle and Finson-Probstein synchrone (sync) analysis at 30, 60, 120, 240, and 360 days for comet 74P.}
\begin{tiny}
\begin{tabular}{ccccccc}

\hline \hline

Date &	PA 	& PA  &	PA  &	PA  &	PA  &	PA  \\ 
	& Tail & Sync& Sync & Sync & Sync & Sync \\
	& measured	& 30 d 	& 60 d 	& 120 d &	240 d &	360 d \\ 
(UT)&	(deg) &	(deg)&	(deg)&	(deg)&	(deg)&	(deg)\\ \hline

2012-Jun-10&	264-264&	262&		261&		261&		261&		261 \\
2012-Jun-26&	260-264&	267&		266&		266&		266&		266 \\
2012-Jul-17&	263-268&	290&		277&		270&		268&		267 \\
2012-Jul-24&	267-269&	315&		287&		274&		269&		268 \\
2012-Aug-18&	268-272&	66&	    55&	    320&		275&		270 \\
2012-Aug-19&	268-272&	67&	    56& 		323&		275&		270 \\
2012-Sep-10&	272-280&	75&	    72&	    53&	    283&		272 \\
2012-Sep-16&	272-280&	77&	    74&	    60&	    285&		272 \\ \hline

\end{tabular}
\end{tiny}
\end{center}
\end{table}

The intensity images for comet 74P are presented in Fig. \ref{fig:Imap_74P}.  No distinct coma structure was found in the processed images indicating no localised activity on the nucleus. Again given the broad wavelength range of the filters used, the coma and tail are mostly representing the dust distribution around the nucleus. Features that could be attributed to gas and plasma are not seen in the images. A dust tail is present throughout the observing period in a westward direction. At the beginning of the observing period this dust tail was also slightly curved towards the south. In late September 2012, and possibly in August 2012, there is a noticeable additional coma extension into the north-western quadrant.\\ \indent
The Finson-Probstein analysis of the dust tail geometry (Table \ref{tab:74P_FIN}) suggests that there is a coma asymmetry in the north-western coma section in September 2012 maybe due to dust grains released during the previous two months before the observations and projected into that coma quadrant as seen from Earth. The westward pointing tail at this time consists of very old dust emitted by the nucleus about a year earlier. The old dust overlaps with the more recent grains during the June and July observing epochs forming a brighter and wider slightly curved dust tail as seen in the images.  

\begin{figure}[tpb!]
\resizebox{\hsize}{!}{\includegraphics[scale=0.5]{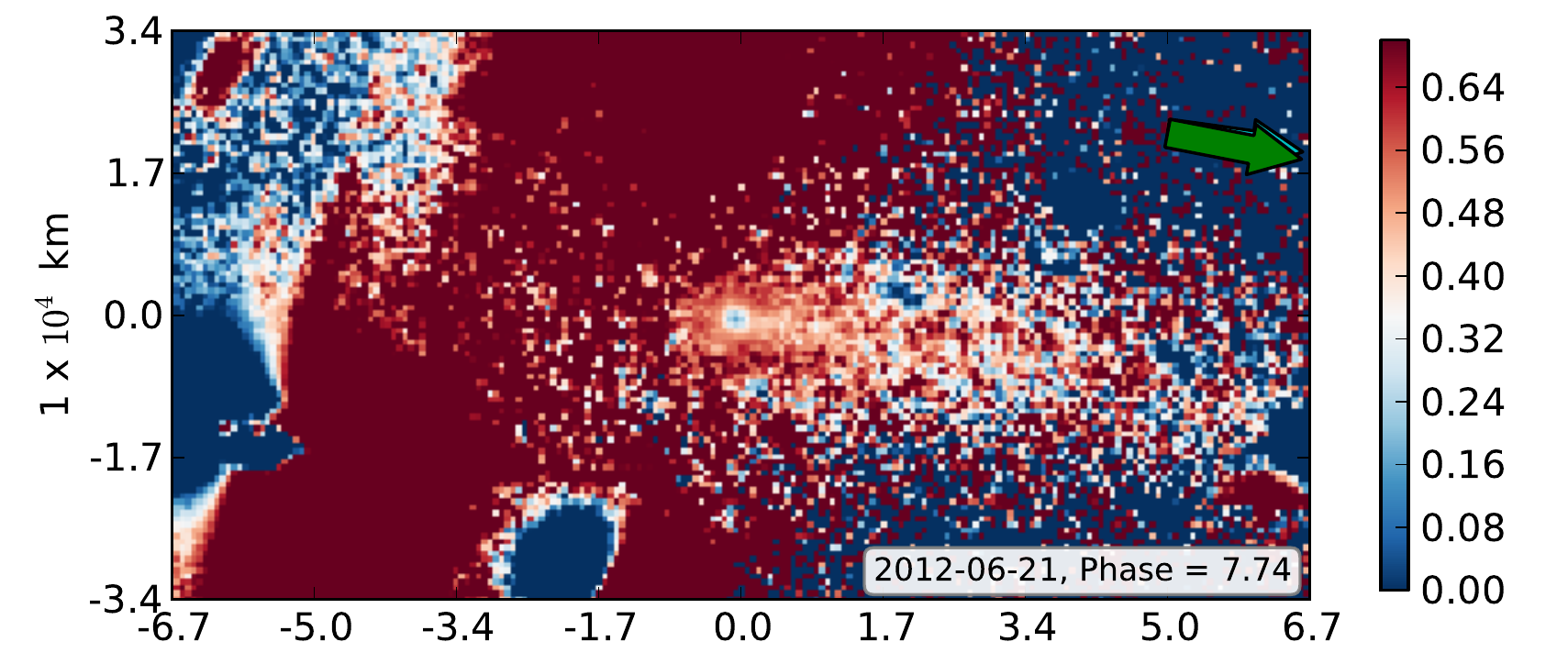}}
\resizebox{\hsize}{!}{{\includegraphics[scale=0.5]{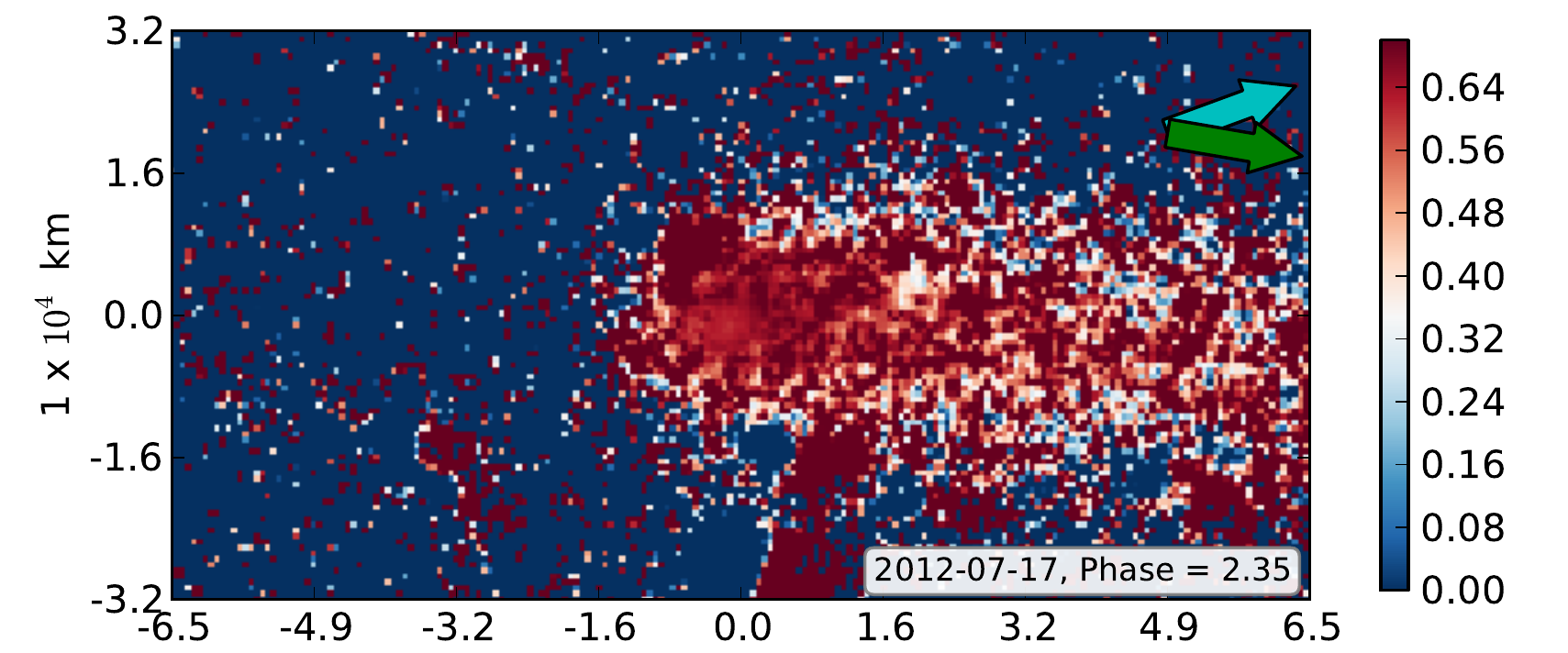}}}
\resizebox{\hsize}{!}{{\includegraphics[scale=0.5]{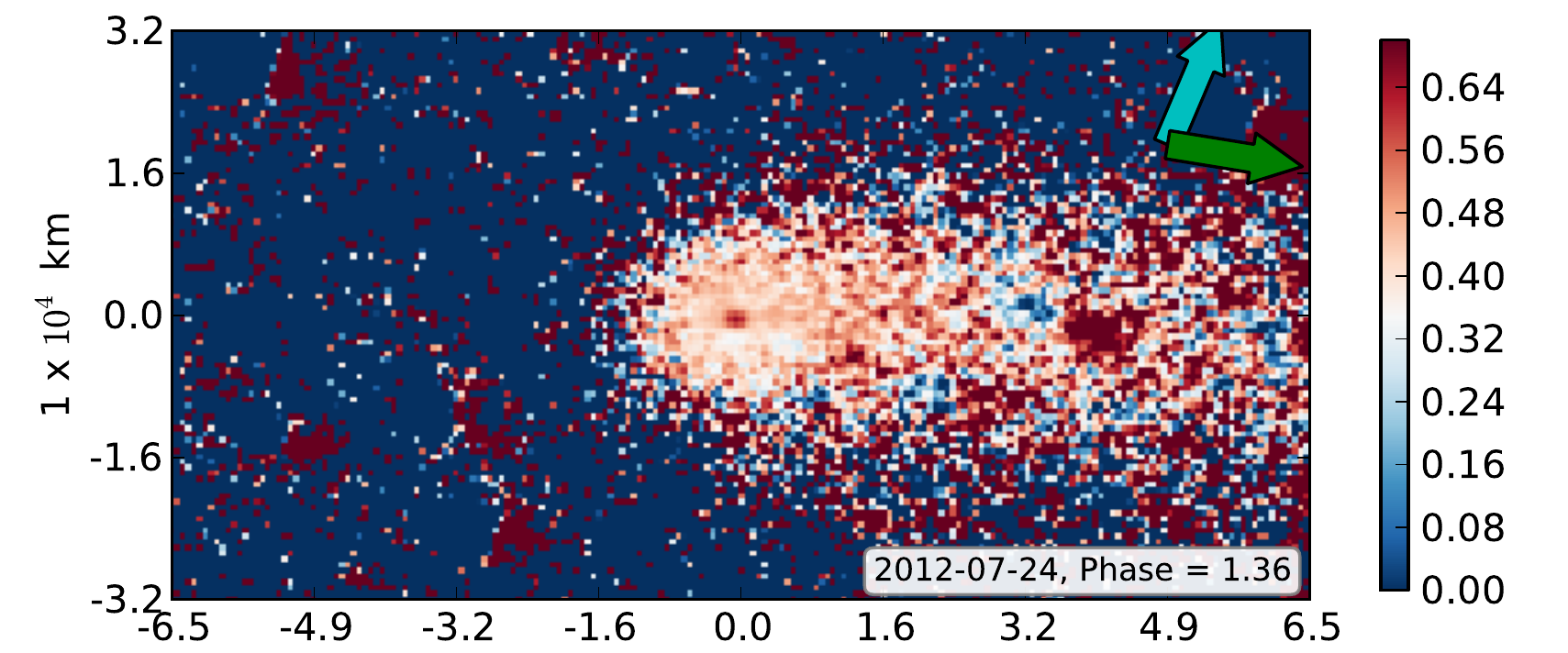}}}
\resizebox{\hsize}{!}{{\includegraphics[scale=0.5]{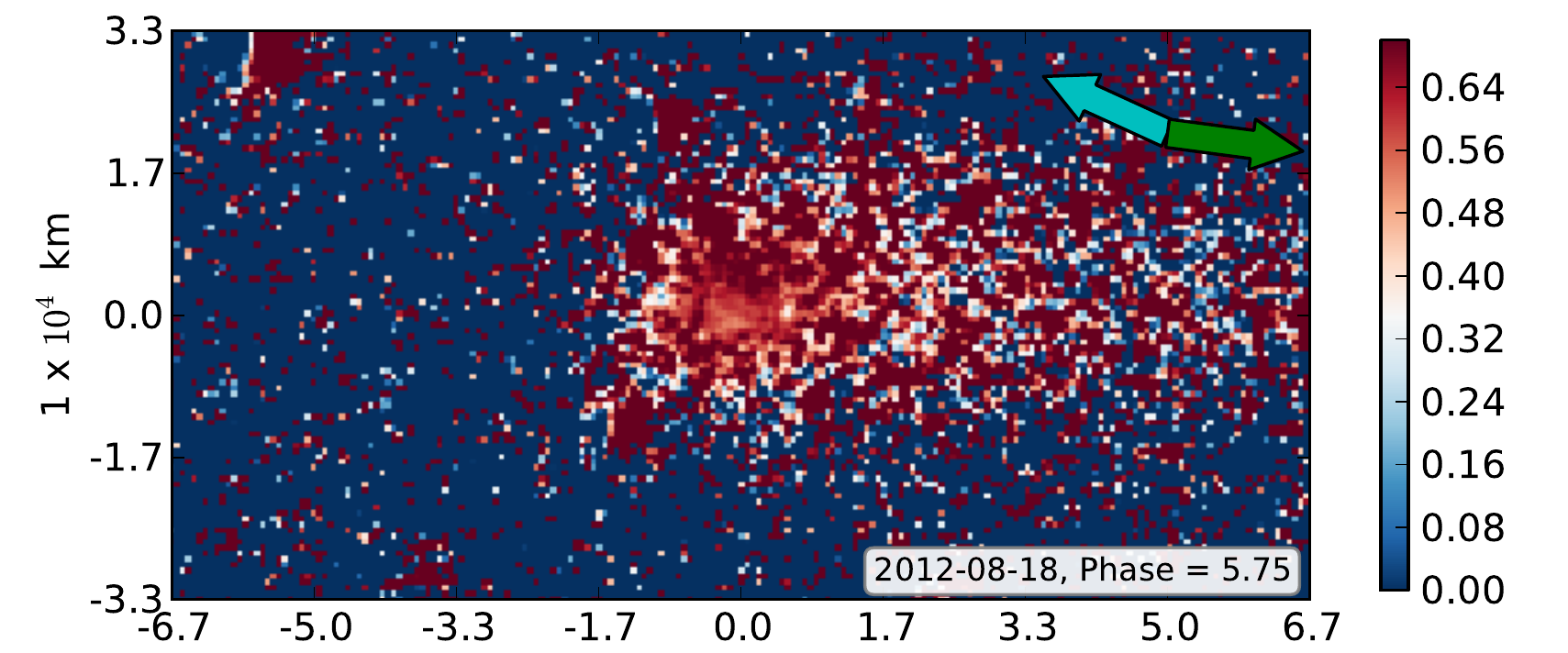}}}
\resizebox{\hsize}{!}{{\includegraphics[scale=0.5]{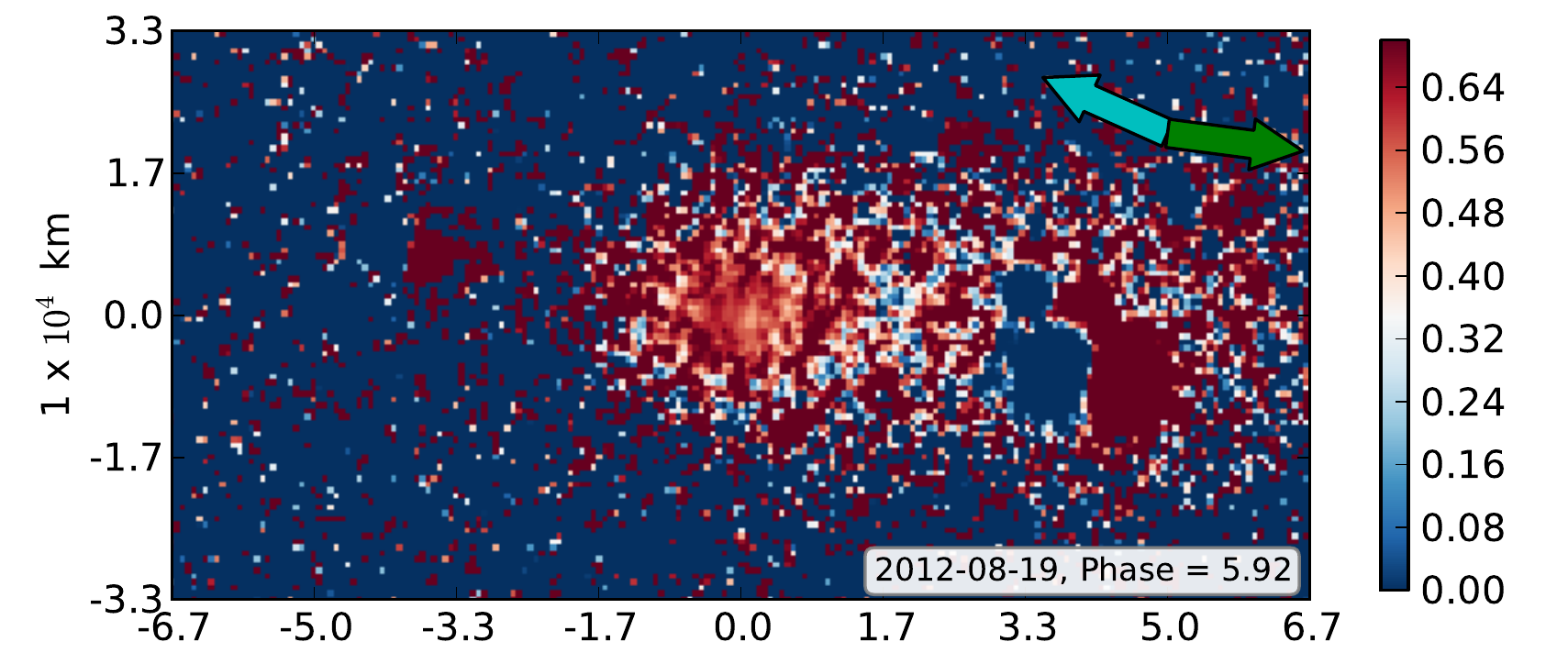}}}
\caption{\textit{V-R} colour map of comet 74P. The green arrow points in the direction of the negative target velocity as seen in the observer's plane of sky. The cyan arrow is the direction of the anti-solar direction. North is up and east is to the left.}
\label{fig:cmap_74P}
\end{figure}

\subsubsection{Colour map}
\label{sec:74P_cmap}

The colour maps produced for comet 74P are shown in Fig. \ref{fig:cmap_74P}. The blue and red spots exhibited on the nights of 21 June and 24 July 2012 are artificial features caused by a large seeing difference between the \textit{V} and \textit{R} images. On the other nights not affected by seeing changes there is no clear evidence of colour variation within the coma or tail region of the comet.

\subsubsection{Polarimetric maps}
\label{sec:74P_polmap}

\begin{figure*}[htbp!]
\resizebox{\hsize}{!}{{\includegraphics[scale=0.5]{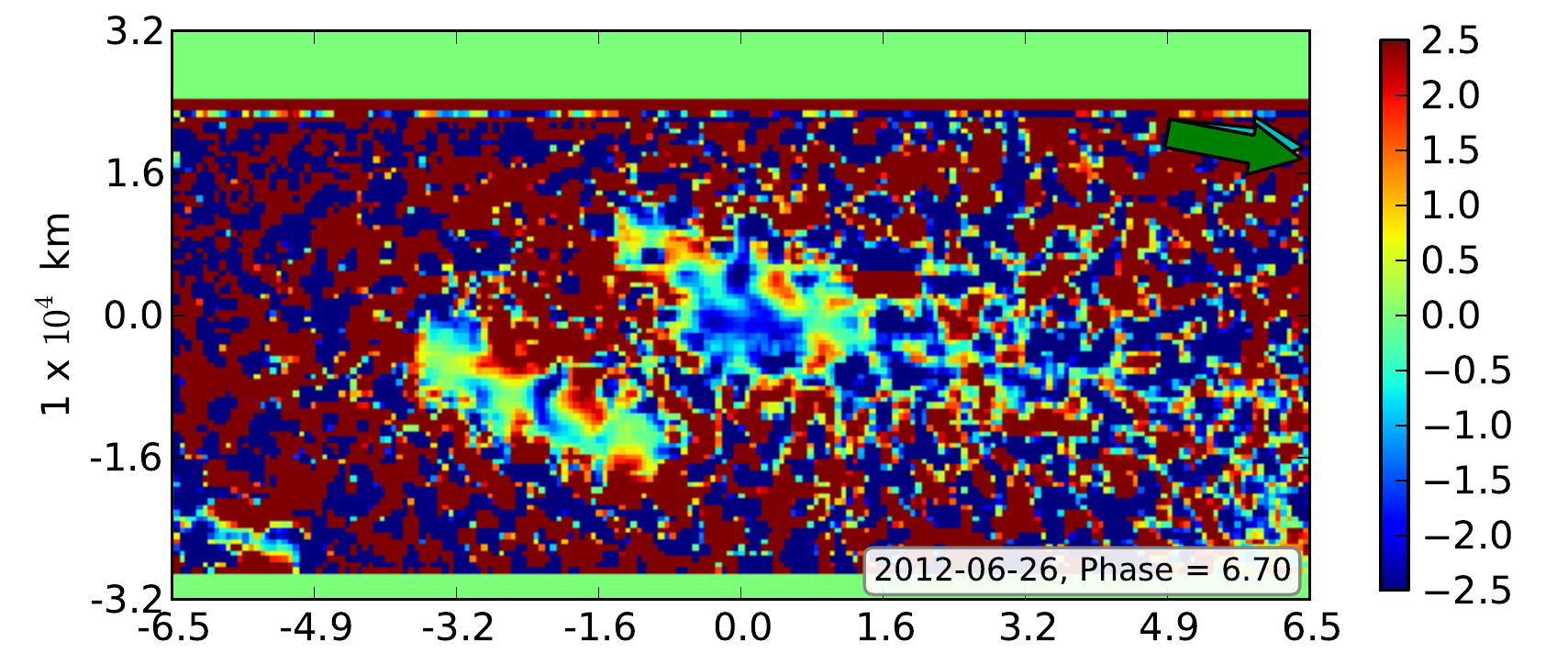}}{\includegraphics[scale=0.5]{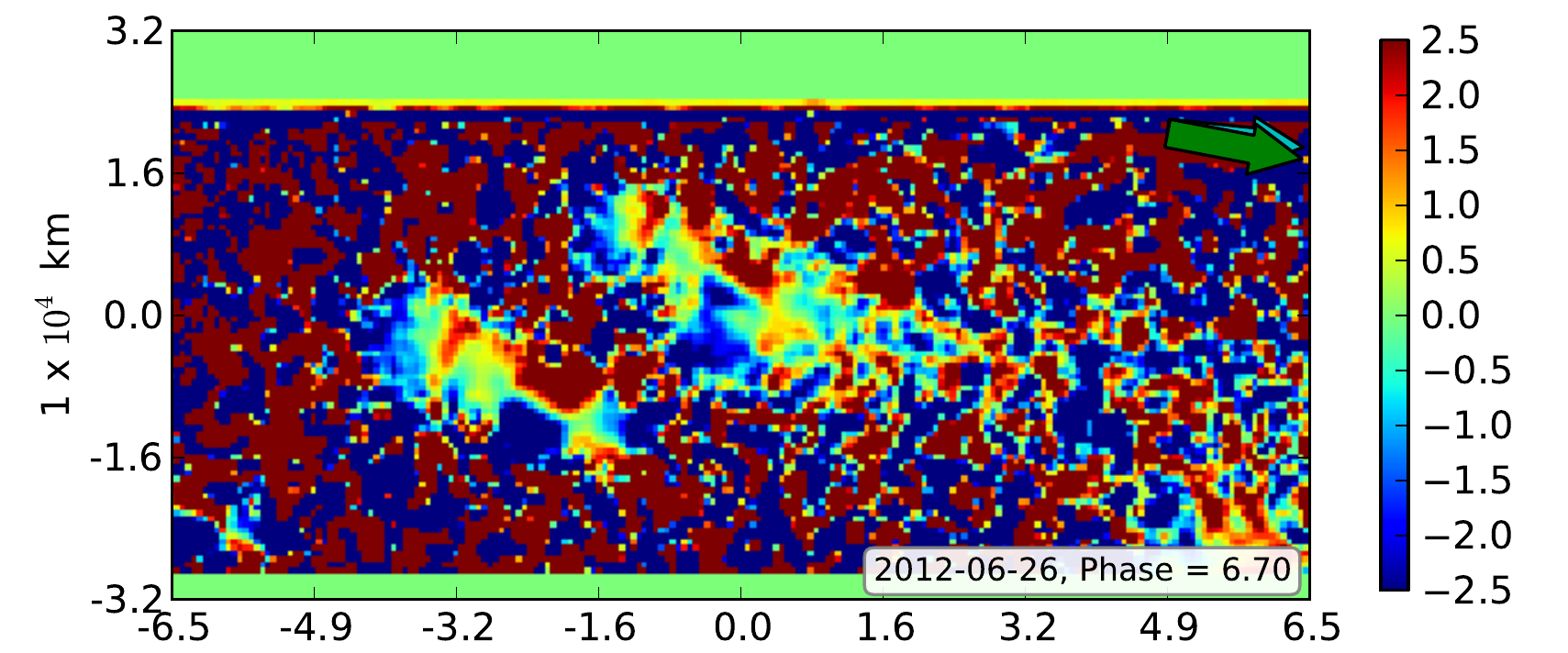}}}
\resizebox{\hsize}{!}{{{\includegraphics[scale=0.5]{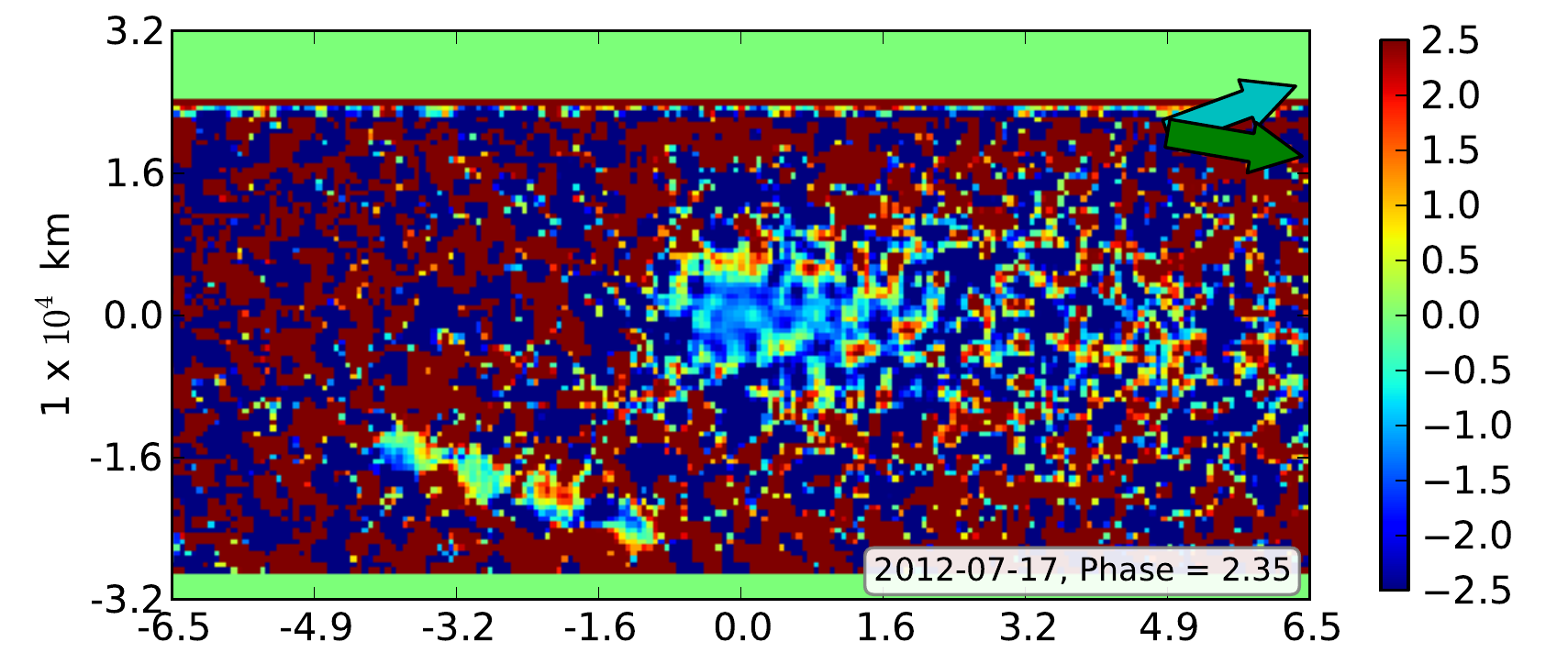}}}{{\includegraphics[scale=0.5]{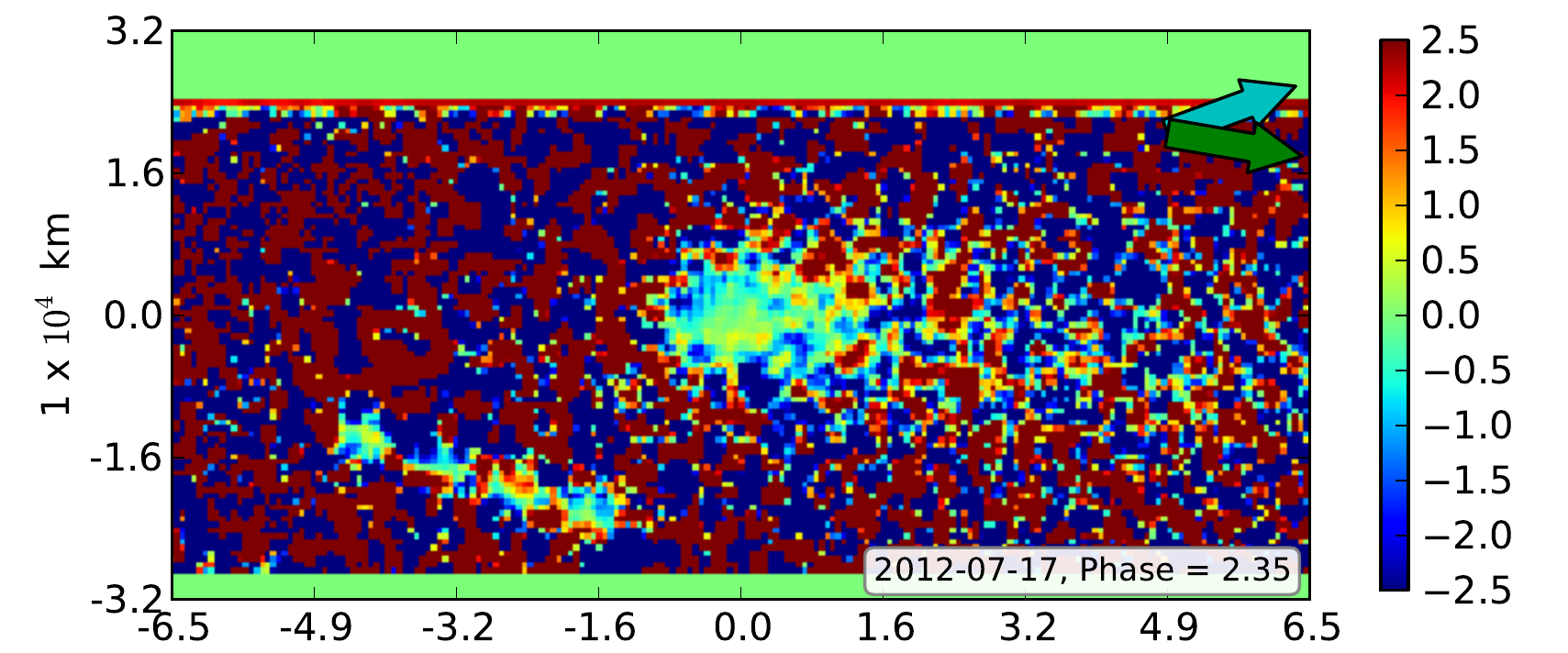}}}}
\resizebox{\hsize}{!}{{\includegraphics[scale=0.5]{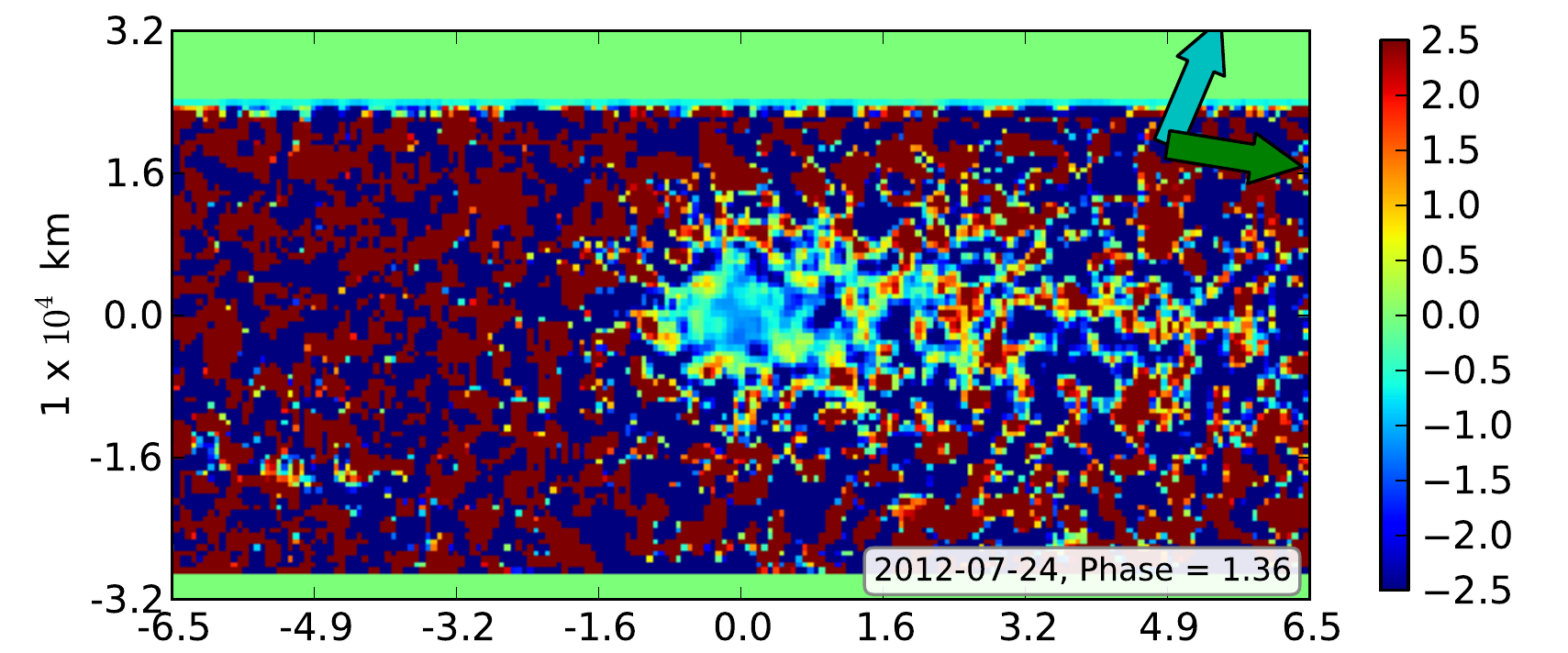}}{\includegraphics[scale=0.5]{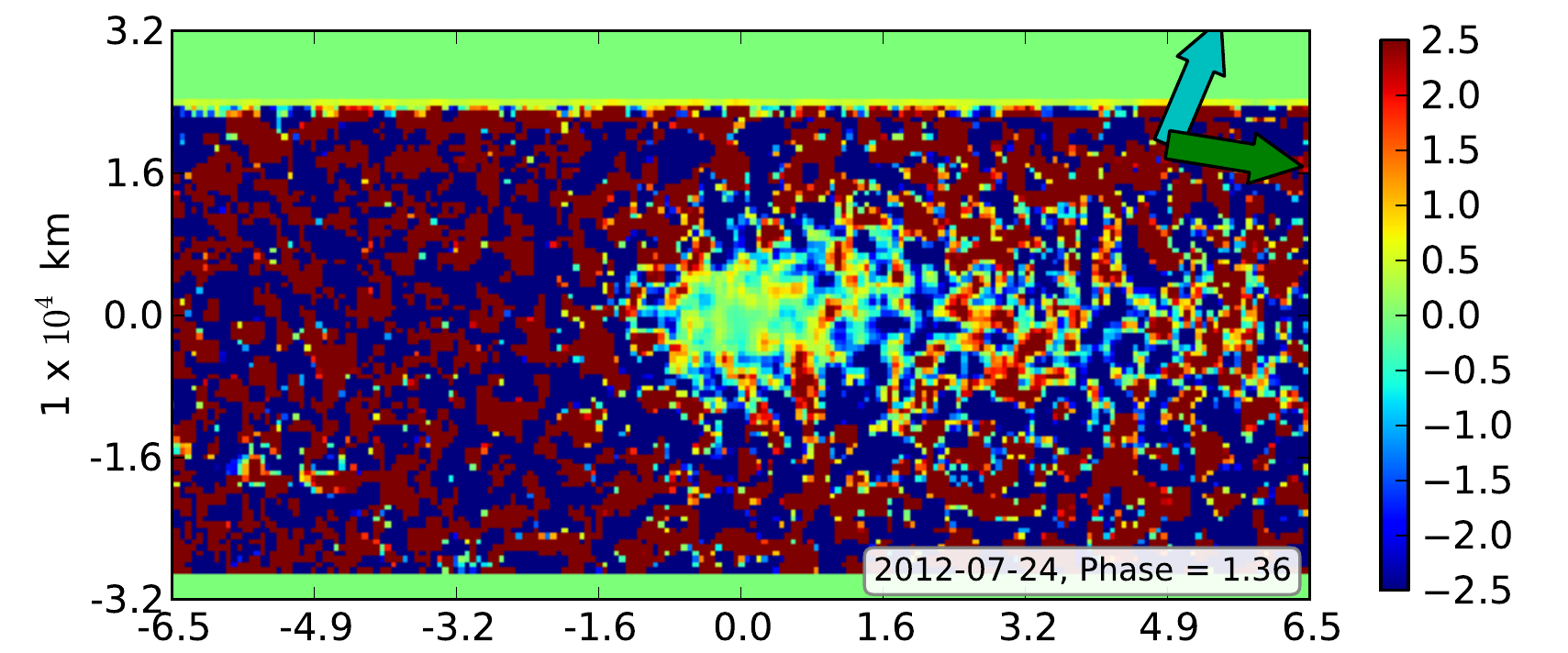}}}
\resizebox{\hsize}{!}{{\includegraphics[scale=0.5]{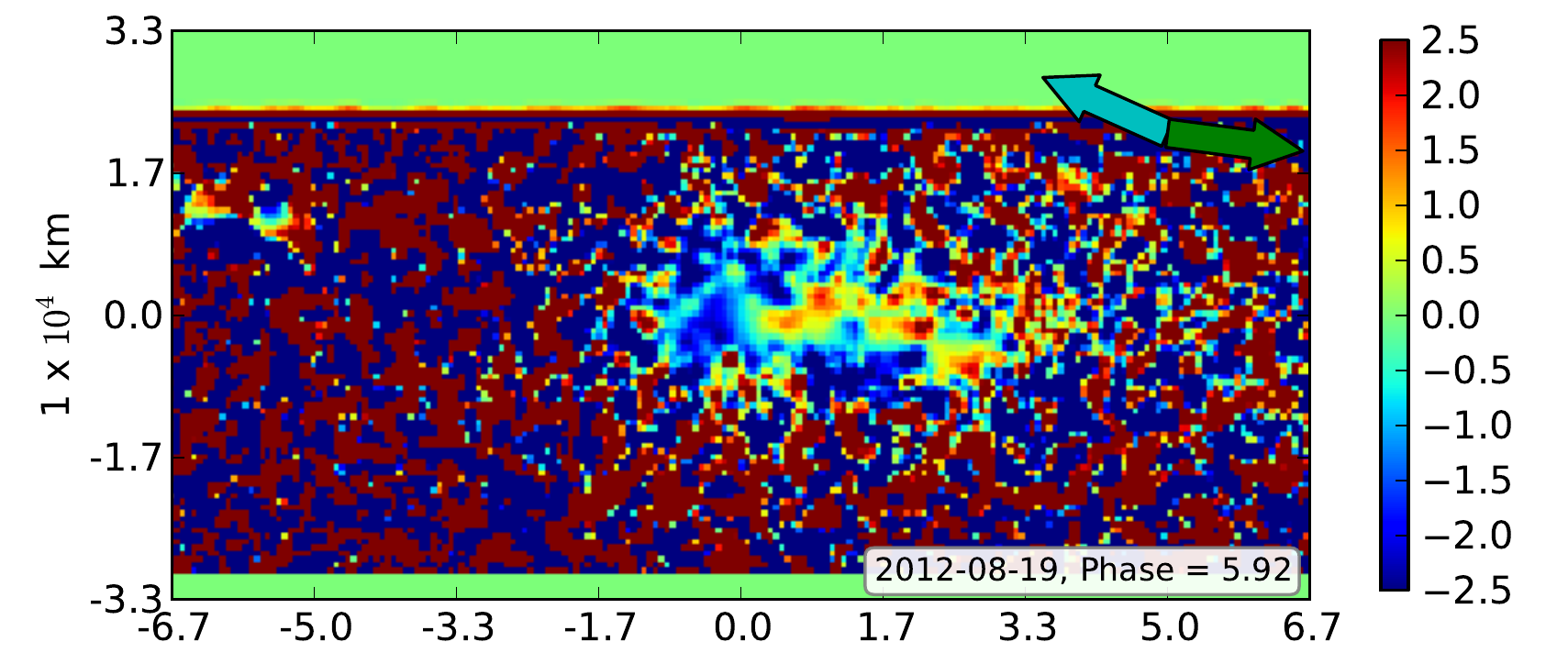}}{\includegraphics[scale=0.5]{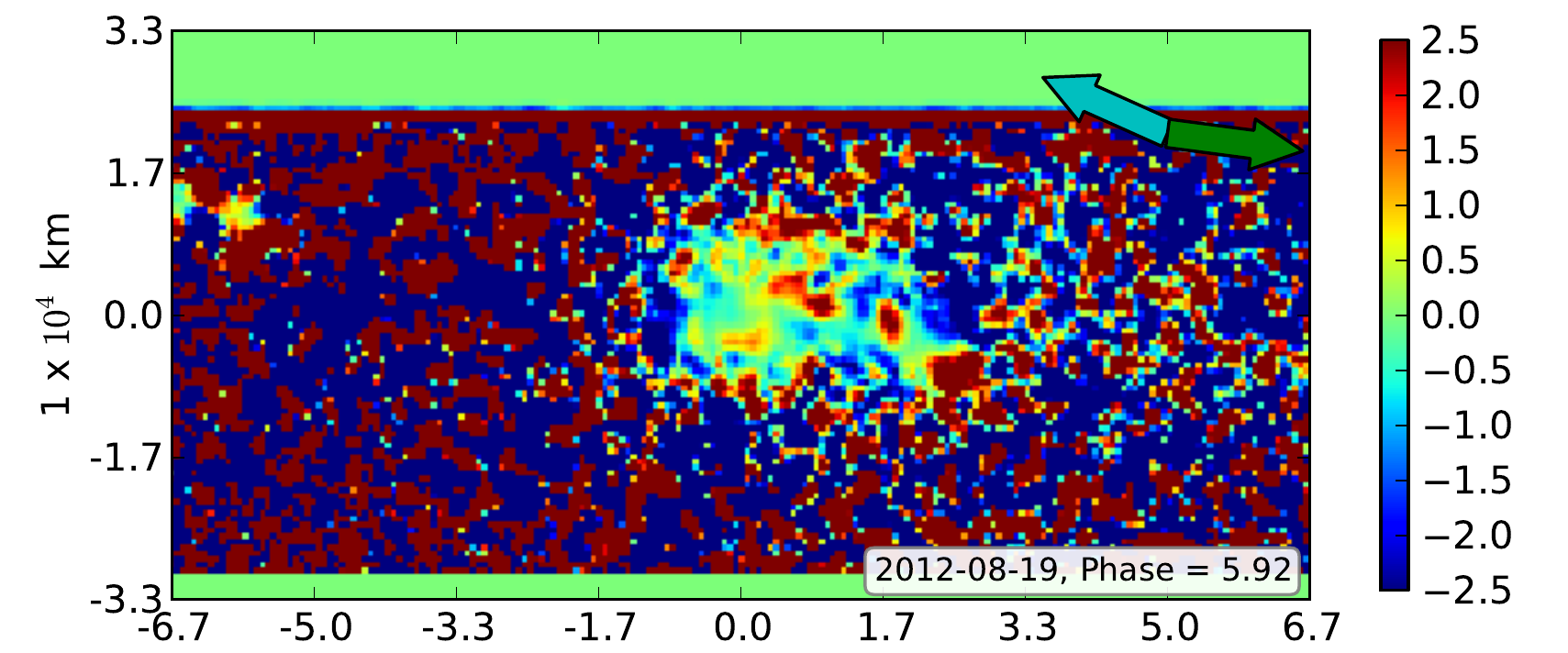}}}
\resizebox{\hsize}{!}{{\includegraphics[scale=0.5]{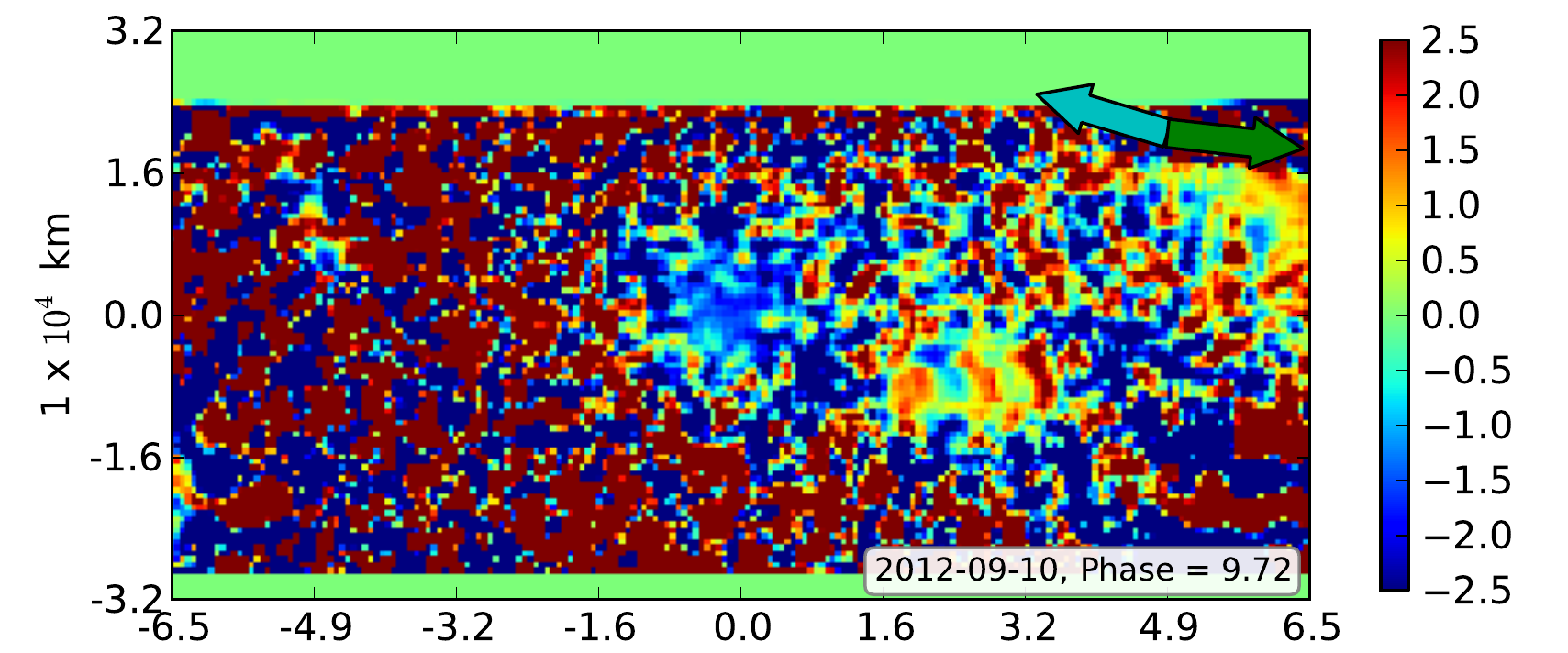}}{\includegraphics[scale=0.5]{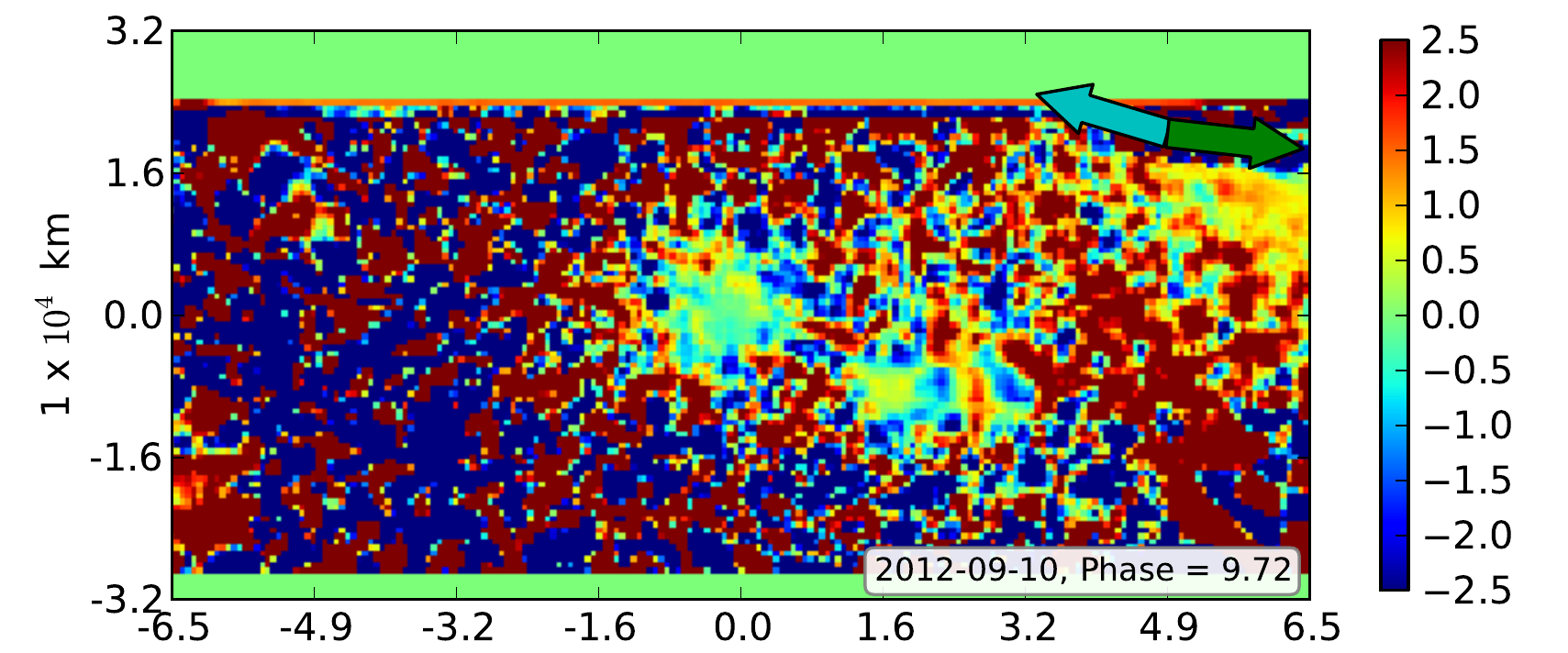}}}
\resizebox{\hsize}{!}{{\includegraphics[scale=0.5]{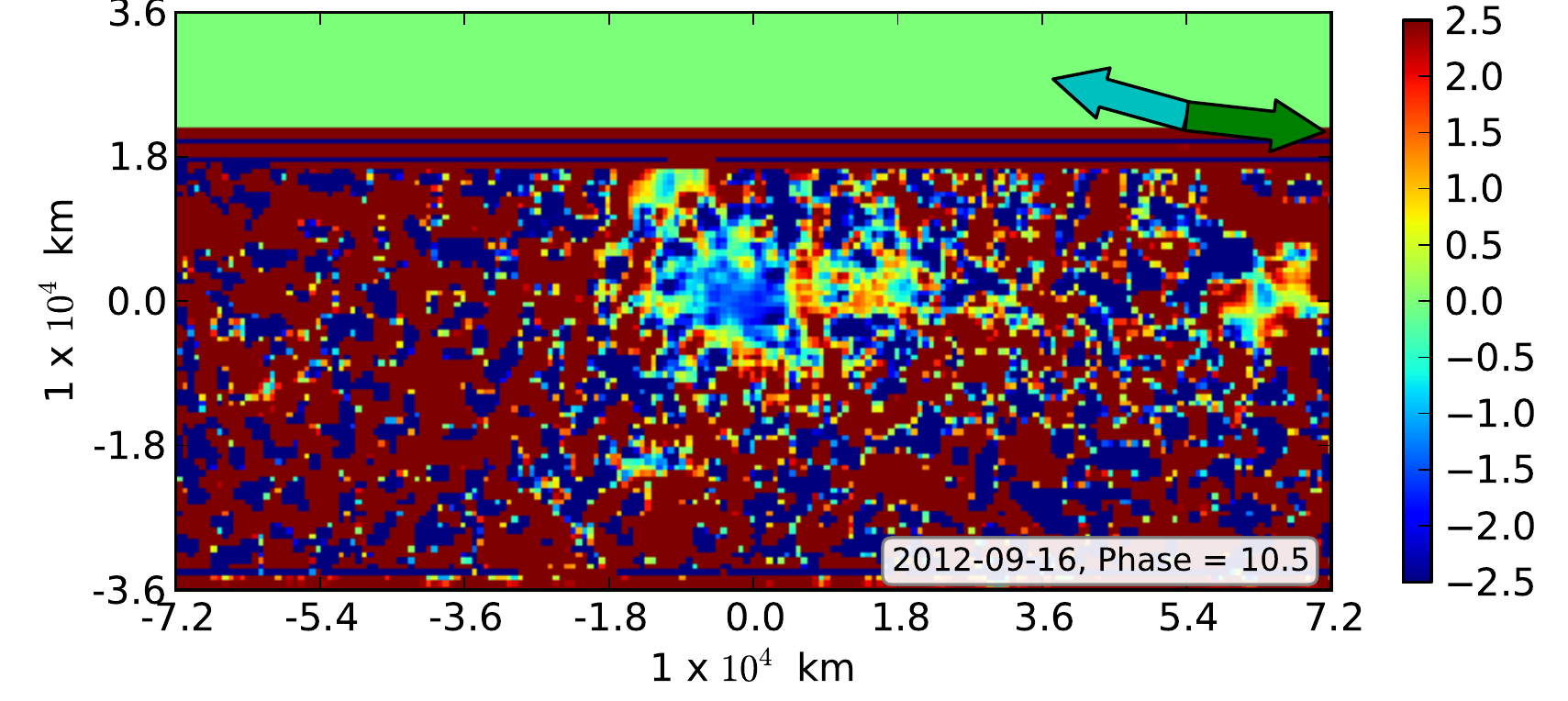}}{\includegraphics[scale=0.5]{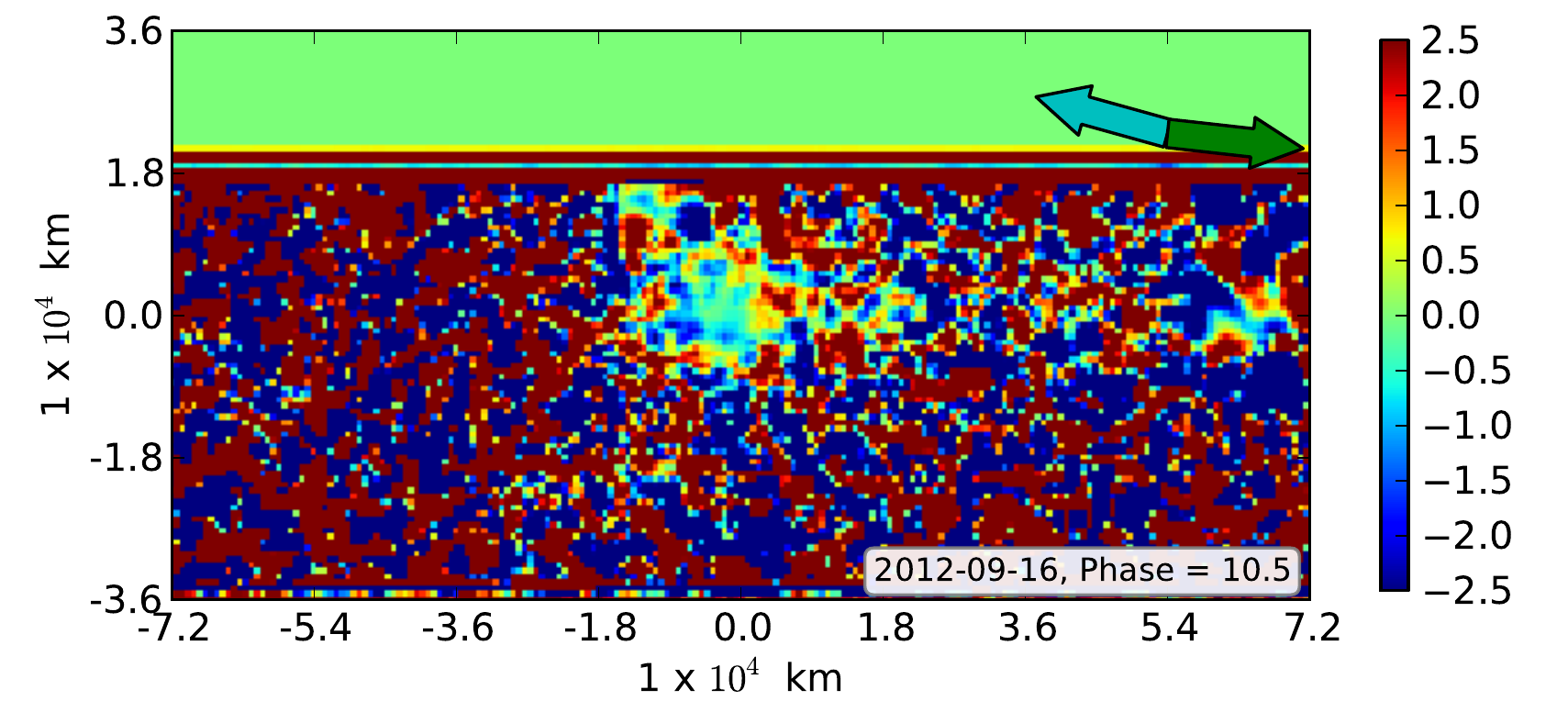}}}
\caption{$P'_Q$ (left) and $P'_U$ (right) polarimetric maps for the comet 74P. Both $P'_Q$  and $P'_U$ are measured in per cent. The green arrow points in the direction of the negative target velocity as seen in the observer's plane of sky. The cyan arrow is the direction of the anti-solar direction. North is up and east is to the left.}
\label{fig:pol_map_74P}
\end{figure*}

In Fig. \ref{fig:pol_map_74P} we present the polarisation maps for 74P. Since 74P was fainter and exhibited a narrow coma and tail the signal-to-noise ratio is lower than for 152P. Additionally the comet passes close to background stars which makes it difficult to search for features in the coma and tail region especially when the $3 \times 3$ box car is applied. The quality of the observations for 74P is lower as our photometric and polarimetric measurements did not always occur on the same night and the presence of background stars changed throughout our polarimetric observations.

On the nights of 26 June, 19 August, and 10 September 2012, the comet passes close to or in front of background stars, which affects our search for structures within the coma and tail of the comet. The least affected nights are the two nights in July 2012. In the $P'_Q$ maps for the observation on 17 and 24 July 2012 (Fig. \ref{fig:pol_map_74P}) we see no clear evidence of structure or change in the amount of polarisation across the coma. In the $P'_U$ maps on both nights there is a small residual polarisation, which, however, shows no definite structure that suggests a jet or an outburst is present. \\ \indent
On the night of 16 September 2012 there was a tracking issue that caused the target to drift toward the edge of the strip during the exposures. Nothing clear can be seen on this night, most likely due to the low signal-to-noise ratio from the data.

\begin{figure}[tpb]
\resizebox{\hsize}{!}{\includegraphics[scale=1.0]{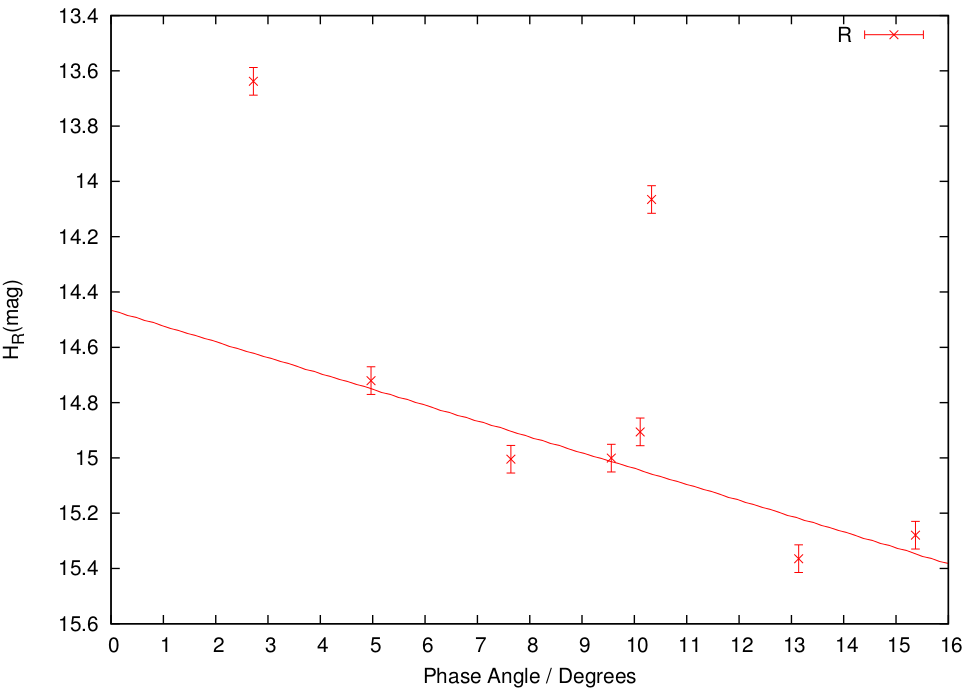}}
\caption{Magnitude corrected for the Sun and Earth distances of comet 67P as a function of phase angle. We note that the points at phase angles 2.7$^\circ$ and 10.3$^\circ$ are contaminated by background sources and are ignored.}
\label{fig:abso_67P}
\end{figure}

\begin{figure}[b!]
\resizebox{\hsize}{!}{\includegraphics[scale=0.4]{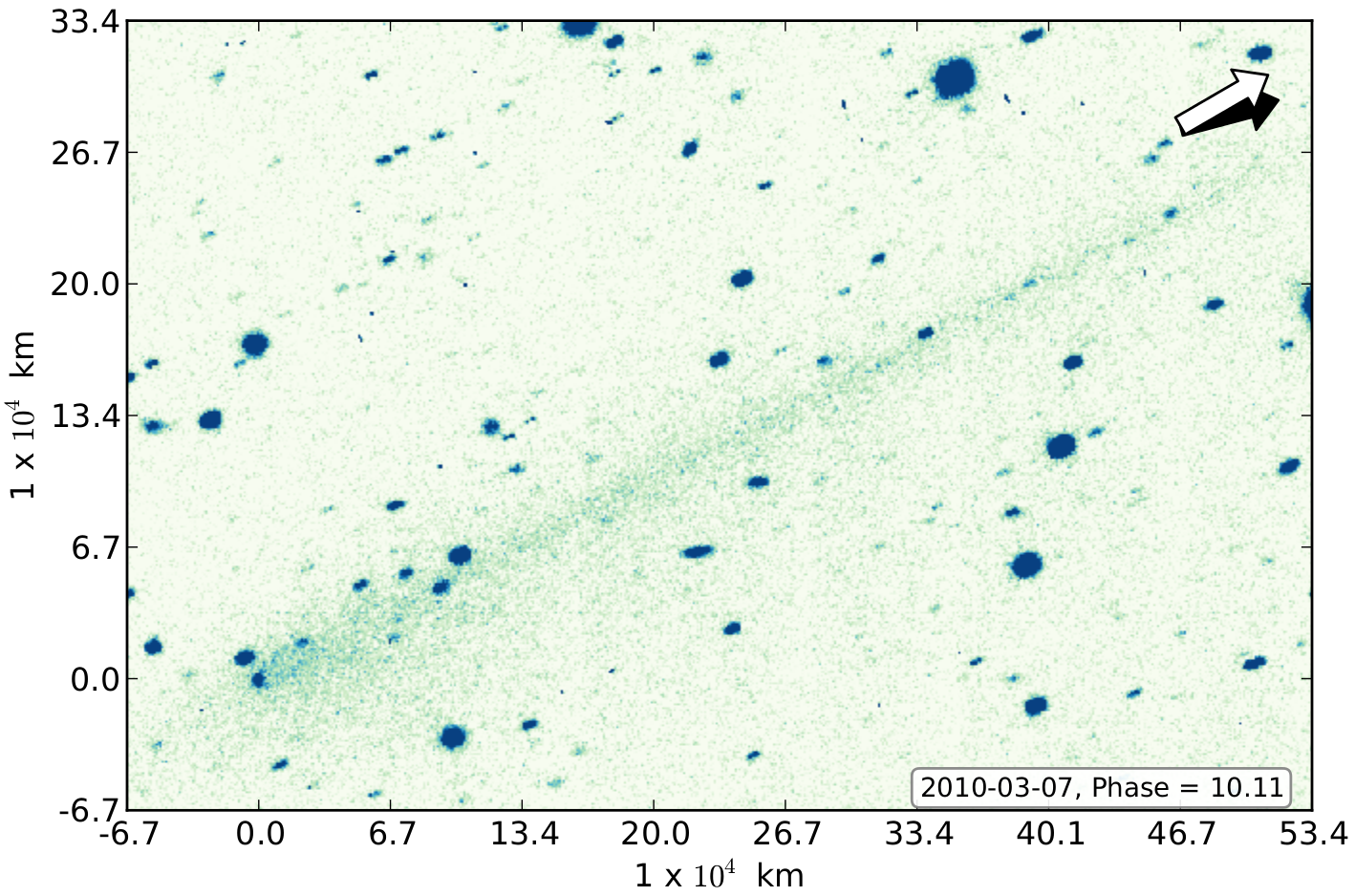}}
\caption{Intensity map of comet 67P. The white arrow points in the direction of the negative target velocity as seen in the observer's plane of sky. The black arrow is the direction of the anti-solar direction. North is up and east is to the left.}
\label{fig:Imap_67P}
\end{figure}
\subsection{Comet 67P/Churyumov–Gerasimenko}
\label{sec:67P}
Since 67P was only observed in the \textit{R} filter we were not able to create \textit{V-R} colour maps. Additionally 67P appeared 2-3 magnitudes fainter than the other comets meaning that the signal-to-noise ratio of the polarimetric measurements was so low that the polarimetric maps generated showed no clear structure.

\begin{figure*}[ht]
\resizebox{\hsize}{!}{\includegraphics[scale=0.5]{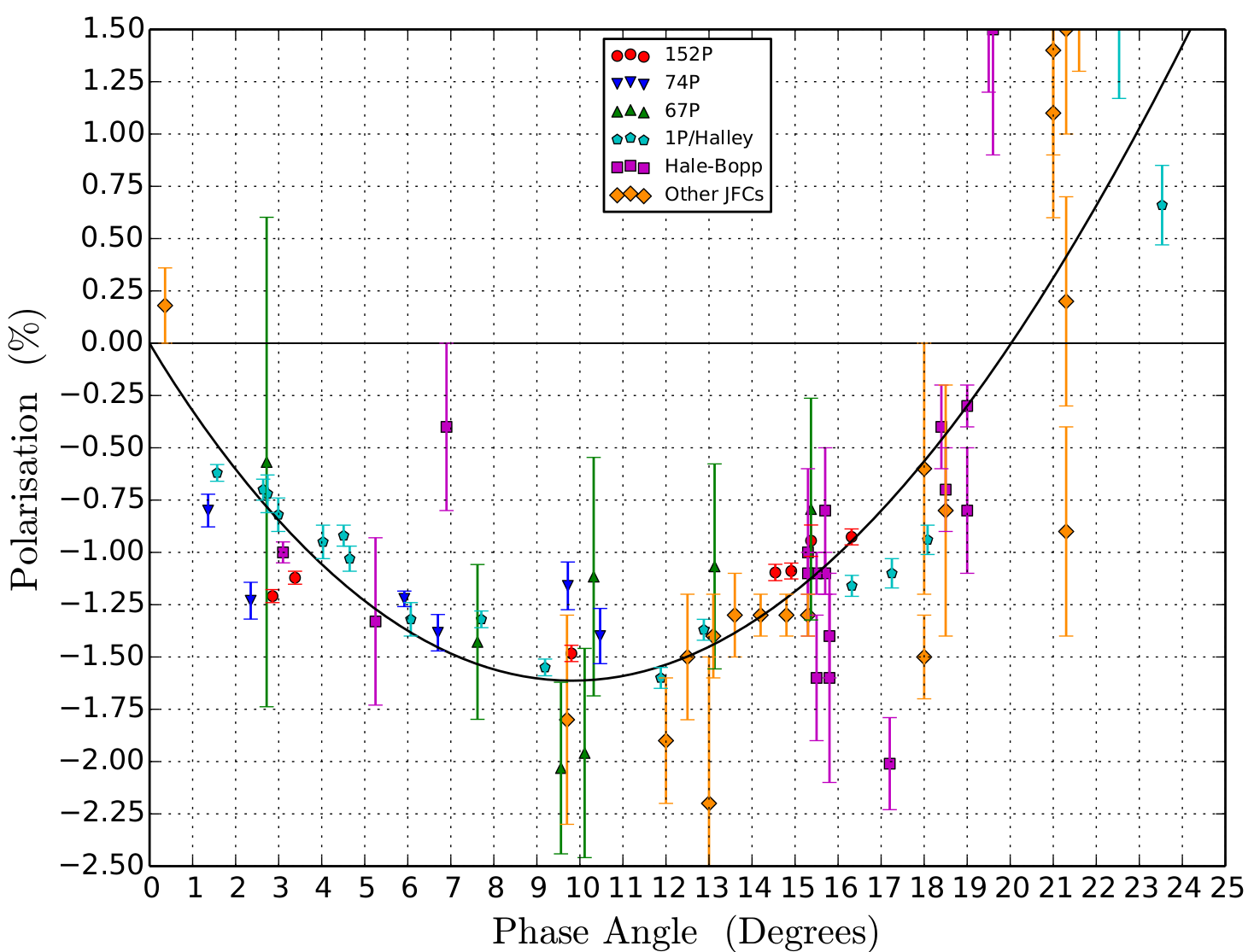}}
\caption{$P'_Q$ as a function of phase angle for comets 67P, 74P, and 152P. The solid black line is a best fit of previously observed comets. Data points shown are 1P/Halley, C/1995 O1 Hale-Bopp, and other Jupiter family comets, which include 22P/Kopff, 49P/Ashbrook-Jackson, observations of 67P/Churyumov-Gerasimenko from 1984, and 81P/Wild 2. These data points can all be found in the comet polarimetric database \citep{Kiselev2006b}.}
\label{fig:pol_phase}
\end{figure*}

\subsubsection{Aperture photometry}
The results for comet 67P are shown in Table 1. In Fig. \ref{fig:abso_67P} we plot the magnitude corrected for the Sun and Earth distances of comet 67P in the \textit{R}-Special filter as a function of phase angle. In this figure there is no evidence of an opposition surge at small phase angles. We note that the points at phase angles  2.7$^\circ$ and 10.3$^\circ$ are contaminated by background sources and are ignored. If we extrapolate, the average brightness at zero phase angle is 15.33 $\pm$ 0.11 in the \textit{R} filter.

Using Eq. \ref{eq:afrho} and the flux extrapolated back to zero phase in the \textit{R}-Special filter and the average $r$ and $\Delta$ distances to the comet yields an $Af\rho$ value of 18.71 $\pm$ 1.80\,cm. This compares to the value of $\leq$ 17.1\,cm measured by \citet{Lowry2003} while the comet was at aphelion.

\label{sec:67P_phot}

\subsubsection{Intensity maps}
\label{sec:67P_struct}

\begin{table}
\begin{center}
\caption{\label{tab:67P_FIN} Comparison between measured position angle and Finson-Probstein synchrone (sync) analysis at 30, 60, 120, 240, and 360 days for comet 67P.}
\begin{tiny}
\begin{tabular}{ccccccc}

\hline \hline

Date &	PA 	& PA  &	PA  &	PA  &	PA  &	PA  \\ 
	& Tail & Sync& Sync & Sync & Sync & Sync \\
	& measured	& 30 d 	& 60 d 	& 120 d &	240 d &	360 d \\ 
(UT)&	(deg) &	(deg)&	(deg)&	(deg)&	(deg)&	(deg)\\ \hline

2010-Feb-09&	298&	290&	291&	293&	295&	297 \\
2010-Feb-22&	297&	288&	290&	293&	295&	297 \\
2010-Mar-06&	295&	286&	288&	292&	295&	297 \\
2010-Mar-07&	296&	286&	288&	292&	295&	297 \\
2010-Mar-09&	296&	285&	288&	291&	295&	297 \\
2010-Mar-16&	295&	284&	287&	291&	295&	297 \\
2010-Mar-25&	295&	276&	284&	290&	295&	297 \\
2010-Mar-29&	296&	272&	283&	290&	295&	297 \\ \hline

\end{tabular}
\end{tiny}
\end{center}
\end{table}

An intensity image for comet 67P is presented in Fig. \ref{fig:Imap_67P}; we only present a single intensity image as all the exposures look quite similar. The coma of 67P does not show any sign of structure. However, we note that the dust coma extends asymmetrically and is larger in the southern direction suggesting ongoing activity in the southern part of the nucleus. There is also a significant but weak $\sim$25" coma peak in the direction of the Sun. The overall appearance of the coma and tail does not change over the course of our observations. The tail orientation is very constant at position angle $\sim$296-297$^\circ$; there is a slight trend to lower position angles with time.\\ \indent
Finson Probstein calculations of the dust tail (Table \ref{tab:67P_FIN}) indicate that the material defining the main tail axis appears to be old and may have been produced by the nucleus about 1/2 to 1 year before observation.

\subsection{Aperture polarimetry of 152P, 74P, and 67P}
\label{sec:pol_phase}

\begin{figure*}[t!]

\resizebox{\hsize}{!}{{\includegraphics[scale=0.5]{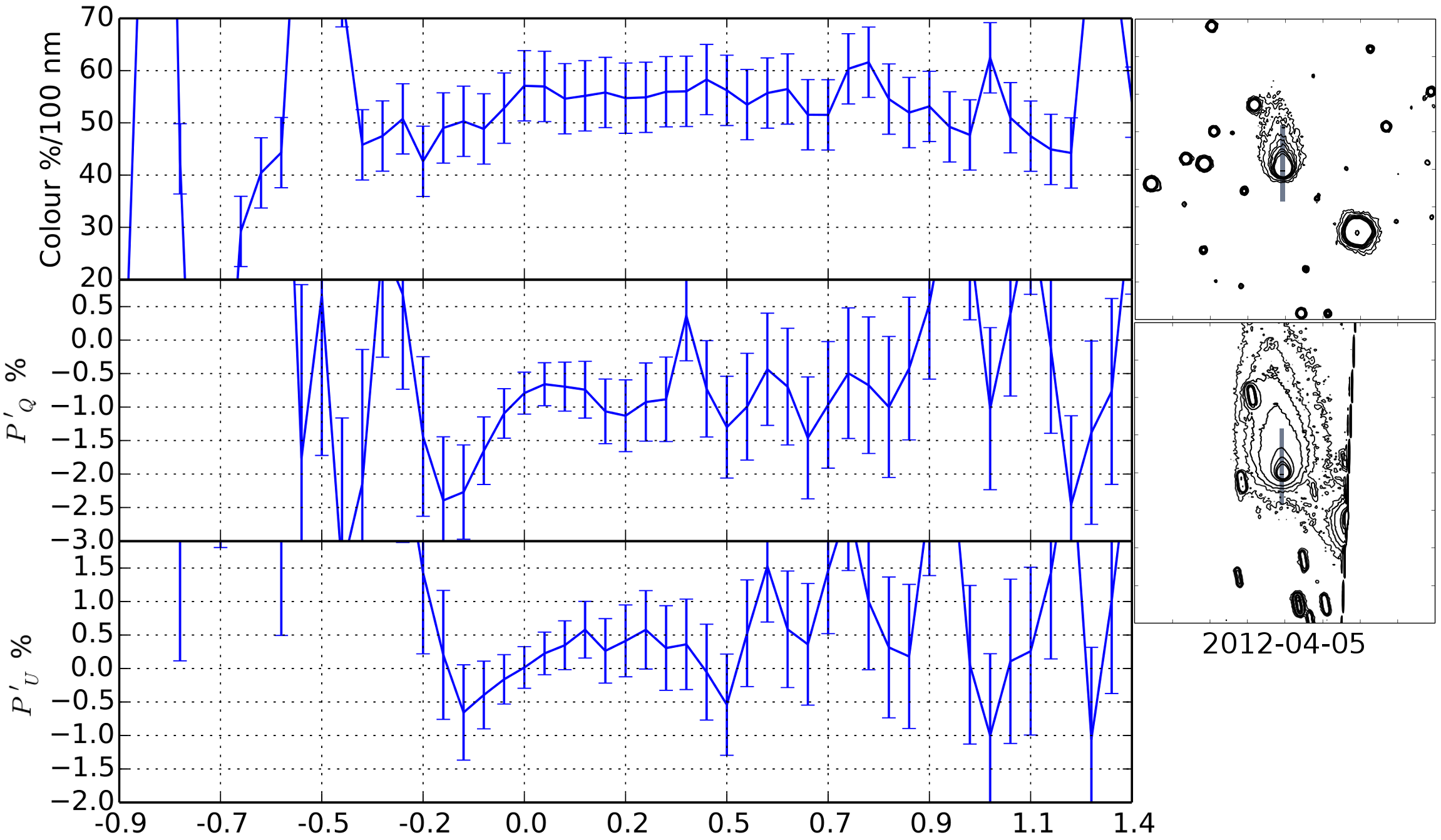}}{\includegraphics[scale=0.5]{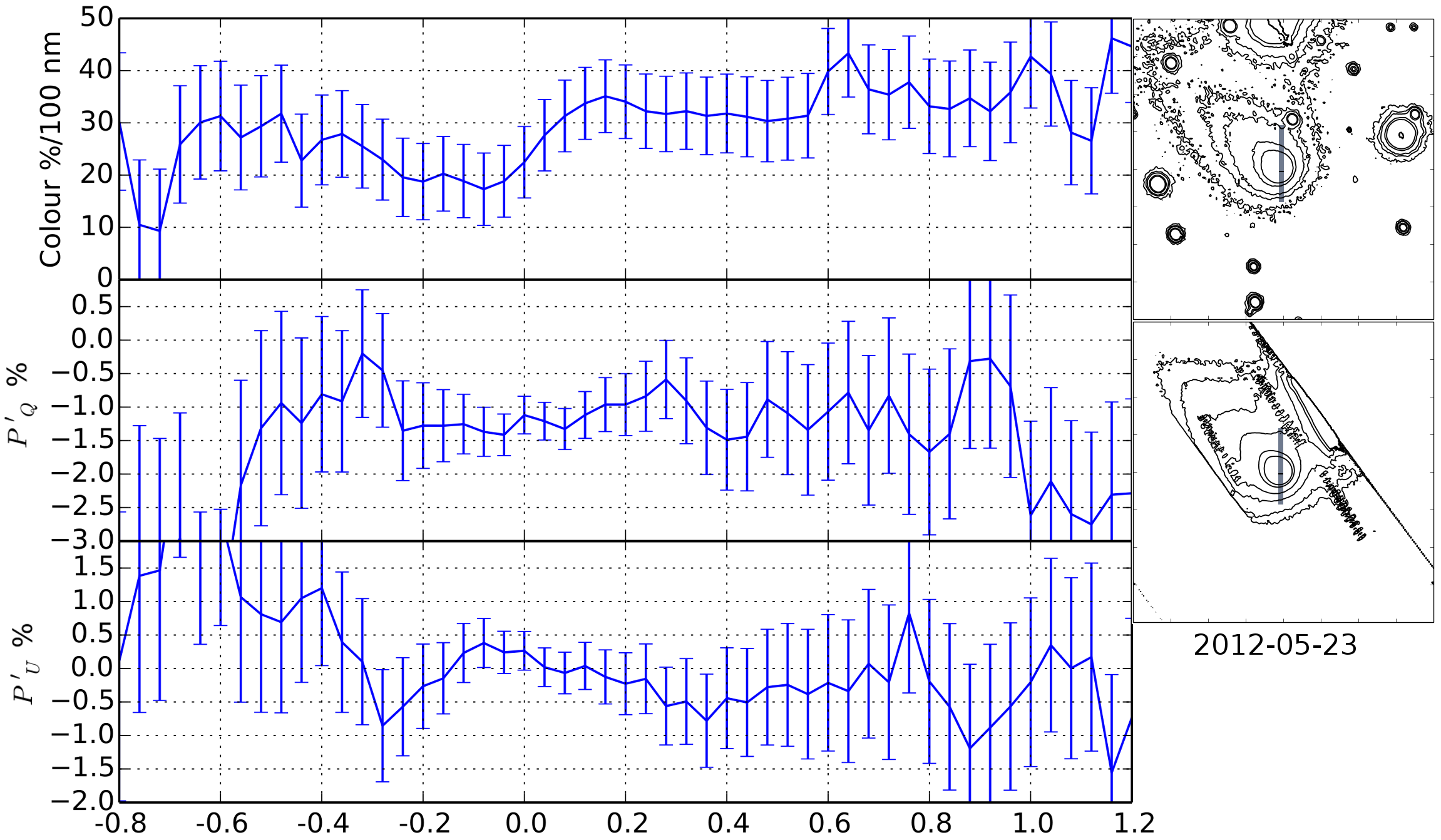}}}
\resizebox{\hsize}{!}{{\includegraphics[scale=0.5]{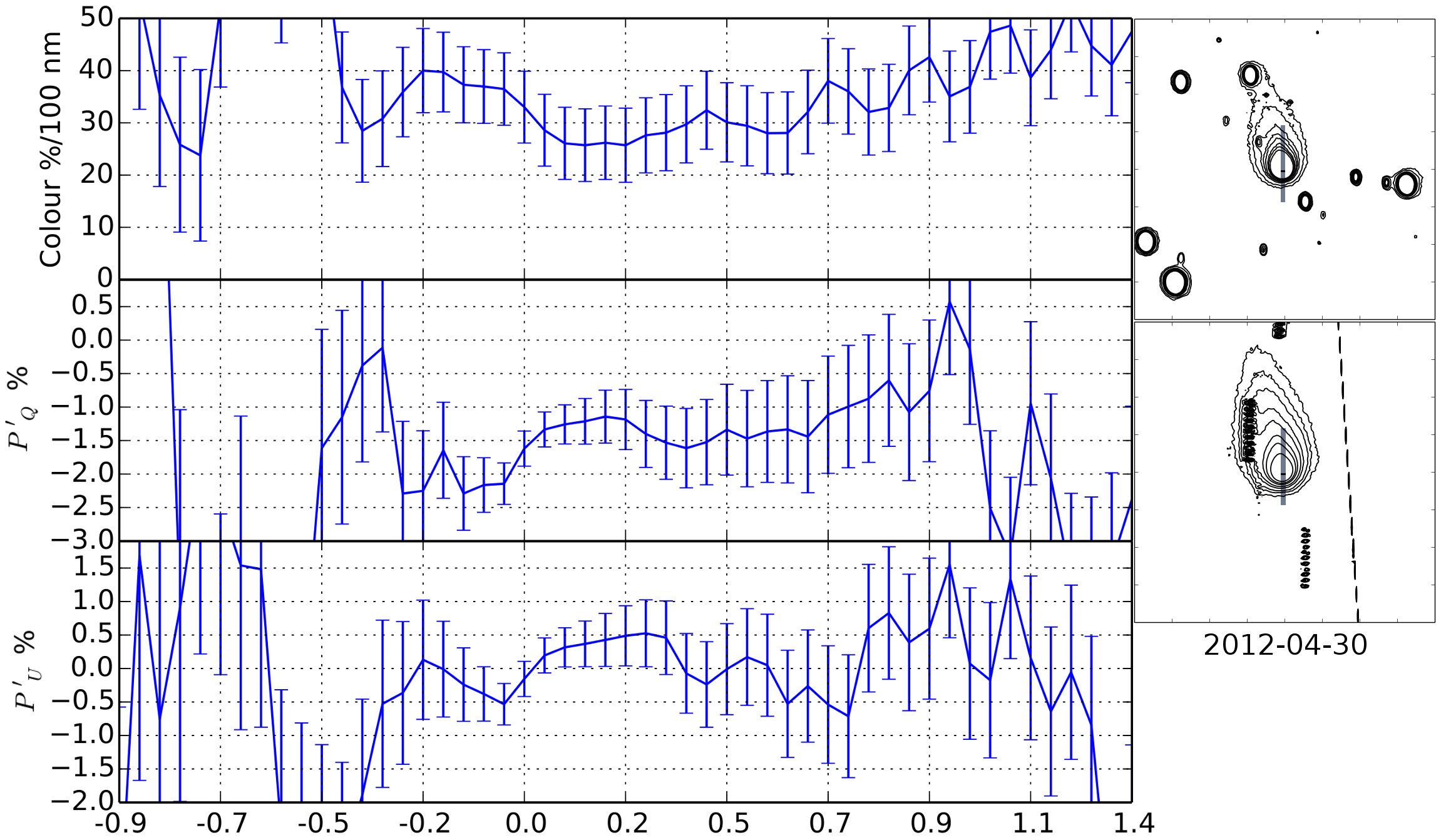}}{\includegraphics[scale=0.5]{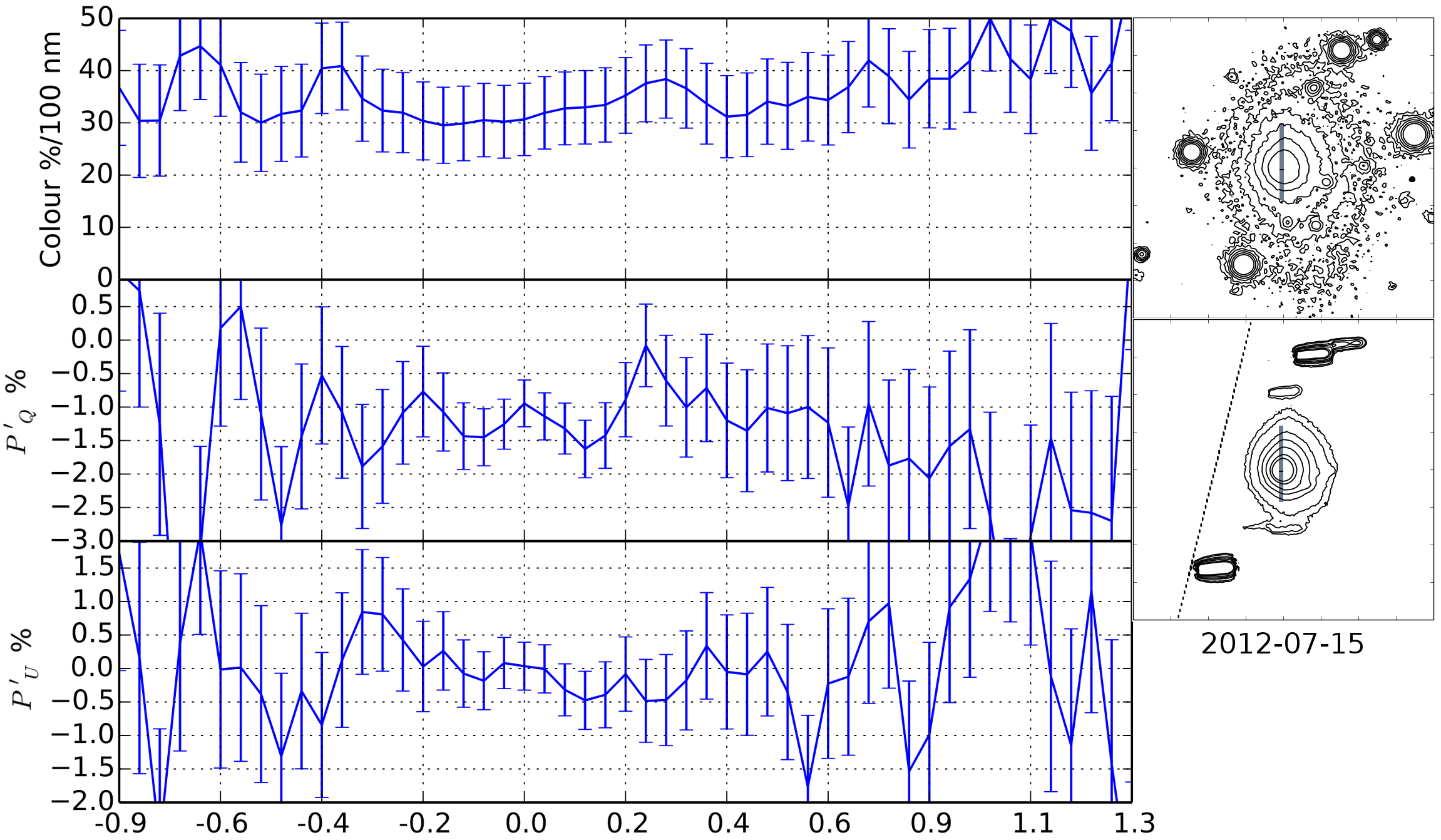}}}
\resizebox{\hsize}{!}{{\includegraphics[scale=0.5]{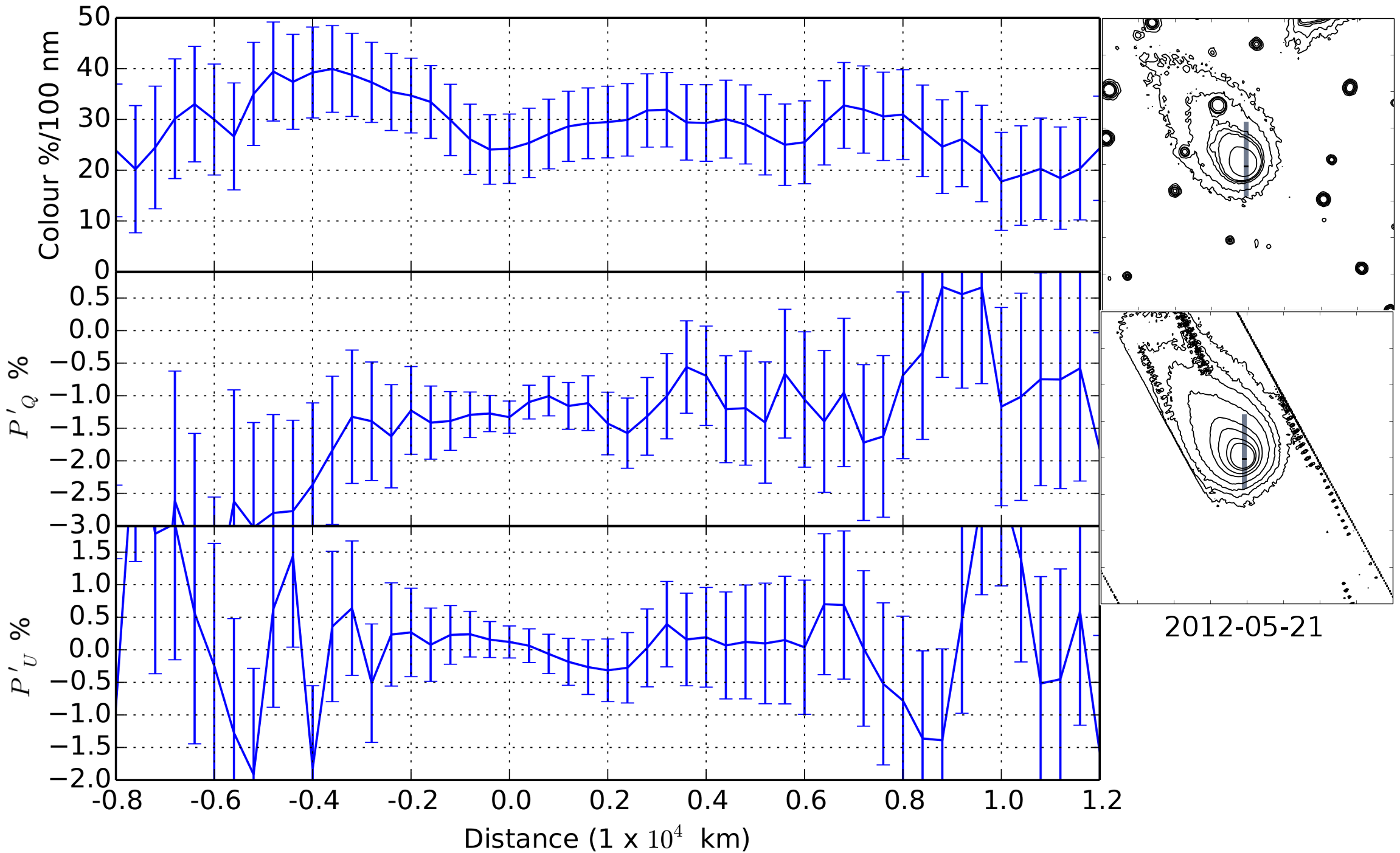}}{\includegraphics[scale=0.5]{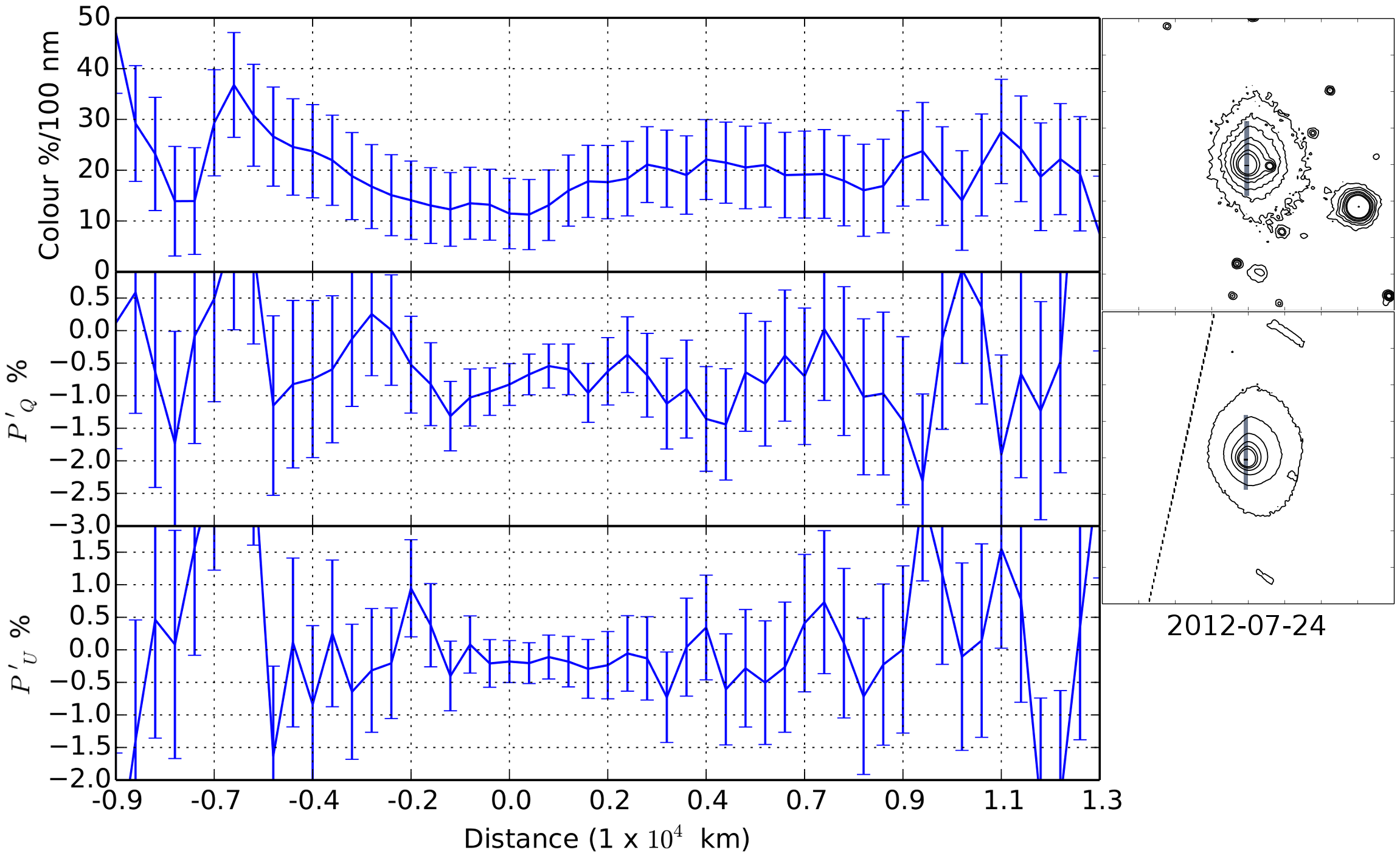}}}

\caption{Scans of the comet 152P colour, $P'_Q$, and $P'_U$ along the solar anti-solar direction. Positive distance is in the anti-solar direction, and
negative distance is in the solar direction. Zero is at the photometric centre of the comet. The upper contour plot shows the intensity of the comet in the photometric images and the lower contour plot shows the photometric intensity of the comet in the polarimetric images; the levels are arbitrary. The small grey shaded area in each contour plot shows the area scanned.}
\label{fig:scan}
\end{figure*}

In Fig. \ref{fig:pol_phase} we present the aperture polarimetry for 67P, 74P, and 152P. The solid black line is a best fit representation of previously observed \textit{R}-band polarimetric data of active comets taken from the polarimetric comet database \citep{Kiselev2006b}. The best fit we used is a trigonometric function that was introduced by \citep{Lumme1993} and outlined by \citet{Penttila2005} and is defined as
\begin{center}
\begin{eqnarray}
P(\alpha) = b\left(\rm{sin}\, \alpha \right)^{c_1}\lbrack \rm{cos} \,(\alpha /2)\rbrack ^{c_2}\, \rm{sin}\left(\alpha - \alpha _0\right)\; ,
\end{eqnarray}
\end{center}
\noindent
where $b$, $c_1$, $c_2$, and $\alpha _0$ are free parameters. Each of these four parameters has an effect on the shape of the fitted phase curve as stated in \citet{Penttila2005}. The parameter \textit{b} is mainly connected to the amplitude of polarisation with a physically acceptable range of values between 0 and 100 if we express the polarisation \textit{P}($\alpha$) in per cent. The parameter \textit{$\alpha _{0}$} is the inversion angle where negative polarisation turns into positive polarisation. This parameter can range between 0 and 180$^\circ$ but typically it is less than 30$^\circ$. The two powers \textit{c$_1$} and \textit{c$_2$} influence the shape of the phase curve. The parameter \textit{c$_1$} mainly affects the position of the minimum, while \textit{c$_2$} has an influence on the maximum and the asymmetry of the curve. Both these parameters should have positive values. This equation can be used for extrapolation only within a phase angle range where well-distributed data points are available. \\ \indent
As we can see from Fig. \ref{fig:pol_phase} all three comets show a very similar polarimetric phase relationship. However polarimetric measurements of comet 67P have a large error owing to a poor signal-to-noise ratio, so it is difficult to make any firm conclusions from this data. However both comets 74P and 152P show a slightly different polarimetric behaviour compared to the best fit curve. They both exhibit an excess in negative polarisation at phase angles < 3$^\circ$. The deviation from the best fit at the small phase angles for comet 152P corresponds to the data points on the 21 and 23 May 2012, which showed the presence of activity in the colour maps (see Fig. \ref{fig:cmap_152P}). We are unable to state the cause for the deviation at small phase angles for comet 74P as we lack both good polarimetric and colour data for these nights, but it is likely due to activity or statistical scatter around the best fit.  Also presented in this graph are other JFCs observed in wavelength range that is similar to our observations. These observations are within the error of both our observations and the best fit line. However there are points that deviate from the best fit and this is likely due to the observations being in slightly different wavelengths or at different heliocentric distances. Therefore, when we compare comet polarimetry in broadband filters we must take the values with a pinch of salt as there are many factors that can influence the amount of polarisation measured (gas contamination, outbursts of activity, etc.).

\section{Discussion and conclusions}

To investigate whether we see any polarimetric and colour trends along the solar anti-solar direction we have taken scans through the photometric centre of the comets. Since comet 74P has few simultaneous photometric and polarimetric observations and 67P has no colour information, we only carry out this analysis for comet 152P. The scans are presented in Fig. \ref{fig:scan} along with a contour plot showing exactly which region of the coma is being scanned.

\begin{figure}[tbp!]
\resizebox{\hsize}{!}{\includegraphics[scale=0.5]{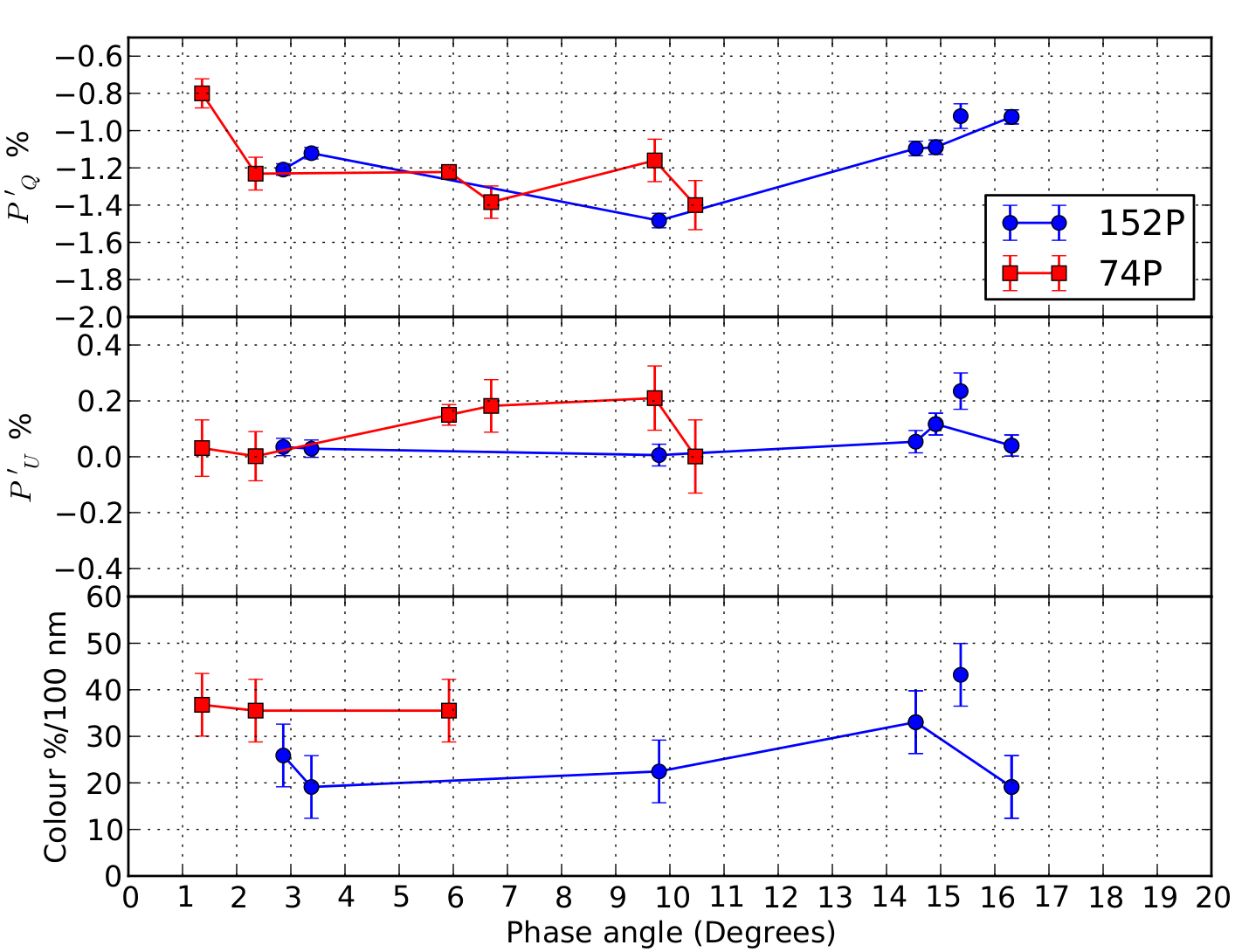}}
\caption{$P'_Q$, $P'_U$, and colour changes as a function of phase angle for comets 74P and 152P.}
\label{fig:152P_pol_col}
\end{figure}

For 152P we do not see any colour or polarisation trends with cometocentric distance (see Fig. \ref{fig:scan}). For the majority of the observations of 152P we see an average colour $\sim$ 30-40$\% / 100\,\rm{nm}$, while the polarisation although it does vary from epoch to epoch due to changing phase angle remains constant across the coma. We fail to see any trend from blue to red that would suggest the sublimation of water ice in our observations which supports our near-infrared observations where we see no water absorption \citep{Kolokolova2001}. The exception is possibly the scan on the 23 May 2012 where we see a change in colour about the photometric centre of the comet; however, we do not see a corresponding trend in polarisation, which instead we expect to see. The reddish colour and the lack of water ice in the coma suggest that the dust is possibly made up of dirty ice or organic particles. Again, if this is the case we do not see any trends that suggest the decomposition or fragmentation of the dust that would show itself as a change in colour from red to blue as the particles get smaller and become more efficient scatterers. The lack of any trends could be a special feature of distant Jupiter family comets. At these large heliocentric distances, sublimation and fragmentation of the dust particles are very slow because the solar radiation is less intense, and this may be the reason why we do not see any gradual cometocentric changes in colour and polarisation. \citet{Beer2006} modelled the lifetime of dust particles at a similar heliocentric distance and found that dirty ice particles live a maximum of 2-3 hours before the ice sublimates. On the other hand, \citet{Beer2006} showed that grains of pure water ice can survive many years before sublimation; however, since the probability of getting pure water ice grains is very small and the fact that we have found no water ice in the near-infrared spectrum suggests that this is not the cause of our lack of variation. \\ \indent
If we consider how the polarisation and colour properties change as a function of phase angle (Fig. \ref{fig:152P_pol_col}) we can see two interesting anomalies for comet 152P. The first occurs at the two small phase angles of 2.9$^\circ$ and 3.4$^\circ$ where we see both an increase in $P'_Q$ in absolute terms and an increase in colour. The increase in $P'_Q$ is the opposite behaviour that we expect to see at these phase angles, indicating that something unusual happened on this night. The combination of this with an increase in colour suggest that either at the phase angle 2.9$^\circ$ the comet produced more ice than usual, or that at a phase angle of 3.4$^\circ$ it produces less ice than usual. The second anomaly occurs at large phase angles between 15 and 24 July 2012 when the colour shows a large decrease. We note that the data point at phase angle 15.4$^\circ$ has to be ignored as it corresponds to observations taken 3 months earlier, which is why this data point is not connected to the others. The decrease in colour is accompanied by a decrease in absolute terms of Stokes $P'_Q$; however, this decrease in polarisation is expected with the change in phase angle. Since the colour maps for this night are influenced by different seeing conditions we are unable to tell if there is a jet or if an outburst has occurred around the time of observation.\\ \indent
Very little can be said about 74P; the lack of quasi-simultaneous colour and polarisation measurements means we cannot draw any firm conclusions. There are two uncharacteristic dips in $P'_Q$ between 5-7$^\circ$ and 9-11$^\circ$. The first is due to the observations being two months apart and are not a good comparison. However the second dip is a little more interesting as the observations are taken a week apart and clearly something unusual occurred in the coma that has caused an increase of $\sim 0.25\,\%$ in absolute terms. This could be caused by smaller particles being emitted from the nucleus or the sublimation of ice \citep{Kolokolova2001}; however without colour information we cannot understand which.\\ \indent
Comparing the $Af\rho$ measurements for these three comets at these large heliocentric distances 74P appears to be the most active comet, closely followed by 152P, with 67P showing the least activity. The large difference in the amount of activity shown by 74P and 152P compared to 67P could suggest that 74P and 152P are relative newcomers to the inner solar system and still have a large reservoir of volatiles present on the surface of their nuclei. 

\begin{acknowledgements}
Part of this work was supported by the COST Action MP1104 ``Polarization as a tool to study the Solar System and beyond''. Silvia Protopapa gratefully thanks NASA's SSO Planetary Astronomy Program (grant $\#$NNX15AD99G) for funding that supported this work. Karri Muinonen's research is supported, in part, by the Academy of Finland grant No. 1257966.

\end{acknowledgements}


\bibliographystyle{aa}
\bibliography{paper}

\begin{thebibliography}{44}
\expandafter\ifx\csname natexlab\endcsname\relax\def\natexlab#1{#1}\fi

\bibitem[{{A'Hearn}(2011)}]{Ahearn2011}
{A'Hearn}, M.~F. 2011, in EPSC-DPS Joint Meeting 2011, 316

\bibitem[{{A'Hearn} {et~al.}(1986){A'Hearn}, {Hoban}, {Birch}, {Bowers},
  {Martin}, \& {Klinglesmith}}]{Ahearn1986}
{A'Hearn}, M.~F., {Hoban}, S., {Birch}, P.~V., {et~al.} 1986, \nat, 324, 649

\bibitem[{{A'Hearn} {et~al.}(1984){A'Hearn}, {Schleicher}, {Millis}, {Feldman},
  \& {Thompson}}]{Ahearn1984}
{A'Hearn}, M.~F., {Schleicher}, D.~G., {Millis}, R.~L., {Feldman}, P.~D., \&
  {Thompson}, D.~T. 1984, \aj, 89, 579

\bibitem[{Appenzeller(1967)}]{Appenzeller1967}
Appenzeller, I. 1967, \pasp, 79, 136

\bibitem[{Appenzeller {et~al.}(1998)Appenzeller, Fricke, F\"{u}rtig,
  G\"{a}ssler, H\"{a}fner, Harke, Hess, Hummel, J\"{u}rgens, Kudritzki, Mantel,
  Meisl, Muschielok, Nicklas, Rupprecht, Seifert, Stahl, Szeifert, \&
  Tarantik}]{Appenzeller1998}
Appenzeller, I., Fricke, K., F\"{u}rtig, W., {et~al.} 1998, The Messenger, 94,
  1

\bibitem[{Bagnulo {et~al.}(2011)Bagnulo, Belskaya, Boehnhardt, Kolokolova,
  Muinonen, Sterzik, \& Tozzi}]{Bagnulo2011}
Bagnulo, S., Belskaya, I., Boehnhardt, H., {et~al.} 2011, \jqsrt, 112, 2059

\bibitem[{{Bagnulo} {et~al.}(2006){Bagnulo}, {Boehnhardt}, {Muinonen},
  {Kolokolova}, {Belskaya}, \& {Barucci}}]{Bagnulo2006}
{Bagnulo}, S., {Boehnhardt}, H., {Muinonen}, K., {et~al.} 2006, \aap, 450, 1239

\bibitem[{Bagnulo {et~al.}(2009)Bagnulo, Landolfi, Landstreet, {Landi
  Degl’Innocenti}, Fossati, \& Sterzik}]{Bagnulo2009}
Bagnulo, S., Landolfi, M., Landstreet, J.~D., {et~al.} 2009, \pasp, 121, 993

\bibitem[{{Beer} {et~al.}(2006){Beer}, {Podolak}, \& {Prialnik}}]{Beer2006}
{Beer}, E.~H., {Podolak}, M., \& {Prialnik}, D. 2006, \icarus, 180, 473

\bibitem[{{Beisser} \& {Drechsel}(1992)}]{Beisser1992}
{Beisser}, K. \& {Drechsel}, H. 1992, \apss, 191, 1

\bibitem[{{Boehnhardt} \& {Birkle}(1994)}]{Boehnhardt1994}
{Boehnhardt}, H. \& {Birkle}, K. 1994, \aaps, 107, 101

\bibitem[{{Boehnhardt} {et~al.}(2008){Boehnhardt}, {Tozzi}, {Bagnulo},
  {Muinonen}, {Nathues}, \& {Kolokolova}}]{Boehnhardt2008}
{Boehnhardt}, H., {Tozzi}, G.~P., {Bagnulo}, S., {et~al.} 2008, A$\&$A, 489,
  1337

\bibitem[{Bonnet {et~al.}(2004)Bonnet, Conzelmann, Delabre, Donaldson, Fedrigo,
  Hubin, Kissler-Patig, Lizon, Paufique, Rossi, Stroebele, \&
  Tordo}]{Bonnet2004}
Bonnet, H., Conzelmann, R., Delabre, B., {et~al.} 2004, 130--138

\bibitem[{{Eaton} {et~al.}(1988){Eaton}, {Scarrott}, \&
  {Warren-Smith}}]{Eaton1988}
{Eaton}, N., {Scarrott}, S.~M., \& {Warren-Smith}, R.~F. 1988, \icarus, 76, 270

\bibitem[{Eisenhauer(2003)}]{Eisenhauer2003}
Eisenhauer, F. 2003, \aap, 4841, 1548

\bibitem[{{Finson} \& {Probstein}(1968)}]{Finson1968}
{Finson}, M.~J. \& {Probstein}, R.~F. 1968, \apj, 154, 327

\bibitem[{{Furusho} {et~al.}(1999){Furusho}, {Suzuki}, {Yamamoto}, {Kawakita},
  {Sasaki}, {Shimizu}, \& {Kurakami}}]{Furusho1999}
{Furusho}, R., {Suzuki}, B., {Yamamoto}, N., {et~al.} 1999, \pasj, 51, 367

\bibitem[{{Hadamcik} \& {Levasseur-Regourd}(2003)}]{Hadamcik2003}
{Hadamcik}, E. \& {Levasseur-Regourd}, A.~C. 2003, \jqsrt, 79, 661

\bibitem[{{Hadamcik} {et~al.}(1997){Hadamcik}, {Levassuer-Regourd}, \&
  {Renard}}]{Hadamcik1997}
{Hadamcik}, E., {Levassuer-Regourd}, A.~C., \& {Renard}, J.~B. 1997, Earth Moon
  and Planets, 78, 365

\bibitem[{{Hadamcik} {et~al.}(2010){Hadamcik}, {Sen}, {Levasseur-Regourd},
  {Gupta}, \& {Lasue}}]{Hadamcik2010}
{Hadamcik}, E., {Sen}, A.~K., {Levasseur-Regourd}, A.~C., {Gupta}, R., \&
  {Lasue}, J. 2010, \aap, 517, A86

\bibitem[{{Jewitt}(2004)}]{Jewitt2004}
{Jewitt}, D. 2004, \aj, 128, 3061

\bibitem[{{Jockers} {et~al.}(2005){Jockers}, {Kiselev}, {Bonev}, {Rosenbush},
  {Shakhovskoy}, {Kolesnikov}, {Efimov}, {Shakhovskoy}, \&
  {Antonyuk}}]{Jockers2005}
{Jockers}, K., {Kiselev}, N., {Bonev}, T., {et~al.} 2005, \aap, 441, 773

\bibitem[{{Kiselev} {et~al.}(2006){Kiselev}, {Velichko}, S., {Rosenbush}, \&
  {Kikuchi}}]{Kiselev2006b}
{Kiselev}, N., {Velichko}, S., {Jockers}, K., {Rosenbush}, V., \& {Kikuchi},
  S., E. 2006, EAR-C-COMPIL-5-DB-COMET-POLARIMETRY-V1.0. NASA Planetary Data
  System

\bibitem[{{Kolokolova} {et~al.}(2001){Kolokolova}, {Jockers}, {Gustafson}, \&
  {Lichtenberg}}]{Kolokolova2001}
{Kolokolova}, L., {Jockers}, K., {Gustafson}, B.~{\AA}.~S., \& {Lichtenberg},
  G. 2001, \jgr, 106, 10113

\bibitem[{Lamy {et~al.}(2011)Lamy, Toth, Weaver, A'Hearn, \& Jorda}]{Lamy2011}
Lamy, P.~L., Toth, I., Weaver, H.~a., A'Hearn, M.~F., \& Jorda, L. 2011,
  \mnras, 412, 1573

\bibitem[{{Levasseur-Regourd} {et~al.}(1996){Levasseur-Regourd}, {Hadamcik}, \&
  {Renard}}]{Levasseur-Regourd1996}
{Levasseur-Regourd}, A.~C., {Hadamcik}, E., \& {Renard}, J.~B. 1996, \aap, 313,
  327

\bibitem[{Lowry \& Fitzsimmons(2001)}]{Lowry2001}
Lowry, S.~C. \& Fitzsimmons, A. 2001, \aap, 365, 204

\bibitem[{Lowry {et~al.}(1999)Lowry, Fitzsimmons, Cartwright, \&
  Williams}]{Lowry1999}
Lowry, S.~C., Fitzsimmons, A., Cartwright, I.~M., \& Williams, I.~P. 1999,
  \aap, 659, 649

\bibitem[{{Lowry} {et~al.}(2003){Lowry}, {Fitzsimmons}, \&
  {Collander-Brown}}]{Lowry2003}
{Lowry}, S.~C., {Fitzsimmons}, A., \& {Collander-Brown}, S. 2003, \aap, 397,
  329

\bibitem[{{Lumme} \& {Muinonen}(1993)}]{Lumme1993}
{Lumme}, K. \& {Muinonen}, K.~O. 1993, LPI Contributions, 810, 194

\bibitem[{{Manset} \& {Bastien}(2000)}]{Manset2000}
{Manset}, N. \& {Bastien}, P. 2000, \icarus, 145, 203

\bibitem[{Markkanen {et~al.}(2015)Markkanen, Penttil\"{a}, Peltoniemi, \&
  Muinonen}]{Markkanen2015}
Markkanen, J., Penttil\"{a}, A., Peltoniemi, J., \& Muinonen, K. 2015,
  Planetary and Space Science, (in press)

\bibitem[{Moehler {et~al.}(2010)Moehler, Freudling, \& M\o~ller}]{Moehler2010}
Moehler, S., Freudling, W., \& M\o~ller, P. 2010, \pasp, 122

\bibitem[{Muinonen(2004)}]{Muinonen2004}
Muinonen, K. 2004, Waves in random Media, 14, 365

\bibitem[{{Muinonen} {et~al.}(2012){Muinonen}, {Mishchenko}, {Dlugach},
  {Zubko}, {Penttil{\"a}}, \& {Videen}}]{Muinonen2012}
{Muinonen}, K., {Mishchenko}, M.~I., {Dlugach}, J.~M., {et~al.} 2012, \apj,
  760, 118

\bibitem[{Myers(1985)}]{Myers1985}
Myers, R. 1985, Icarus, 216, 206

\bibitem[{Myers \& Nordsieck(1984)}]{Myers1984}
Myers, R. \& Nordsieck, K. 1984, Icarus, 439

\bibitem[{Penttil\"{a} {et~al.}(2005)Penttil\"{a}, Lumme, Hadamcik, \&
  Levasseur-Regourd}]{Penttila2005}
Penttil\"{a}, A., Lumme, K., Hadamcik, E., \& Levasseur-Regourd, A.-C. 2005,
  Astronomy and Astrophysics, 432, 1081

\bibitem[{{Protopapa} {et~al.}(2014){Protopapa}, {Sunshine}, {Feaga}, {Kelley},
  {A'Hearn}, {Farnham}, {Groussin}, {Besse}, {Merlin}, \& {Li}}]{Protopapa2014}
{Protopapa}, S., {Sunshine}, J.~M., {Feaga}, L.~M., {et~al.} 2014, \icarus,
  238, 191

\bibitem[{{Renard} {et~al.}(1996){Renard}, {Hadamcik}, \&
  {Levasseur-Regourd}}]{Renard1996}
{Renard}, J.-B., {Hadamcik}, E., \& {Levasseur-Regourd}, A.-C. 1996, \aap, 316,
  263

\bibitem[{{Renard} {et~al.}(1992){Renard}, {Levasseur-Regourd}, \&
  {Dollfus}}]{Renard1992}
{Renard}, J.~B., {Levasseur-Regourd}, A.~C., \& {Dollfus}, A. 1992, Annales
  Geophysicae, 10, 288

\bibitem[{{Rosenbush} {et~al.}(1994){Rosenbush}, {Rosenbush}, \&
  {Dement'ev}}]{Rosenbush1994}
{Rosenbush}, V.~K., {Rosenbush}, A.~E., \& {Dement'ev}, M.~S. 1994, \icarus,
  108, 81

\bibitem[{{Yang} \& {Sarid}(2010)}]{Yang2010}
{Yang}, B. \& {Sarid}, G. 2010, \iaucirc, 9139, 2

\bibitem[{Zubko {et~al.}(2015)Zubko, Videen, Hines, Shkuratov, Kaydash,
  Muinonen, Knight, Sitko, Lisse, Mutchler, Wooden, Li, \&
  Kobayashi}]{Zubko2015}
Zubko, E., Videen, G., Hines, D.~C., {et~al.} 2015, Planetary and Space
  Science, (in press)

\end{thebibliography}
%

\end{document}